\documentclass[fleqn,usenatbib]{mnras}

\usepackage[T1]{fontenc}

\usepackage{eso-pic}

\usepackage{graphicx}	
\usepackage{amsmath}	
\usepackage{amssymb}	
\usepackage[shortlabels]{enumitem}

\usepackage{multicol}

\title[MaNGA: stellar gradients in ellipticals]{Galaxy properties as revealed by MaNGA. I. Constraints on IMF and M$_*$/L gradients in ellipticals}

\author[ Dom\'inguez S\'anchez et al.]{
\parbox{\textwidth}{
  H.~Dom\'{i}nguez S{\'a}nchez$^1$\thanks{E-mail: \texttt{\rm \texttt{helenado@sas.upenn.edu}}}, M.~Bernardi$^{1}$\thanks{\texttt{\rm \texttt{bernardm@sas.upenn.edu}}}, J.~R.~Brownstein$^2$, N.~Drory$^3$ and R.~K.~Sheth$^{1}$}\\
 \vspace{0.cm}\\~\\
$^{1}$ Department of Physics and Astronomy, University of Pennsylvania, Philadelphia, PA 19104, USA\\
$^{2}$ Department of Physics and Astronomy, University of Utah, 115 S. 1400 E., Salt Lake City, UT 84112, USA\\
$^{3}$ McDonald Observatory, The University of Texas at Austin, 1 University Station, Austin, TX 78712, USA\\
}

\pubyear{2019}

\begin{document}
\label{firstpage}
\pagerange{\pageref{firstpage}--\pageref{lastpage}}
\maketitle

\begin{abstract}
We estimate ages, metallicities, $\alpha$-element abundance ratios and stellar initial mass functions of elliptical (E) and S0 galaxies from the MaNGA-DR15 survey.  We stack spectra and use a variety of single stellar population synthesis models to interpret the absorption line strengths in these spectra.  We quantify how these properties vary across the population, as well as with galactocentric distance. This paper is the first of a series and is based on a sample of pure elliptical galaxies at $z \le 0.08$. We confirm previous work showing that IMFs in Es with the largest luminosity ($L_r$) and central velocity dispersion ($\sigma_0$) appear to be increasingly bottom heavy towards their centres.  For these galaxies the stellar mass-to-light ratio decreases at most by a factor of 2 from the central regions to $R_e$. In contrast, for lower $L_r$ and $\sigma_0$ galaxies, the IMF is shallower and $M_*/L_r$ in the central regions is similar to the outskirts, although quantitative estimates depend on assumptions about element abundance gradients.
 Accounting self-consistently for these gradients when estimating both $M_*$ and $M_{\rm dyn}$ brings the two into good agreement: gradients reduce $M_{\rm dyn}$ by $\sim 0.2$ dex while only slightly increasing the $M_*$ inferred using a Kroupa IMF. This is a different resolution of the $M_*$-$M_{\rm dyn}$ discrepancy than has been followed in the recent literature where $M_*$ of massive galaxies is increased by adopting a Salpeter IMF throughout the galaxy while leaving $M_{\rm dyn}$ unchanged. A companion paper discusses how stellar population differences are even more pronounced if one separates slow from fast rotators.
\end{abstract}




\section{Introduction}\label{sec:intro}
The spectrum of a galaxy is a linear combination of the spectra of its stars.  Stellar spectra depend on mass, age and chemical composition, so a galaxy's spectrum encodes information about its stellar mass, the mean age of its stars, more detailed information about its star formation history (single-burst, episodic, time-scales), its chemical composition (metallicity, $\alpha$-element abundance ratios, dust content) and the IMF (which describes the mix of stars formed in each episode).  Decades of work have shown how to decode this `fossil record' \citep{Worthey1994, Trager1998, Kauffmann2003, Bernardi2006, Panter2007, Graves2009}.  We now know that the mix of stars in galaxies varies across the galaxy population, over and above the obvious variations with morphology across the Hubble sequence.

In this and the following papers of this series, we focus almost exclusively on early-type galaxies (Es and S0s) since late-type galaxies (Spirals) have gas and dust which complicate the spectral analysis.  In fact, in this paper, we only study Es (we study S0s in paper III -- in prep.). However, even amongst Es, the stellar populations depend on other global properties such as velocity dispersion, luminosity, size, etc. \citep{Trager1998,Thomas2005,Bernardi2005,Bernardi2006,Bernardi2011}.

In addition to varying across the early-type galaxy population, there are stellar population gradients even within a single galaxy.  While color gradients have been seen for some time \cite[][and references therein]{Wu2005, LaBarbera2012}, spectroscopic gradients \cite[e.g.][]{Davies1993} have begun to receive more attention.  Recent studies of gradients in small samples of galaxies used long-slit spectroscopy \cite[e.g.][and references therein]{SanchezBlazquez2007, Spolaor2009, Spolaor2010, Koleva2011}, finding strong color gradients mostly driven by metallicity, in agreement with N-body hydrodynamical simulations \citep{Tortora2011}. 

Following these pioneering studies we are now on the cusp of a revolution in the study of gradients.  This is because of the advent of Integral Field Units (IFUs) which provide spatially resolved spectroscopy for galaxies.  The SAURON \citep{Emsellem2004} and ATLAS$^{\rm 3D}$ \citep{Cappellari2011} surveys of a decade ago provided estimates of kinematic gradients (i.e., rotation curves and velocity dispersion profiles) and stellar population gradients in tens to hundreds of early-type galaxies, each sampled by tens to hundreds of spaxels. Stellar population gradients have been also studied in more recent IFU surveys such as CALIFA \cite[CALAR Alto Integral Field Area;][]{Sanchez2012} and SAMI \cite[Sydney Australian Astronomical Observatory Multi-object IFS;][]{Croom2012}, among others. At the moment, the MaNGA survey (Mapping Nearby Galaxies at Apache Point Observatory; \citealt{Bundy2015, Law2015, Wake2017, Westfall2019}) provides this, as well as sufficiently high quality spectra to determine chemical abundance gradients, for about two thousand early-types, each sampled by hundreds to thousands of spaxels.

Regarding the stellar populations of  early-type galaxies derived from IFUs, there is a general agreement on the existence of strong negative metallicity gradients \cite[e.g.][]{Scott2009, GonzalezDelgado2014, GonzalezDelgado2015, Greene2015,Boardman2017,vdSande2018,Li2018, Parikh2019,Zhuang2019,Zibetti2019,Ferreras2019}, while the consensus about age gradients is less robust -- the majority of the results show flat to mild negative profiles ($\sim$0.1 dex or 25$\%$ change), but some authors show evidence of stronger gradients \cite[e.g.][]{GonzalezDelgado2015} or even positive gradients  \citep[e.g.][]{Kuntschner2010, Zibetti2019}. Different analysis methods make a fair direct comparison difficult. 

Why do gradients matter?  Perhaps the crudest measure of the inhomogeneous distribution within a galaxy is the correlation between stellar population and distance from the center: the stellar population gradient.  This is expected to constrain and separate its star formation history from its assembly history \cite[e.g. inside-out or outside-in?  in-situ or ex-situ?][]{Larson1974,White1980,Carlberg1984,Pipino2006}.  But, most importantly, stellar population gradients affect how we estimate the stellar mass of a galaxy (\citealt{Bernardi2018b} and references therein).  Stellar masses are the bricks which build the bridge that connects galaxy formation models to dark matter halos and hence to cosmology.  Reliable stellar mass estimates are crucial for reconciling the stellar mass density today with that inferred from the integrated star formation rate.  They also impact discussions of the efficiency of feedback from active galactic nuclei in the quasar and/or radio modes, and the response of the dark matter halo to galaxy formation.

There are currently two distinct methods for estimating the mass in stars.  One exploits the fact that the light from a galaxy is simply a linear combination of the light from its stars.  So, by finding that linear combination of stellar spectra -- each with its own mass-to-light ratio (e.g., in the optical, young massive stars have small mass-to-light ratios) -- which best-fits the observed spectrum, one can constrain the overall mass-to-light ratio of the galaxy.  We will refer to this as $M_*/L_r$.  In this approach, the stellar mass is obtained by multiplying the estimated $M_*/L_r$ by the observed $L_r$ to yield $M_*$.

The other method uses the motions of the stars to constrain their collective mass.  Typically, this estimate is based on the Jeans equation, and requires some assumption about the distribution of dark matter, which is expected to dominate the mass far from the center, and some knowledge of the orbital anisotropies.  We will refer to this mass estimate as $M_*^{\rm dyn}$.  The most widely cited dynamical mass estimates \citep{Cappellari2013a, Cappellari2013b} are based on the additional assumption that the shape of the light profile traces the shape of the stellar mass profile -- i.e., that the stellar mass-to-light ratio is constant (the total mass-to-light ratio is not constant, of course, because dark matter dominates on large scales).  This is an assumption of convenience -- it has no physical motivation.  In this approach, the value of $M_*^{\rm dyn}/L_r$ is determined by matching the observed velocity dispersion, rather than by matching detailed features of the spectrum.

Thus, roughly speaking, $M_*^{\rm dyn}$ depends on the shape of the fitted light profile, but not on its amplitude, whereas the stellar population based estimate $M_*$ depends more on the total light $L_r$ than on the detailed profile shape.  In this respect, comparing $M_*$ and $M_*^{\rm dyn}$ is attractive, since it nicely separates out two distinct sources of uncertainty.  In addition, $M_*$ depends on assumptions about the dust content, the IMF of the stellar population and so on, whereas $M_*^{\rm dyn}$ (in principle) does not.  On the other hand, $M_*$ does not depend on the dark matter distribution or orbital anisotropies, whereas $M_*^{\rm dyn}$ does.  These two estimates are thought to provide two distinct routes -- albeit with rather different systematic biases -- to the same underlying physical quantity.

It has been known for some time that, if one assumes the same IMF within a galaxy and across the population, then $M_*^{\rm dyn}/M_*^{\rm SP}$ varies across the early-type population \cite[e.g.][]{Bender92, Bernardi2003, SB09, Cappellari2013b, Li17, Bernardi2018a}. This discrepancy between the $M_{\rm dyn}$ and $M_*$ estimates has driven many to conclude that the IMF is Salpeter, or even super-Salpeter, in massive galaxies \cite[e.g.][]{Dutton2012,Wegner2012,Tortora2013,Lasker2013,Cappellari2013b,Tortora2014, Li17, Bernardi2018a}.  More generally, the relation between  $M_*^{\rm dyn}$ and $M_*^{\rm SP}$ has been used to constrain the form of the IMF in galaxies \cite[e.g.][]{Smith2014, Lyubenova2016}. In addition, because strong lensing measurements provide estimates of the total mass, combining them with $M_*^{\rm dyn}$ should provide complementary constraints on the IMF \cite[e.g.,][]{Treu2010,Thomas2011,Newman2017A,Oldham&Auger2018}.  However, these studies suffer from degeneracies between the assumed dark matter profile and the stellar mass-to-light ratio which can bias the IMF estimate.  Indeed, recent work shows that when information from weak gravitational lensing is included, then the previously claimed evidence for bottom heavy IMFs from such studies is weakened dramatically \citep{Sonnenfeld2018}.

Of course, if gradients are important, then the stellar light and matter profiles have different shapes. This implies that the Jeans equation-based $M_*^{\rm dyn}$ estimates currently in the literature are incorrect, because they are based on the assumption that the true stellar mass-to-light ratio is independent of distance from the center.  While this has been known for some time, the conventional wisdom had been that this is a small effect.  This is based on analyses which assume that the IMF within a galaxy is fixed, and in this case the stellar population derived $M_*/L_r$ is about 20\% larger in the center than it is within $R_e/2$ (i.e. half the projected half-light radius).  However, recent work suggests that if the IMF is also allowed to vary when fitting the spectrum, then the $M_*/L_r$ difference may be as large as a factor of 3 \citep{vD2017}.  As \cite{Bernardi2018b} note, if $M_*/L_r$ gradients really are this large, then $M_*^{\rm dyn}$ estimates currently in the literature must be revised downwards.

It is not obvious that a factor of 3 is realistic.  The \cite{vD2017} analysis was based on a handful of objects. This is comparable to the sample sizes in other recent studies of IMF gradients \citep{MN2015, MN2015b, MN2015c, LaBarbera2016, Vaughan2018a, Vaughan2018b, Sarzi2018, Oldham&Auger2018}. While all consistently find that the central regions of galaxies favour a bottom-heavy IMF, it is important to extend these analysis to larger samples in order to have more robust statistics. The samples are small in part because determining the IMF is not an easy task: changes in the IMF only lead to rather subtle effects on the spectrum \citep{CvD2012, LaBarbera2013, MN2015, LaBarbera2016, TW2017}, some of which are degenerate with other stellar population differences (e.g., star formation histories, chemical abundances, etc.). High signal-to-noise spectra are required to disentangle IMF gradients from these other effects. 

The MaNGA survey \citep{Bundy2015} provides an IFU sample that is an order of magnitude larger compared to what was previously available. \cite{Parikh2018} and \cite{Zhou2019} describe a first attempt at estimating IMF gradients in the MaNGA ETGs sample based on a stacking analysis of the spectra.  However, the sample size available at the time, roughly half of the the current MaNGA data release, meant that they were only able to measure rather approximate trends across the population.  They divided the sample in bins of stellar mass, without separating Es and S0s or taking into account redshift evolution effects.  Moreover, about a quarter of the objects in the \cite{Parikh2018} sample are not classified as early-types in the  MaNGA Deep Learning Morphology Value Added Catalogue  (MDLM-VAC; \citealt{Fischer2019}).  Therefore, the main goal of the present study is to estimate gradients in ages, metallicities and abundance ratios when one allows for radial variations in the IMF, and to use these to estimate gradients in $M_*/L_r$ in a larger and cleaner sample of MaNGA Es, which we divide into bins based on the absolute magnitude $M_r$ and central velocity dispersion $\sigma_0$ before stacking their spectra. A companion paper (\citealt{BDS2019}, hereafter Paper~II) extends this analysis by further sub-dividing the sample based on $R_e$ and kinematics (i.e. slow from fast rotators).  A third paper in this series studies the properties of S0 galaxies -- in prep. 

This paper is organized as follows. Section~\ref{sec:data} describes the dataset, and the gradients we measure in the spectra.  Section~\ref{sec:results} interprets these measurements in terms of ages, metallicities, $\alpha$-element abundance ratios, stellar initial mass functions and $M_*/L_r$ gradients using stellar population synthesis models. Section~\ref{sec:MdMs} compares stellar population and dynamical mass estimates.  A final section summarizes our findings.  The Appendix describes some tests of the robustness of our findings by using other SSPs models, IMF parametrizations and IMF indicators.

\section{Data}\label{sec:data}

\subsection{MaNGA survey}
The MaNGA survey (\citealt{ Bundy2015}) is a component of the Sloan Digital Sky Survey IV (\citealt{Blanton2017}; hereafter SDSS IV). MaNGA uses integral field units (IFUs) to measure multiple spectra across $\sim$ 10000 nearby galaxies \cite[see][for the sample selection]{Wake2017}. The IFU observations enable the construction of detailed kinematic and chemical composition maps of each galaxy \citep{Westfall2019}. In this work, we use the MaNGA DR15 \citep{Aguado2019}, which provides observations for $\sim 4600$ galaxies. MaNGA DR15 includes datacubes with spectral information in the wavelength range 360-10000 nm and spatial sampling of 1-2 kpc thanks to an observational strategy which includes dithering (see \citealt{Westfall2019} for more details). Apart from the observed spectra, DR15 also provides kinematic maps (stellar velocity and velocity dispersion), as well as emission and absorption line estimates for each spectrum.

The photometric parameters used throughout this work come from the PyMorph Photometric Value Added Catalogue (MPP-VAC) presented in \citet{Fischer2019}. The MPP-VAC provides photometric parameters from S\'ersic and S\'ersic + Exponential fits to the 2D surface brightness profiles of the MaNGA DR15 galaxy sample (4672 entries for 4599 unique galaxies) in the SDSS $g$, $r$, and $i$ bands. In addition to total magnitudes, effective radii, S\'ersic indices, axis ratios $b/a$, etc., MPP-VAC also includes a flagging system (FLAG$\_$FIT) which indicates the preferred fit model (S\'ersic or S\'ersic + Exponential). In this work, for each galaxy, we use the best-fit parameters in the SDSS $r$-band for the model indicated by FLAG$\_$FIT. When  FLAG$\_$FIT = 0 -- i.e.,  no preference between S\'ersic or S\'ersic + Exponential fits -- we use the values returned by the latter.

The MPP-VAC provides two estimates of the total magnitudes and sizes:  One corresponds to integrating the best fitting surface brightness profiles to infinity, and the other to truncating these profiles at $7R_e$.  (The scale which contains half this `truncated' light is slightly smaller than the `untruncated' $R_e$.)  For most of the analysis in this paper, we use the `truncated' manitudes and sizes.  

\begin{figure}
  \centering
  \includegraphics[width=0.9\linewidth]{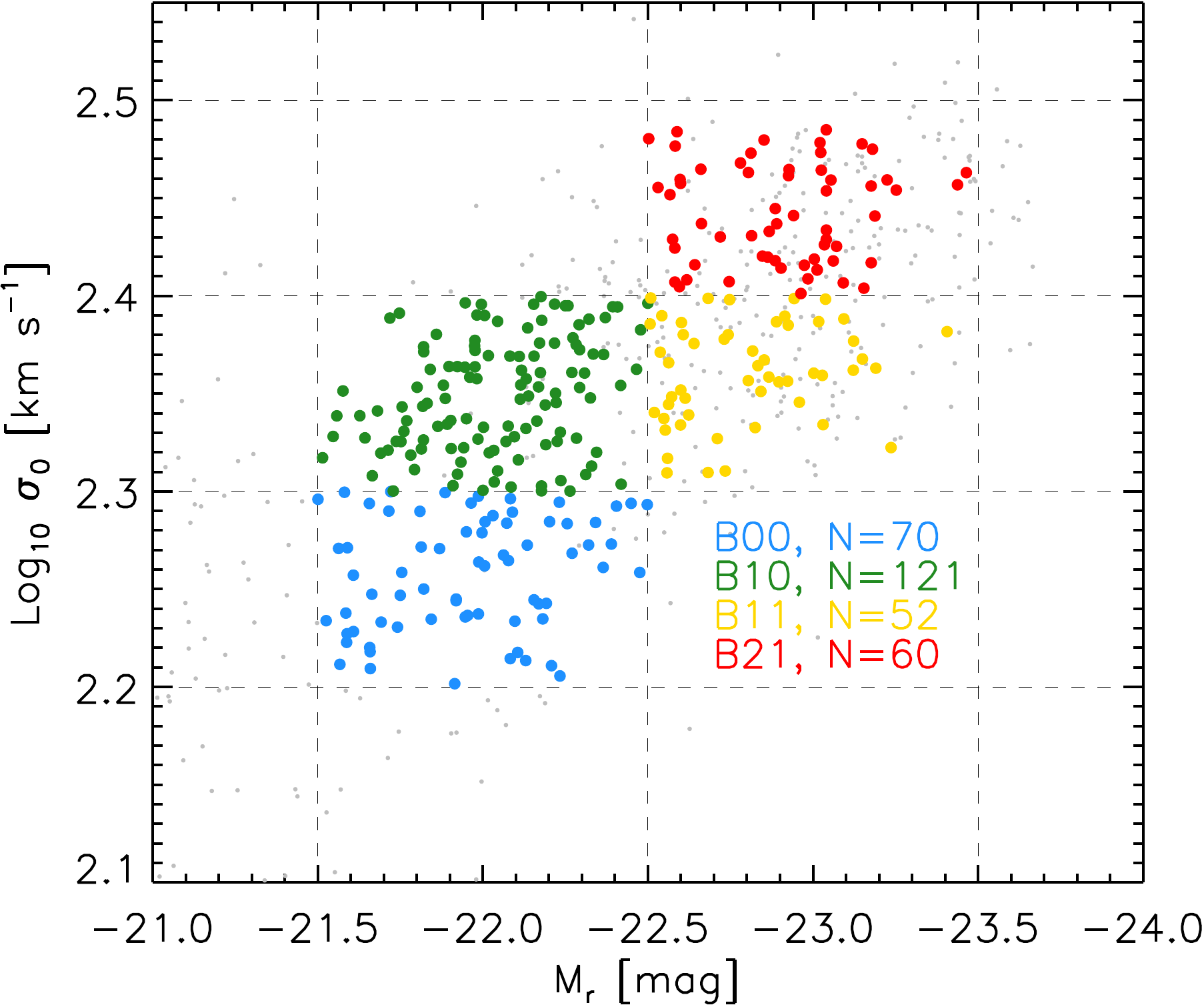}
  \caption{Distribution of central velocity dispersion $\sigma_0$ and absolute magnitude $M_r$ in our sample (grey dots).  Blue, green, yellow and red dots show the objects with $z\le 0.08$ which we assign to bins B00, B10, B11 and B21 described in Table~\ref{tab:bin}.} \label{fig:sample}
\end{figure}

\begin{table}
\centering
SELECTION OF GALAXIES\\
\begin{tabular}{ccc}
  \hline
 Condition & Observations & Galaxies \\
 \hline
 \hline
       Es                        &  1052  &  1028  \\
       FLAG$\_$FIT $\ne$ 3       &  1002  &  982   \\
       No contamination          &   814  &  797   \\ 
 \hline
 \hline
 \end{tabular}
\caption{The number of galaxies in our sample of Es, as described in the text.}
\label{tab:sample}
\end{table}

\begin{table}
\centering
BINNING OF GALAXIES\\
\begin{tabular}{ccccc}
  \hline
  Bin &  M$_r$ & Log$_{10}$ $\sigma_0$ & Galaxies & Galaxies \\
      &  [mag] &     [km s$^{-1}$]    &   all $z$  &   $z\le 0.08$    \\  
 \hline
 B00    & $-21.5, -22.5$  &  2.20, 2.30  & 74  & 70 \\
 B10    & $-21.5, -22.5$  &  2.30, 2.40  & 133 & 121 \\
 B11    & $-22.5, -23.5$  &  2.30, 2.40  & 138 & 52 \\
 B21    & $-22.5, -23.5$  &  2.40, 2.50  & 164 & 60 \\
\hline
\hline
\end{tabular}
\caption{Number of galaxies in each bin with and without redshift cut.}
\label{tab:bin}
\end{table}

\subsection{Sample selection and binning}
\label{sect:sample}

\begin{figure}
  \centering
  \includegraphics[width=0.9\linewidth]{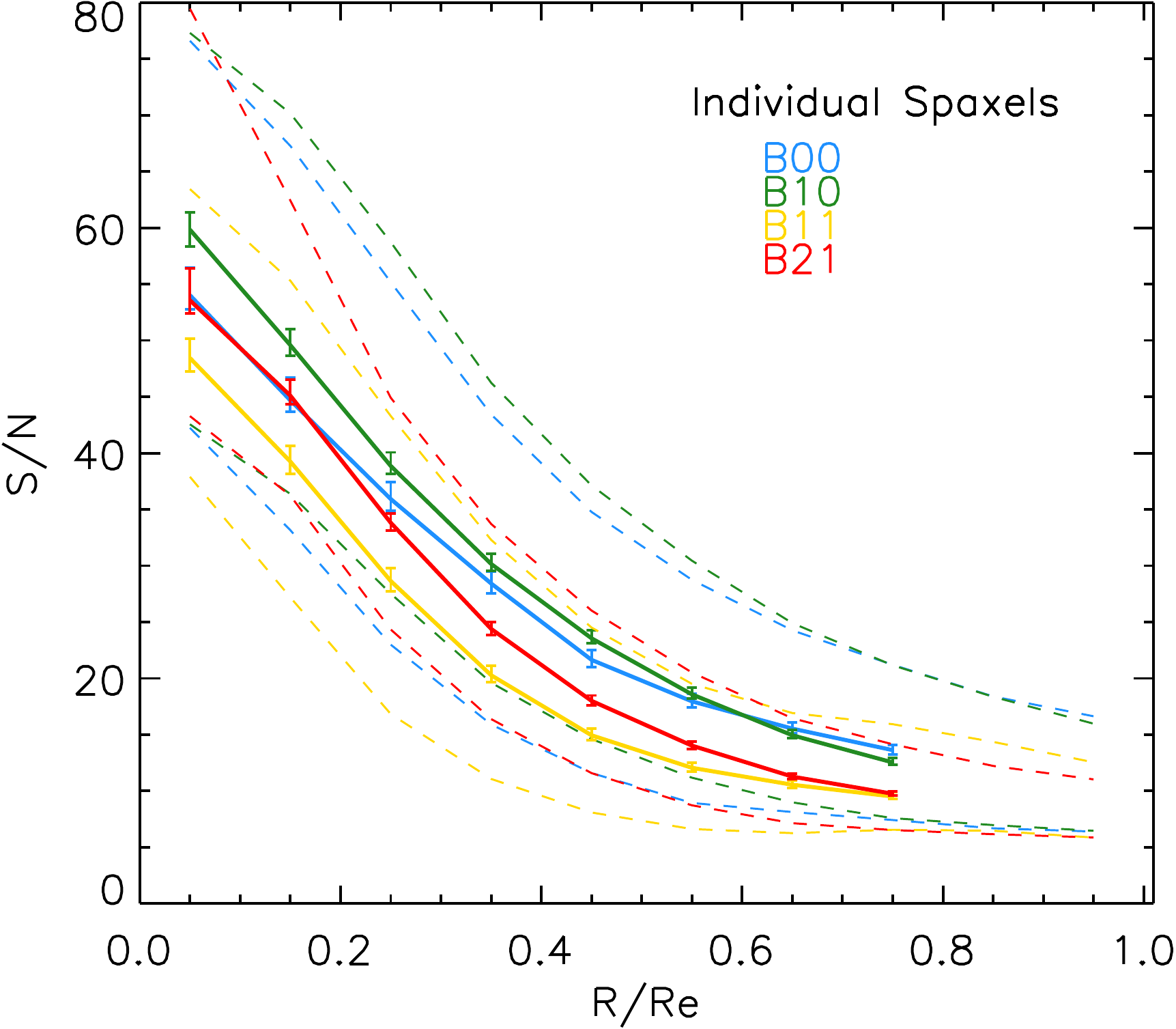}
  \caption{The median signal-to-noise of the spaxels included in each bin (solid lines with error bars) decreases monotonically with distance from the center, for the four bins defined in Table~\ref{tab:bin}. Dashed lines show the range which includes 68\% of the spaxels around the median. The typical S/N per spaxel is less than 100, which is why a stacking analysis is necessary.}
  \label{fig:SNindiv}
\end{figure}

\begin{figure}
  \centering
  \includegraphics[width=0.9\linewidth]{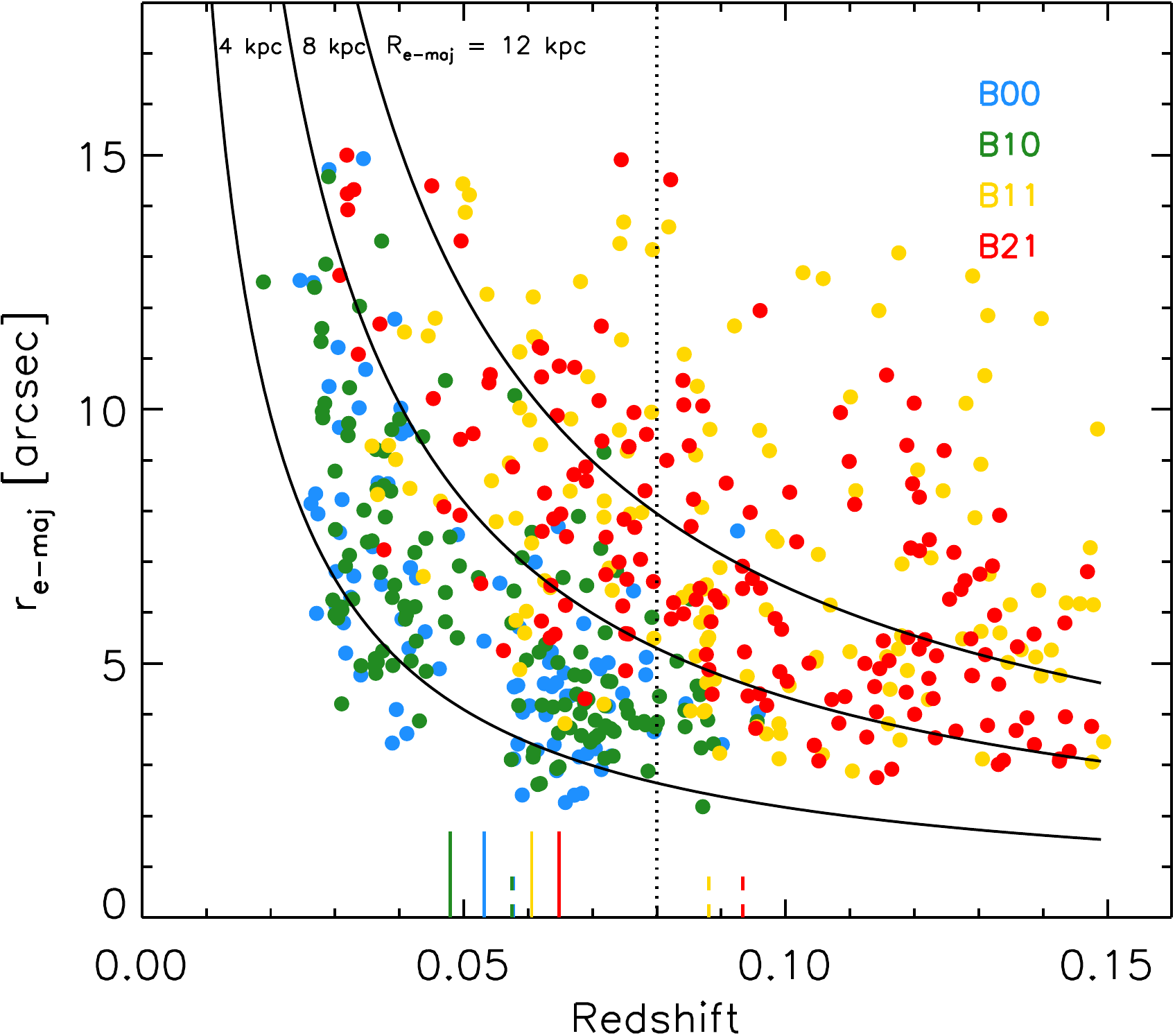}
  \caption{Joint distribution of angular size and redshift for the Es in bins B00, B10, B11 and B21.  Small vertical lines show the median $z$ for each bin if we include all objects (dashed), or if we restrict the sample to $z\le 0.08$ (solid).  There are very few B00 or B10 objects at $z\ge 0.08$.  The same galaxy, if placed at $z\le 0.05$, will be covered by many more spaxels than if it is at higher $z$.  For the largest galaxies, the IFU may not cover the entire region within $R_e$; this is particularly a concern for bins B11 and B21.}
  \label{fig:rez}
\end{figure}

This paper and its companion (Paper~II) use a sample of pure E galaxies to study their properties as a function of global parameters (i.e. absolute magnitude, central velocity dispersion and half-light radius) as well as galactocentric distance (i.e. their gradients). Since Es have neither complex star formation histories nor multiple structural components (such as spiral arms or bars) we assume that they can be well approximated by a single stellar population (SSP).

To select a pure sample of Es we use the companion morphological catalog  to MPP-VAC \citep[MDLM-VAC]{Fischer2019}. The MDLM-VAC provides morphological properties (e.g., TType, presence of bar, edge-on galaxies, etc.) derived from supervised Deep Learning algorithms based on Convolutional Neural Networks. Details on the Deep Learning model architecture, training  and testing procedures are given in \citet{DS2018}. We require
$$
 {\rm TType}\le 0   \quad \textrm{and} \quad  P_{\rm S0}\le 0.5
$$
to select our sample of Es from the MDLM-VAC (which includes 4672 observations for 4599 unique galaxies).  The first condition selects early-type galaxies (1948 observations for 1908 galaxies) as opposed to late-type galaxies, and the second selects Es (1052 observations for 1028 galaxies) rather than S0s (see Table \ref{tab:sample}). About $\sim 22$\% of the  MaNGA DR15 galaxies are Es.

Since we require of accurate photometry and kinematics, we remove from our sample galaxies with FLAG$\_$FIT=3 from MPP-VAC (i.e., no available photometric parameters), as well as galaxies with unreliable spectra due to contamination by neighbors (removed after visual inspection). We also limited our sample selection to galaxies with $z\le 0.08$, for the reasons we discuss in Section~\ref{sec:methods}.

Figure~\ref{fig:sample} shows the relation between central velocity dispersion $\sigma_0$\footnote{$\sigma_0$ is the value of the velocity dispersion at 0.25 arcsec (corresponding approximately to less than 0.1Re for the galaxies in our sample). This value is obtained by interpolating the velocity dispersion profiles (the median value of the velocity dispersion of the spaxels in circularized radial bins) at 0.25 arcsec.} and absolute magnitude $M_r$ for the whole E sample (small grey dots). We divide the low redshift ($z\le 0.08$) E sample into four bins based on $M_r$ and $\sigma_0$ (colored symbols).
To measure absorption features accurately, spectra must have high SN ($\ge 100$).  As we discuss in Section~\ref{sec:methods}, this requires that we create stacked spectra.  Our bin sizes were chosen with this requirement on SN in mind. The bin limits, as well as the number of galaxies in each bin are given in Table \ref{tab:bin}. For galaxies with repeated observations we only use the best S/N observation for each one.

Note that $M_r=-22.5$ is close to the critical luminosity at which various scaling relations change slope \citep{Bernardi2011}.  This corresponds to a stellar mass of $2\times 10^{11}M_\odot$ if the IMF is \cite{Chabrier2003}.  In Section~\ref{sec:MdMs}, we show that the IMF is not Chabrier, and use our results to provide a better estimate of the translation from $L$ to $M_*$.  

\begin{figure}
  \centering
  \includegraphics[width=0.9\linewidth]{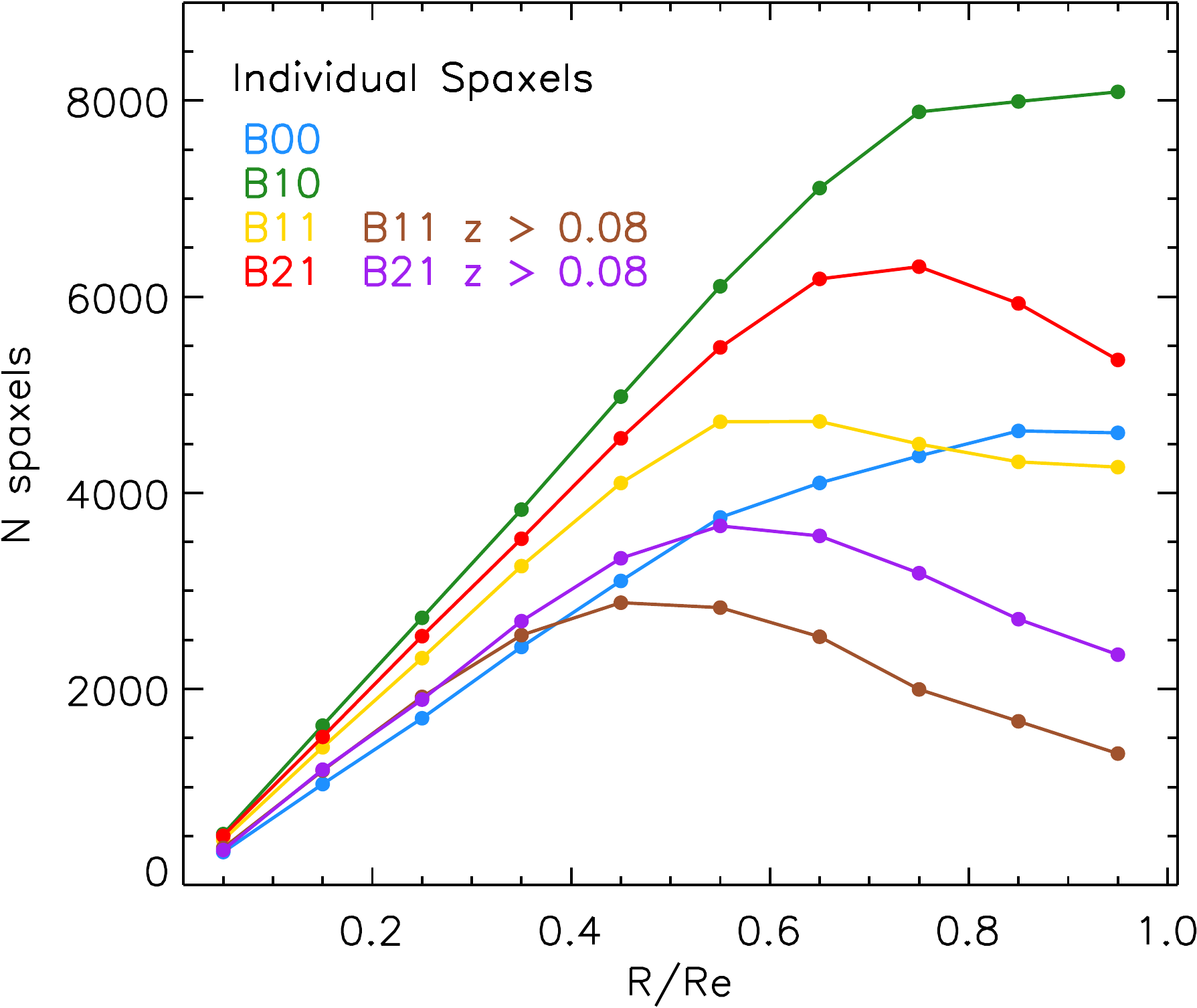}
  \caption{Number of spaxels in each radial bin which contribute to our results for the four bins in $M_r$ and $\sigma_0$ defined in Table~\ref{tab:bin}, and have $z\le 0.08$ (top four curves) and from galaxies at $z>0.08$ (bottom two curves; there are very few B00 or B10 galaxies at this higher $z$).  Some of the curves decrease at large $R$ because, for the largest galaxies, the IFU may not cover the entire region within $R_e$.}
  \label{fig:Nspx}
\end{figure}

\subsection{Stacked spectra and Lick indices}\label{sec:methods}

\begin{table*}
\centering
\begin{tabular}[width=\linewidth]{cccccc}
 \hline
 Lick index & Blue continuum  & Feature & Red continuum & Units & Source \\
 \hline
 H$_\beta$    &    4827.875 -- 4847.875  &  4847.875 --   4876.625 &   4876.625 --   4891.625 &   \AA   &  1 \\
 Mg$b$       &   5142.625  --  5161.375   &  5160.125  --  5192.625   &  5191.375  --  5206.375    &  \AA   & 1 \\
 Fe5270   &    5233.150  --  5248.150   &  5245.650   --  5285.650   &  5285.650  -- 5318.150   &  \AA   & 1 \\
 Fe5335   &    5304.625 --  5315.875   & 5312.125  -- 5352.125  &  5353.375  --  5363.375  & \AA  & 1 \\
 TiO2$_{\rm SDSS}$  &   6066.625  --  6141.625  &  6189.625  --  6272.125  &  6422.000 --  6455.000  &  mag & 2 \\
 TiO2    &     6066.625  --  6141.625   & 6189.625   -- 6272.125  &  6372.625  -- 6415.125   & mag & 1  \\
 TiO1     &    5816.625   --  5849.125  &  5936.625  -- 5994.125   & 6038.625  -- 6103.625   &  mag  & 1 \\
 \hline
 \hline
 \end{tabular}
 \caption{Lick indices used in this work and their corresponding definitions: (1) \citet{Trager1998}, (2) \citet{LaBarbera2013}.  TiO1 and TiO2 are discussed in the Appendix.}
 \label{tab:Lick}
\end{table*}

\begin{figure*}
  \begin{minipage}{0.6\linewidth}
    \centering
    \includegraphics[width=\linewidth]{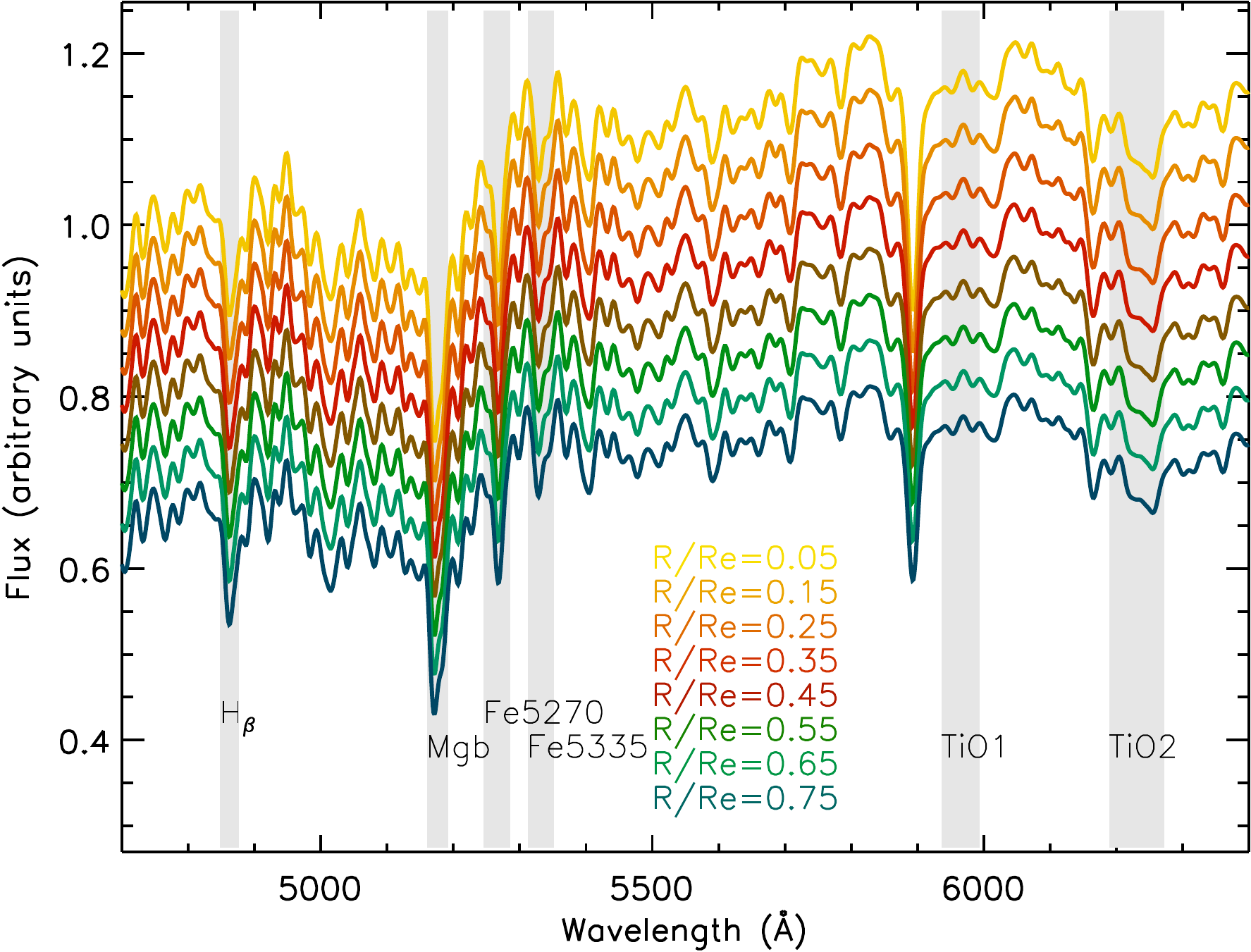}
    \label{fig:spectra}
  \end{minipage}
  \hspace{0.cm}
  \begin{minipage}{0.35\linewidth}
    \centering
    \includegraphics[width=\linewidth]{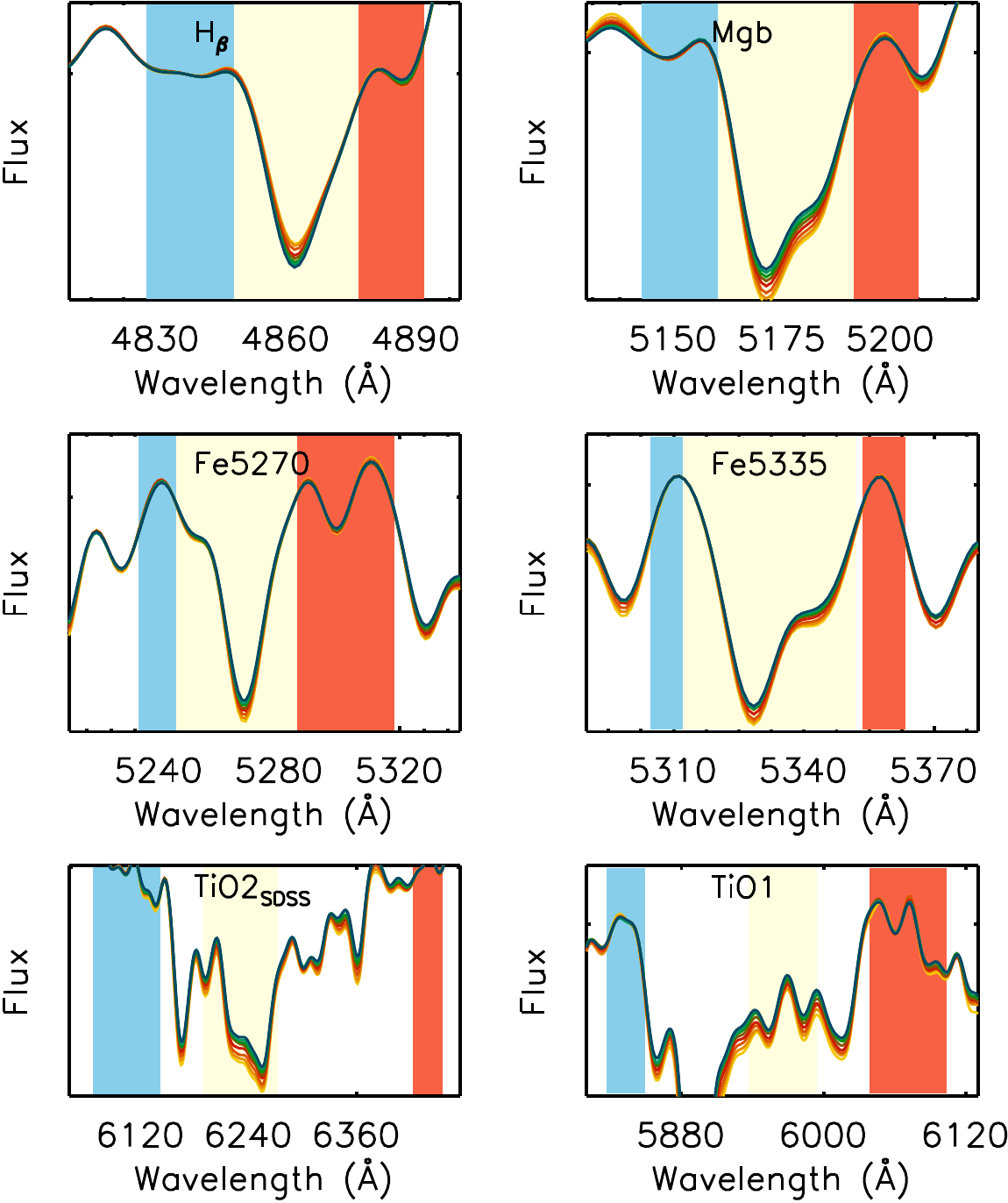}
    \label{fig:zoom}
  \end{minipage}
	 \caption{Left: Stacked spectra from the objects in bin B10 (see Table~\ref{tab:bin}) for a range of distances from the center $R/R_e$.  Spectra are normalized and have been offset vertically for clarity. Right: Same as previous figure, but now showing zoom-ins around the Lick indices which play an important role in this paper. The blue, red and yellow  shaded regions correspond to the  continuum and spectral features as  defined in Table \ref{tab:Lick}.} \label{fig:spec}
\end{figure*}

We would like to measure radial gradients of Lick indices reliably.  This requires SN greater than 100.  As Figure~\ref{fig:SNindiv} shows, the typical S/N in a spaxel lies well-below this value.  Therefore, we must work with stacked spectra.  We generate these by stacking together the spectra of galaxies in the same bin (see Table~\ref{tab:bin}). For each galaxy, we use all the spaxels from the MAPS-VOR10-GAU-MILESHC files that have SN $\ge$ 5.  However, as we are interested in measuring gradients from the central regions out to about $R_e$ for each bin, we would like to make stacks for a narrow range in projected distance $R$ for each bin.  Figure~\ref{fig:rez} shows the joint distribution of angular size and redshift for the Es in our four bins before any redshift cut.  This shows that the same galaxy, if placed at $z\le 0.05$, say, will be covered by many more spaxels than if it is at higher $z$ -- since the spaxel angular size is fixed.  As a result, for the largest galaxies, the IFU may not quite cover the entire region within $R_e$; this is particularly a concern for bins B11 and B21.  We will return to this point later.  

Figure~\ref{fig:rez} shows that there are almost no B00 or B10 objects at $z\ge 0.08$ (see also Table \ref{tab:bin}).  For these, the sample is essentially volume limited.  The small vertical lines in Figure~\ref{fig:rez} show the median $z$ for each bin if we include all objects (dashed), or if we restrict the sample to $z\le 0.08$ (solid).  Figure~\ref{fig:Nspx} shows the number of spaxels which contribute to each $R/R_e$ stack, again separated into the contribution from $z\le 0.08$ and $z>0.08$.  Notice that the contribution from $z>0.08$ is small, especially at large $R_e$,  even though more than half the B11 and B21 objects lie at these higher redshifts.  This suggests that we will not suffer a significant loss in SN if we simply restrict the sample to $z\le 0.08$.  Doing so has the added benefit of reducing the lookback time spanned by our sample.  The lookback time to the median redshift of samples B00 and B10 is about 0.8~Gyrs, and is relatively unchanged by restricting to $z\le 0.08$, whereas for bins B11 and B21 the median looktime is reduced from 1.2 to 0.9~Gyrs.  While this does not seem dramatic, note that the lookback time to $z=0.15$ is 1.9~Gyrs.  Thus, volume-limiting the sample to $z=0.08$, reduces lookback time systematics significantly. 

We also limit our analysis to 0.8$R/R_e$, where the number of spaxels  starts decreasing significantly (see Figure \ref{fig:Nspx}), to avoid any bias in our results due to different galaxy sizes (i.e., the stacks in the outer regions would be dominated by the contribution of either galaxies with large angular sizes and/or small physical size $R_e$). We have tested that the results presented in this paper are robust out to  0.8 $R/R_e$, while different limits need to be taken into account when splitting the sample by size or rotation (see Paper~II).

\begin{figure}
  \centering
  \includegraphics[width=0.9\linewidth]{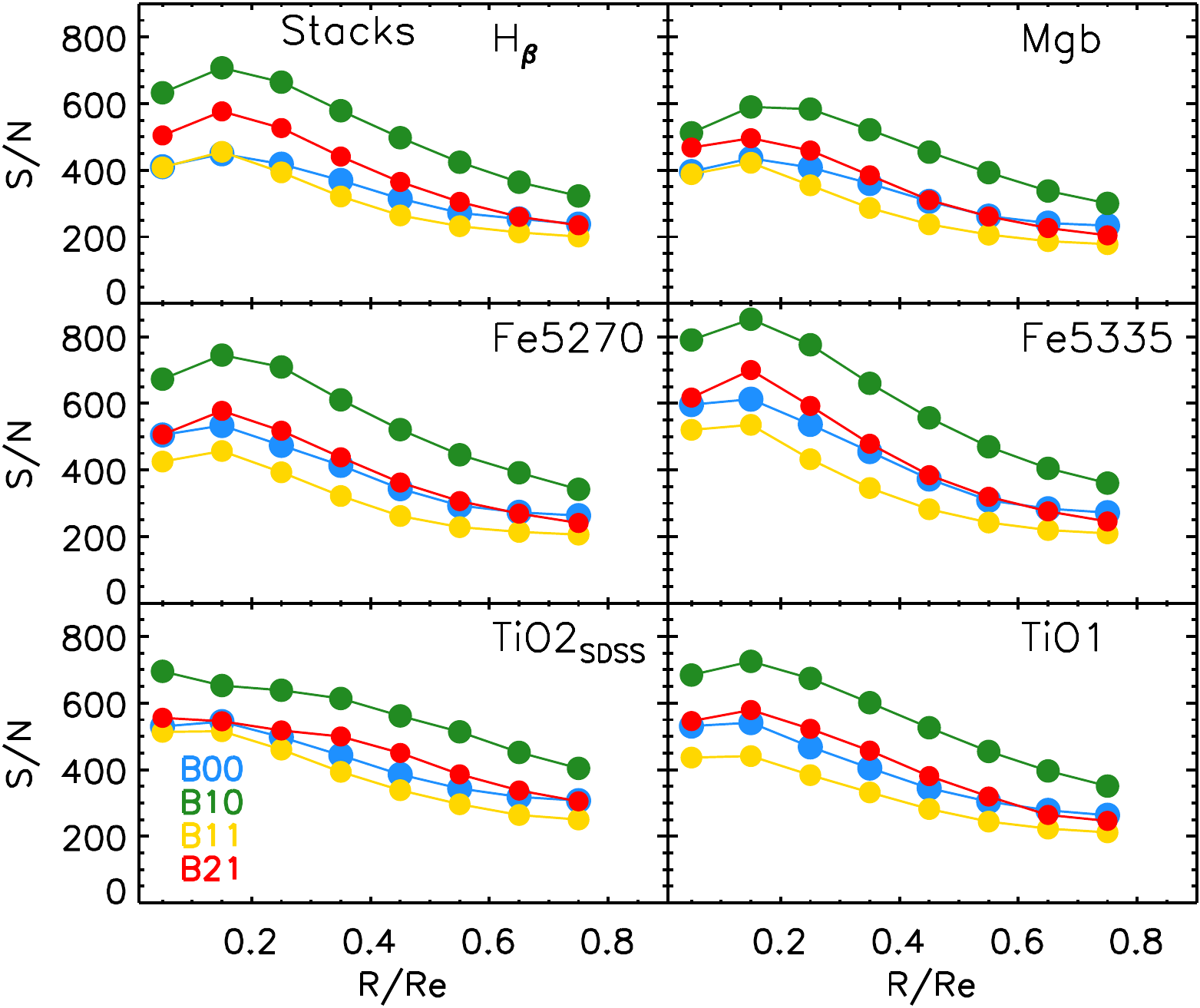}
  \caption{Signal-to-noise profiles for the Lick indices measured from the stacked spectra of galaxies in the bins defined in Table~\ref{tab:bin}.  Our stacks have S/N $\gg 100$ on all scales we explore in this paper.}
  \label{fig:SN}
\end{figure}

\subsection{Comparing stacks with median spaxel values}

In what follows, we work with emission line-corrected spectra.  This means that we have subtracted the emission line component (reported in the LOGCUBE-HYB10-GAU-MILESHC files, obtained by the standard MaNGA pipeline) from the observed flux.  These spectra are then rest-frame corrected by setting
\begin{equation}
  \lambda_{\rm rest} = \frac{\lambda_{\rm obs}}{(1+z)(1 + v/c)},
  \label{eq:lambdarest}
\end{equation}
where $v$ is the rotational velocity, and the $\lambda$s are converted to air wavelength units following the standard convention \citep{Morton1991}. (The MaNGA datacubes are given in vacuum units, while the Lick indices and the stellar population models are defined using air wavelength units. We use  air wavelength units throughout the paper.)

Our stacking procedure is as follows: For each bin, we stack the spaxels in radial bins $R/R_e$, where $R_e$ is the circularized effective radio of each galaxy ($R_e = R_{e,{\rm maj}}\sqrt{b/a}$;  $R_{e,{\rm maj}}$ is the truncated semimajor axis and $b/a$ is the ratio between the semimajor and semiminor axis from the best-fit indicated by FLAG$\_$FIT in the MPP-VAC). The radial steps are $0.1R/R_e$ in size, reaching out to $\sim 0.8\, R/R_e$.  (The IFUs do not cover beyond $R_e$ for a significant fraction of the larger galaxies in our sample.)  At each wavelength, we define the stacked spectrum as having the median value of the $3\sigma$ clipped normalized flux  (at each wavelength) of the spaxels belonging to that radial bin. We computed the error in the median flux as 1.25$\sigma \,N^{1/2}$, with $\sigma$ the standard deviation and $N$ the number of spaxels.  We also accounted for the correlation between spaxels, by multiplying the error by $1+1.62\,\log(N_{\rm spx-gal})$, where $N_{\rm spx-gal}$ is the number of spaxels per galaxy \citep{Westfall2019}. Skylines or masked wavelength regions are not used in the stacking.

Before stacking, spectra should be normalized. \cite{Parikh2018} fix the normalization region to be 6780-6867\AA.  Instead, we normalize each Lick index separately using the median value of the pseudo-continuum region.  Therefore, we do not create a single long stacked spectrum and then measure indices in it.  Rather we create a stack for each of the Lick indices.  This reduces the impact of variation in the shape of the spectral energy distribution (but not in the absorption feature). The returned Lick index values are within a few percent, but  the effect on the S/N  is significant (increasing up to a factor of 2, since it is proportional to the standard deviation of the stacked spectra).

We  tried an alternative stacking procedure: constructing radial stacks for each galaxy and then stacking the galaxies corresponding to each bin described in Table \ref{tab:bin} together. With this methodology, each galaxy contributes equally to the final stack; however, it has the disadvantage of penalizing the galaxies with the larger SN (larger number of spaxels or higher surface brightness).  The results presented in the following sections (based on the stacks derived using all the available spaxels) are consistent with the results obtained by stacking the radial stacks of individual galaxies, with the latter  being slightly more noisy.

Once we have created the stacks, we smooth them to a resolution of 300~km~s$^{-1}$ and we measure the Lick indices. Table \ref{tab:Lick} lists the Lick indices which play an important role in this paper.  To illustrate the quality of the stacks, Figure~\ref{fig:spec} shows the stacked spectra for the objects in bin B10 for a range of $R/R_e$. The spectra have been offset vertically for clarity. The right panels of Figure~\ref{fig:spec} show zoom-ins around the Lick indices discussed in the following sections.  These panels show clear trends of index-strength with distance from the center, which we quantify shortly.  Figure~\ref{fig:SN} shows the signal-to-noise profiles for the Lick indices measured from the stacked spectra. Clearly our stacks have S/N $\gg 100$ on all scales we explore in this paper.

\subsubsection{Velocity dispersion profiles}

\begin{figure} 
   \centering
   \includegraphics[width=0.9\linewidth]{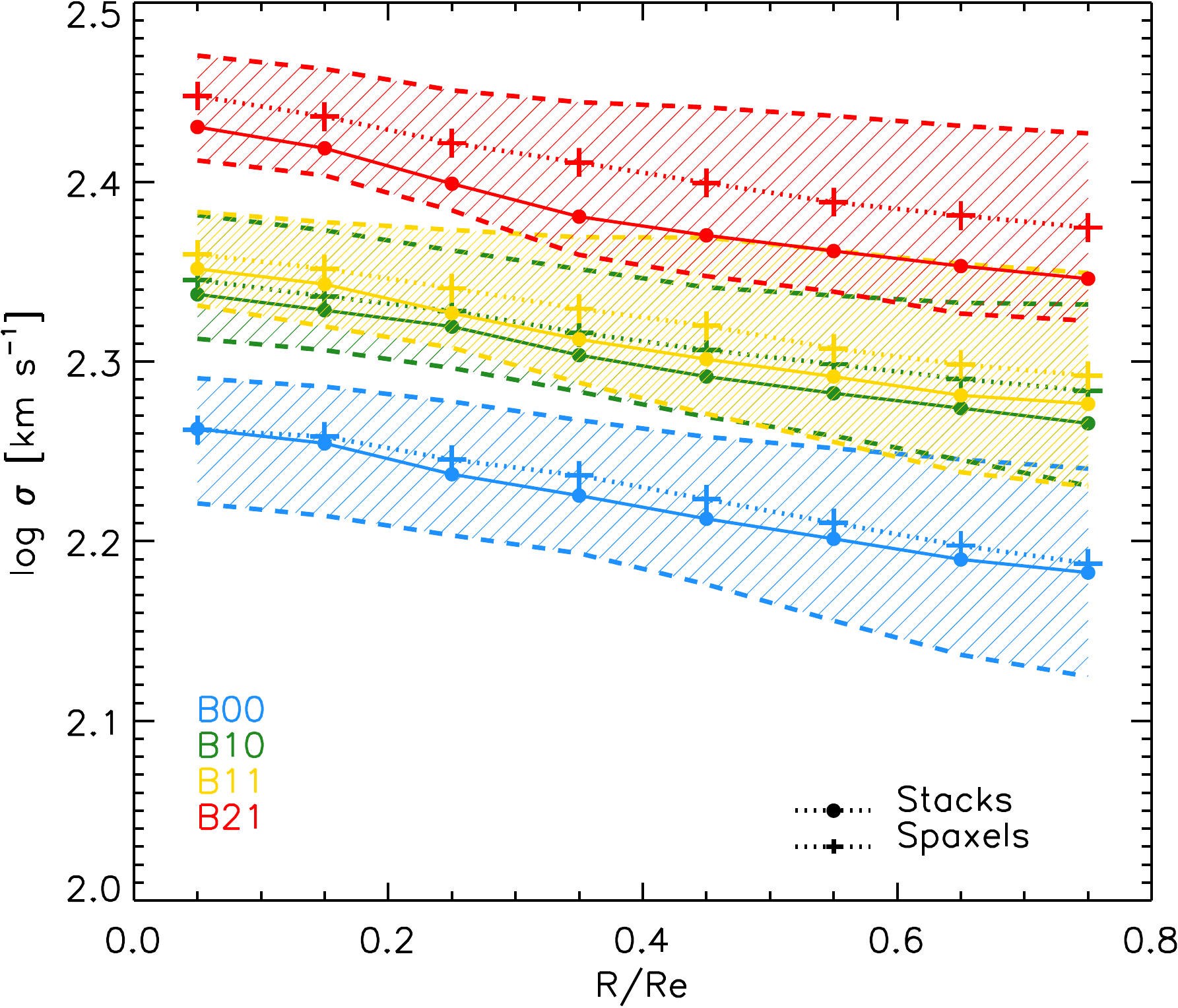}
   \includegraphics[width=0.9\linewidth]{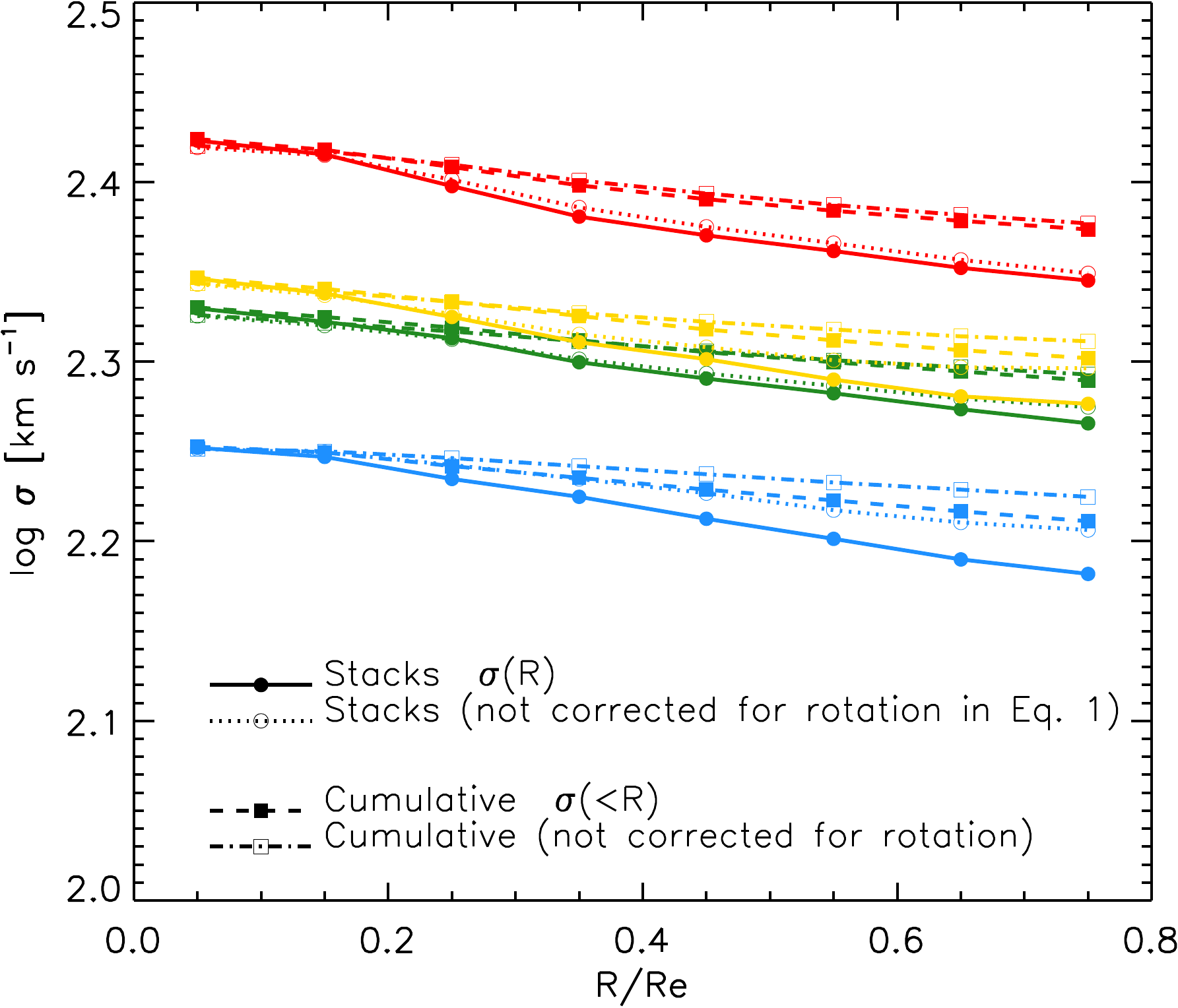}
   \caption{Top: Velocity dispersion profiles for the four bins defined in Table~\ref{tab:bin}, estimated from our stacked spectra (circles connected by solid lines) and from the median of the velocity dispersion measurements made on the individual spaxels themselves (crosses connected by dotted lines; the hashed region shows the range which includes 68\% of the objects around the median). Bottom: Velocity dispersion profiles estimated from our stacked spectra in radial bins before (open circles) and after (filled circles) correcting the individual spaxels for rotation (i.e. using equation~\ref{eq:lambdarest} with $v=0$ and $v\ne 0$, respectively) and from the cumulative stacks (i.e., $\sigma(<R)$) obtained  before (open squares) and after (filled squares) correcting for rotational velocity (i.e. $v\ne0$ in equation~\ref{eq:lambdarest}). Note that velocity dispersions quoted in this work are not corrected by seeing effects.} \label{fig:sigmaR}
\end{figure}

Figure~\ref{fig:sigmaR} shows velocity dispersion profiles for our four $\sigma_0$ and $M_r$ bins.  In both panels,  filled circles connected by a solid line show the value from each stack (note that here the stacks are not smoothed to 300 kms$^{-1}$).  In the top panel, the thinner lines with crosses show the median of the individual spaxels, with the hashed region showing the range which includes 68\% of the objects around the median.  The two agree to within about 10\%.  Aside from showing the expected trend that $\sigma$ increases from blue to yellow to red, with green being similar to yellow (see Figure~\ref{fig:sample}), both estimates show clearly that $\sigma$ decreases approximately as $\sigma(R)\propto R^{-0.1}$.  The estimated $\sigma(<R)$, shown as filled squares in the bottom panel of figure~\ref{fig:sigmaR}, is slightly shallower, $\sigma(< R)\propto R^{-0.06}$, and agrees with previous work on individual spectra (e.g. \citealt{Jorgensen1995}).  In contrast, \cite{Parikh2018} report flat  profiles for their stacks.  We get flatter profiles if we neglect to subtract the effects of rotation from the spaxels before stacking (empty symbols in the bottom  panel), but strongly believe that rotation should be removed before doing any analysis.  

\subsubsection{Line-index strength profiles} 

\begin{figure}
  \centering
  \includegraphics[width=0.98\linewidth]{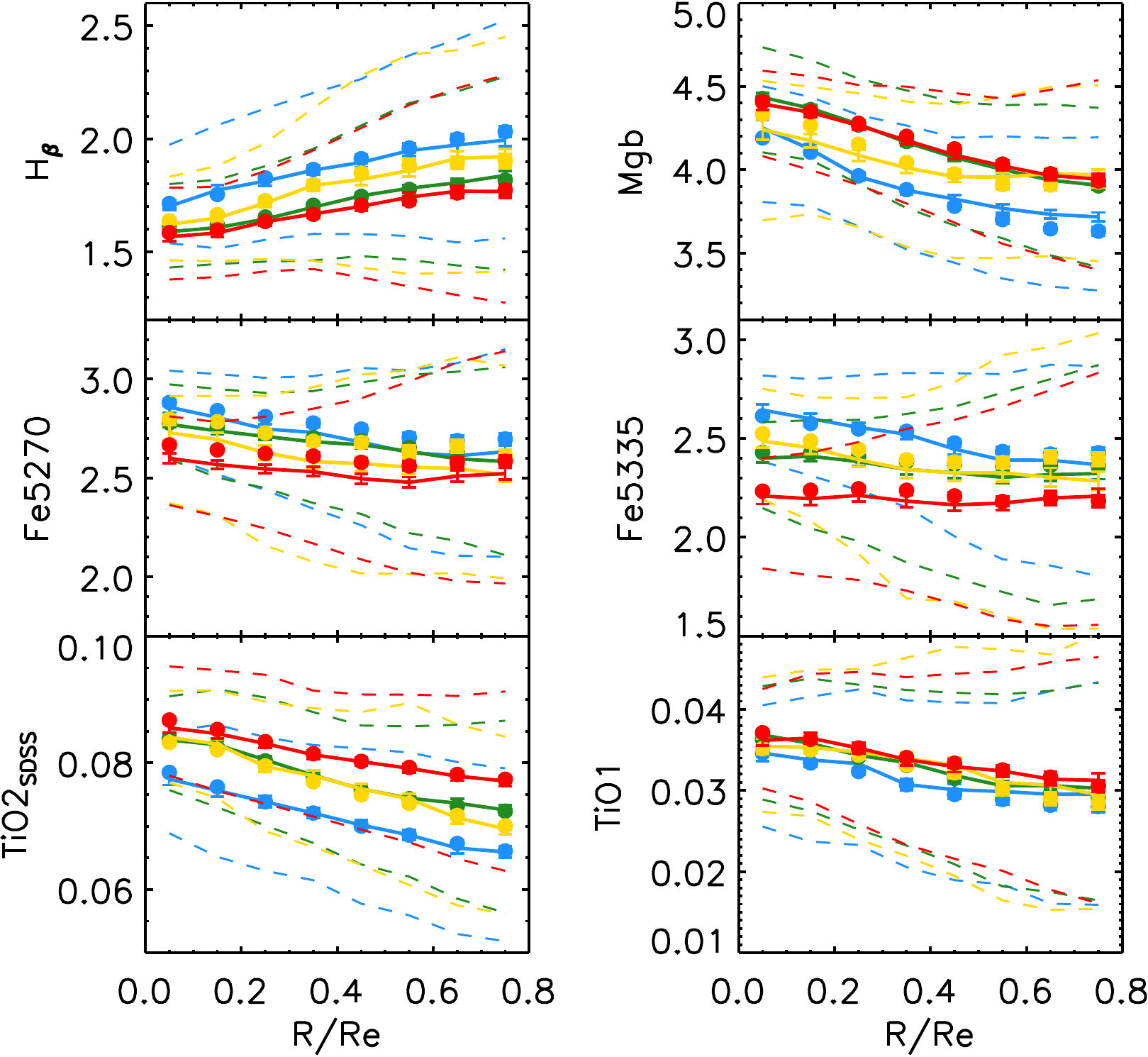}
  \caption{Line-index strength as a function of projected radius for the four bins defined in Table~\ref{tab:bin}, estimated from our stacked spectra (symbols), and from the median of the measurements made on the spaxels themselves (solid lines).  Dashed lines show the region which encloses 68\% of the spaxels at each $R/R_e$. }
 \label{fig:lickR}
\end{figure}

\begin{figure}
  \centering
  \includegraphics[width=0.9\linewidth]{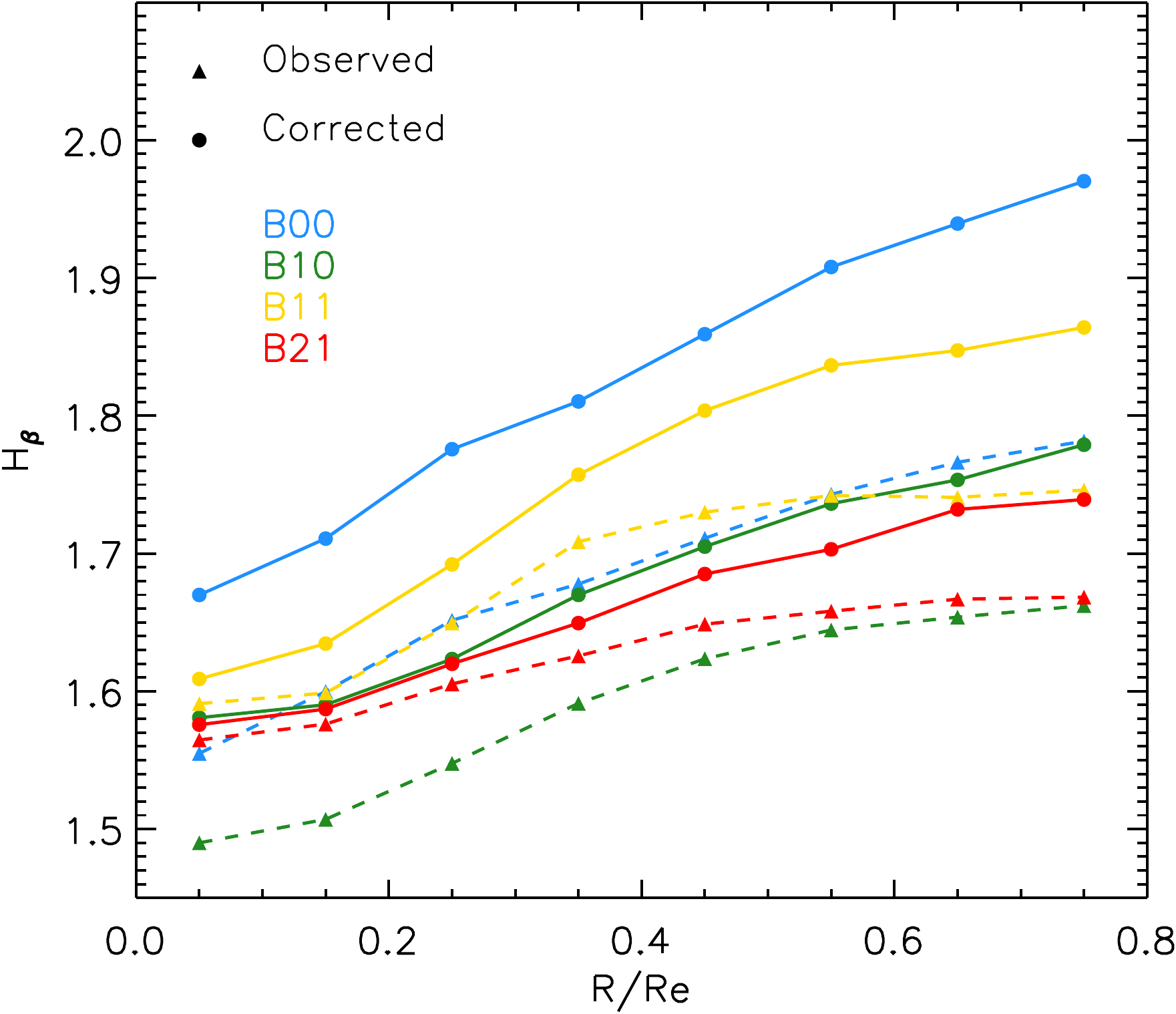}
  \caption{Median observed (dashed) and emission-corrected (solid) H$_\beta$ absorption line strength in our stacked spectra.  We use the emission corrected lines throughout.}
  \label{emis}
\end{figure}

Figure~\ref{fig:lickR} shows a similar analysis to Figure~\ref{fig:sigmaR} but for the line-indices in our four bins.  I.e. for each $R$ and bin, symbols show the Lick index strength measured from the stacked spectrum, and solid lines show the median of the line-index measurements for the individual spaxels (here too the stacks are not smoothed to 300 kms$^{-1}$).  We detect clear changes in line strength with $R/R_e$:  Whereas H$_\beta$ decreases towards the center, the Fe, Mg and TiO lines all increase. In what follows, we present results based on TiO2$_{\rm SDSS}$, and comment on TiO2 and TiO1 in the Appendix.

Before moving on, it is worth making three points. First, because bin B10 has similar $\sigma_0$ to bin B11, and similar $L_r$ to bin B00, it is curious that it has H$_\beta$ and Mg$b$ more similar to bin B21 (see top panels of Figure~\ref{fig:lickR}).  At least for H$_\beta$, one might worry that something may be systematically wrong with our emission correction for this bin.  To address this, Figure~\ref{emis} shows the observed and emission corrected line strengths in our four bins (smoothed to 300~km~s$^{-1}$).  The correction does not appear to be systematically different for B10 than for the other bins:  the correction for B10 is smaller than for B00 but larger than for B11 and B21.  We conclude that the anomalously weak H$_\beta$ for B10 is not an artifact.  

Second, for Fe, we found we had to treat bin B11 slightly differently from the others.  Whereas the peak of the distribution of Fe$>0$ values measured in the spaxels is in good agreement with the value we measure from our stack (so this is the value we use when plotting the yellow solid line), there are a large number of spaxels in which Fe $\sim$ 0.  This number is large enough that the median is biased low if we include the Fe $\sim$ 0 values.  We are not sure what causes this, but believe that the measurement from the stacked spectrum is more reliable. Third, while the agreement between the symbols and the solid lines is reassuring (for all bins and scales), even small differences matter quite a lot.  E.g., as we will see below, even differences in Fe of 0.1 \AA\ matter.  In general, the discrepancies are larger in the outer regions, where the S/N of the individual spaxels is smaller, so we have more confidence in the values measured from the stacks.

\section{Comparison with stellar population models}\label{sec:results}

\begin{figure}
  \centering
  \includegraphics[width=0.9\linewidth]{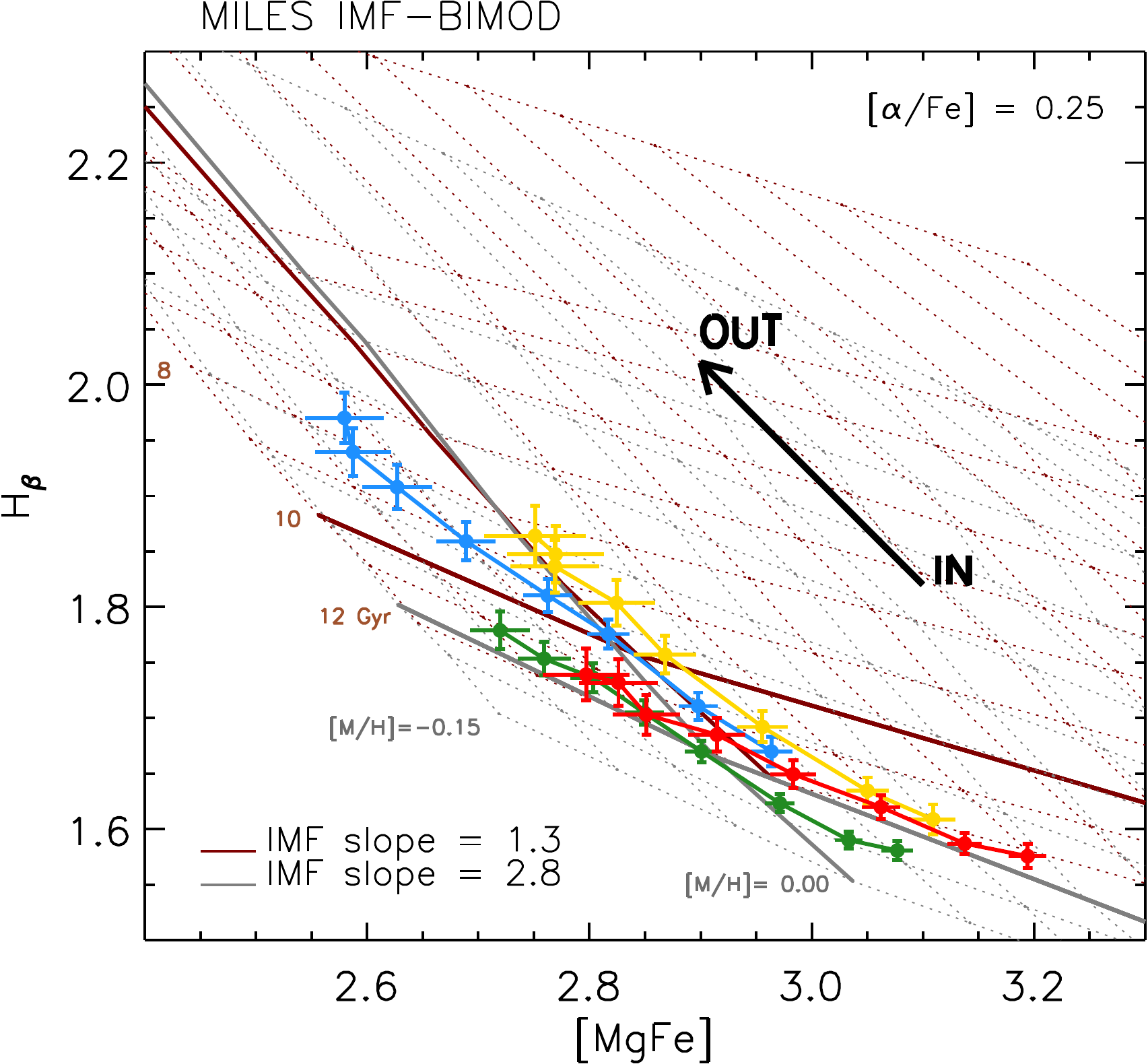}
  \includegraphics[width=0.9\linewidth]{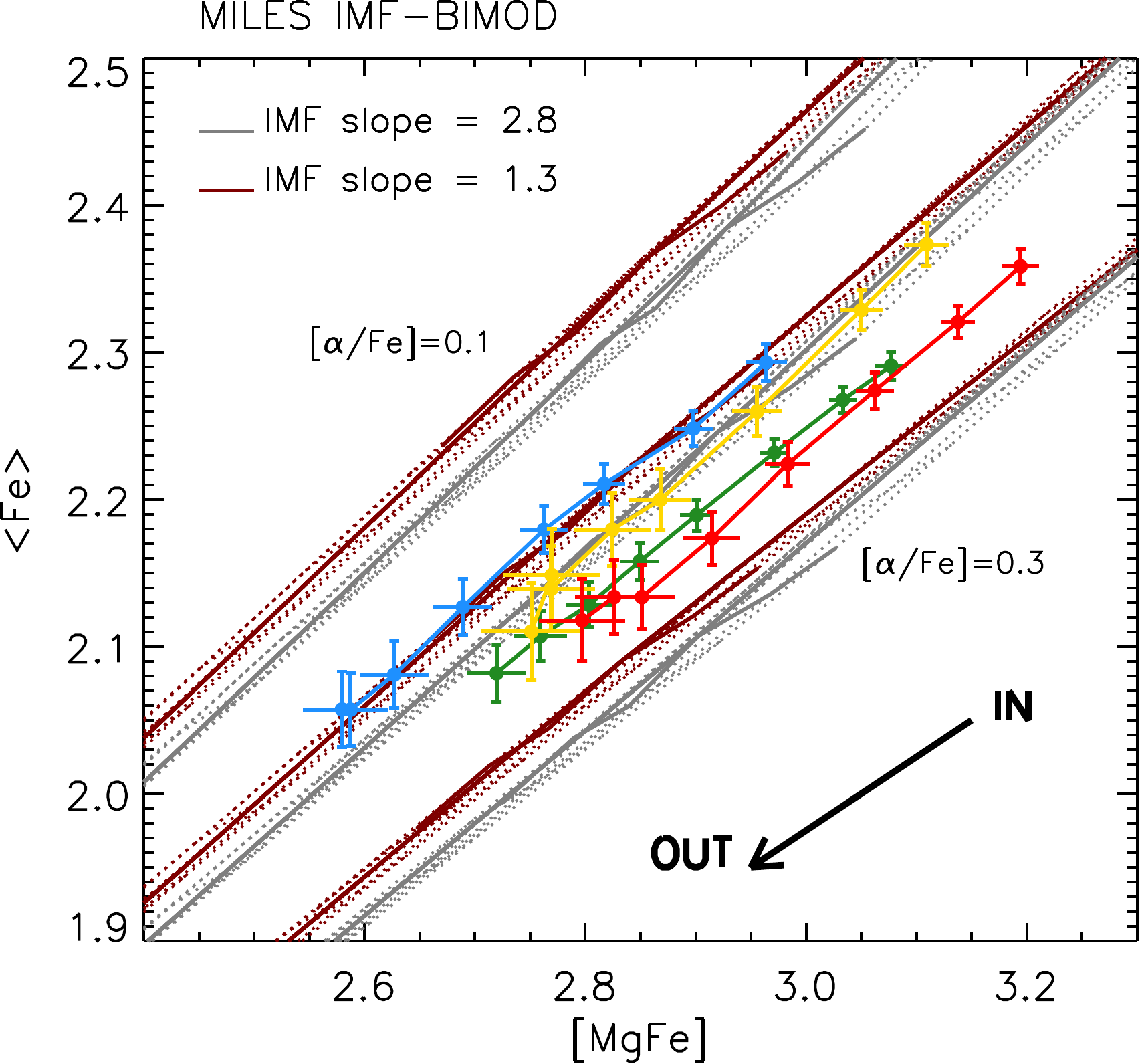}
  \caption{H$_\beta$-[MgFe] and <Fe>-[MgFe] index diagrams.  Symbols show measured gradients in line strength in each bin (thick bold arrow indicates the direction of increasing galactocentric distance), and differences between bins.  Dotted lines show age-metallicity grids for two SSP models, which have very different IMFs, to illustrate how such diagrams can be used to estimate IMF-dependent age, metallicity and enhancement factors. Both the data and the models have been smoothed to a common resolution of 300 km s$^{-1}$. In the top panel, the grids depend very weakly on [$\alpha$/Fe]; we set [$\alpha$/Fe] = 0.25, which the bottom panel indicates is reasonable. Grids have spacing intervals of 1 Gyr, 0.05 and 0.1 for the age, [M/H] and [$\alpha$/Fe], respectively. To guide the eye, the two thick solid lines for each model show lines of fixed age (10~Gyrs) and metallicity (solar).} \label{MilesBI}
\end{figure}

In this section we use the Lick indices listed in Table~\ref{tab:Lick} that we measured in our stacked spectra.  As is conventional, we work with  
$${\rm <\!\!Fe\!\!>} \equiv \rm{(Fe5270+Fe5335)/2}$$
  and
$${\rm [MgFe]} \equiv \sqrt{{\rm Mg}b\, {\rm <\!\!Fe\!\!>}}.$$
  We find clear differences in line-index strength between the $\sigma_0$ and $L_r$ bins, as well as strong gradients within each bin (Figure~\ref{fig:lickR}).  Rather than studying these individually, we instead work with Lick-index pairs.  This is because it has long been known that, for a fixed IMF, a plot of H$_\beta$-[MgFe] is a good age-metallicity indicator \citep{Worthey1994}, whereas <Fe>-[MgFe] is a diagnostic of the [$\alpha$/Fe] $\alpha$-enhancement ratio \citep{Trager1998}. We have also studied the effect of replacing H$_\beta$ with H$_{\beta 0}$ of \cite{Cervantes2009}: while this results in small quantitative differences, they are not large enough to warrant showing them as a separate series of figures.  The Appendix describes results based on TiO1 and TiO2 instead.

\subsection{Stellar population models}
In what follows, we will use the MILES-Padova models with BiModal IMFs to interpret our measurements. Our results depend somewhat on a number of assumptions such as the choice of SSP models, the IMF parametrization or the IMF indicators. The Appendix discusses why we chose the MILES-Padova models with BiModal IMFs, and describes what happens if we use a number of alternatives:  MILES-BaSTI, MILES-Padova with UniModal IMFs,  as well as models from \cite{TW2017} and \cite{TMJ2011}. There we also comment on the use of TiO2 and TiO1 as IMF sensitive indices.

The MILES models \citep{Vazdekis2010} use the MILES stellar library to provide SSPs for the full optical spectral range (3540-7409~\AA) at high resolution (FWHM=2.51~\AA, \citealt{FalconBarroso2011}) for a wide range of ages, metallicities and IMFs. The model spectra are stored in air wavelengths. There are two sets of isochrones available:  Padova00 \citep{Girardi2000} and BaSTI \citep{Pietrinferni2004,Pietrinferni2006}.  The Padova00 isochrones are given at base [$\alpha$/Fe]\footnote{The base [$\alpha$/Fe] is the $\alpha$-enhancement given by the `base' models defined in \citet[][section 3.1]{Vazdekis2015}, which employ solar isochrones and for which [M/H] is assumed to be equal to [Fe/H].}, while the BaSTI isochrones provide three different [$\alpha$/Fe] = (base, solar and 0.4). The scaled-solar spectra (i.e. [$\alpha$/Fe]=0) have abundances from \cite{Grevesse1998}, whereas the $\alpha$-enhanced spectra ([$\alpha$/Fe]$= +0.4$) assume that [X/Fe]$= +0.4$ for the elements O, Ne, Mg, Si, S, Ca and Ti, and that the other elements have solar abundances.

Unfortunately, the BaSTI isochrones return unrealistic ages for our measurements (see Appendix), so in the following analysis we use the Padova00 isochrones. The Padova00 isochrones also include the later stages of stellar evolution, using a simple synthetic prescription for incorporating the thermally pulsing AGB regime to the point of complete envelope ejection. The range of initial stellar masses extends from 0.15 to $7M_\odot$. We use the Lick indices provided by the webtool server\footnote{http://www.iac.es/proyecto/miles/pages/webtools/tune-ssp-models.php} at 300~km~s$^{-1}$ resolution.
To account for [$\alpha$/Fe], we derive an $\alpha$-enhancement correction from the BasTI isochrone models which we then apply to the Padova00 MILES models \footnote{Note that there are systematic differences in the ages and metallicities of BaSTI and Padova00 isochrones which could have a (likely minor) effect on this correction.}:

\begin{align}
  I_{\rm Padova}({\rm age, Z, IMF,} \alpha) &=
  I_{\rm Padova}({\rm age, Z, IMF, baseFe})\nonumber\\
  & \quad\times \frac{I_{\rm BaSTI}({\rm age, Z, IMF}, \alpha)}{I_{\rm BaSTI}({\rm age, Z, IMF, baseFe})}.
\end{align}

The available IMFs for the MILES models used include the Unimodal and Bimodal shapes described in \cite{Vazdekis1996}, the Kroupa-universal and revised- \cite{Kroupa2001} and the \cite{Chabrier2003} IMFs. The unimodal IMF is a power-law function characterized by its slope $\Gamma_u$.  The standard \cite{Salpeter1955} IMF is obtained when $\Gamma_u$=1.3. The bimodal IMF is similar to the unimodal for stars with masses above $0.6M_\odot$, but has fewer lower mass stars, which is parametrized by transitioning to a shallower slope at lower masses.  Its slope $\Gamma_b$ is the only free parameter (as in the unimodal case). The IMF slopes range from 0.3 to 3.5.  We use the range  of $\Gamma_u=[0.8 - 2.5]$ for the unimodal and $\Gamma_b=[0.8-3.5]$ for the bimodal IMFs.

\subsection{Age, metallicity and [$\alpha$/Fe] given an IMF}

The two panels in Figure~\ref{MilesBI} show the age-metallicity diagnostic plots. In this and following figures, the grids in age, metallicity and $\alpha$-enhancement are refined by interpolating the models. In short, for each index, we fix all the parameters (age, Z, IMF, [$\alpha$/Fe]), except the one we want to refine, and then we interpolate the index value to the new sampling. We use spacing intervals of 1 Gyr for the age, 0.05 for [M/H] and 0.1 for [$\alpha$/Fe]. To guide the eye, the two thick solid lines for each model show lines of fixed age (10~Gyrs) and metallicity (solar). In what follows, we will only use a combination of Fe and Mg lines to constrain the $\alpha$-enhancement (see bottom panel in Figure \ref{MilesBI}) without taking individual element ratios into account. Therefore,  the [$\alpha$/Fe] we report is  in practice based entirely on the [Mg/Fe] abundance.

In each panel, each of the four $\sigma_0$-$L_r$ bins are represented by eight points:  these show measurements of the index strength in bins of 0.1R/$R_e$ from the center.  There are clear well-defined radial gradients in the measured H$_\beta$, <Fe> and [MgFe] line strengths:  the central regions have smaller H$_\beta$, larger [MgFe] and larger <Fe>.  As we noted before, the trend with $\sigma_0$ and $L_r$ is less clear.  Whereas the blue, yellow and red samples, (B00, B11, and B21) which have successively larger $\sigma_0$, are also ordered in H$_\beta$, the green symbols (B10) have substantially smaller H$_\beta$.  For now, we wish to use SSPs to interpret these gradients before turning to global trends with $\sigma_0$ and $L_r$. The anomalous behavior of the green symbols (B10) is studied in more details in our companion paper (Paper~II).

Each panel of Figure~\ref{MilesBI} also shows two different MILES-Padova SSP model grids having BiModal IMFs as labelled.  For reference, the BiModal IMF with $\Gamma_b$=1.3 is very similar to a Kroupa IMF.  These grids show that, for a given IMF, the central regions seem to be older and more metal-rich, whatever the value of $\sigma_0$ and $L_r$, while the [$\alpha$/Fe] gradients are less evident.  The age estimates depend weakly on the IMF; the metallicity estimates less-so.  Moreover, the relative age differences between bins depend even less on IMF.

We quantify this more precisely as follows.  
For each IMF, we find the triple of age, metallicity and [$\alpha$/Fe] which best describes the measured H$_\beta$, [MgFe] and <Fe> line strengths.  Here `best' means that we minimize a distance between measured and predicted index strengths.  Whereas the usual procedure (e.g. \citealt{LaBarbera2013, MN2015}) normalizes each separation by the associated measurement error for the index, this is not quite appropriate.  E.g., if models span a large range of values in one index and only a small range in another, then the best-fit distance would be determined by the index with the largest range of values.  We account for this by first normalizing distances by the typical range spanned by models -- fortunately, in practice, this does not vary strongly between indices (except for the TiO1 and TiO2 indices).  Strictly speaking, because we have sampled the models on a grid, this procedure has merely found the nearest triple to our measurements:  we then search for the other seven models which define the cube that encloses our measurements, and use (tri-)linear interpolation from these values to determine the age, metallicity and [$\alpha$/Fe] values which we use below.

\subsection{Allowing IMF variations}\label{sec:IMF}

\begin{figure}
  \centering
  \includegraphics[width=1\linewidth]{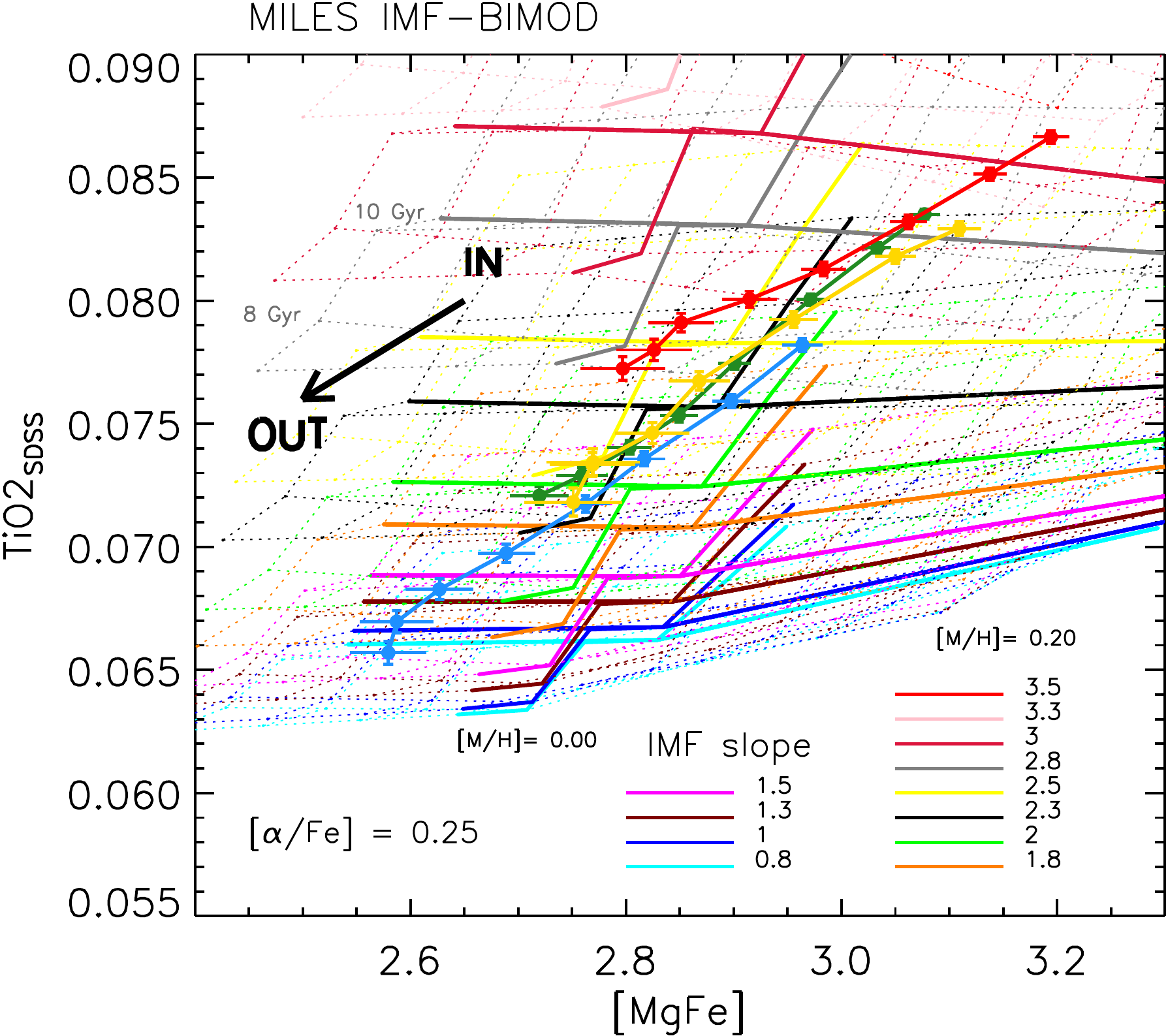}
  \caption{TiO2$_{\rm SDSS}$-[MgFe] index diagram, showing gradients in line strength in each bin (thick bold arrow indicates the direction of increasing galactocentric distance), and differences between bins.  Dotted lines show age-metallicity grids for [$\alpha$/Fe]=0.25 and a range of IMF slopes, to illustrate how this diagram can be used to discriminate between IMFs.} \label{TiO2MilesBI}
\end{figure}

\begin{figure}
  \centering
  \includegraphics[width=0.9\linewidth]{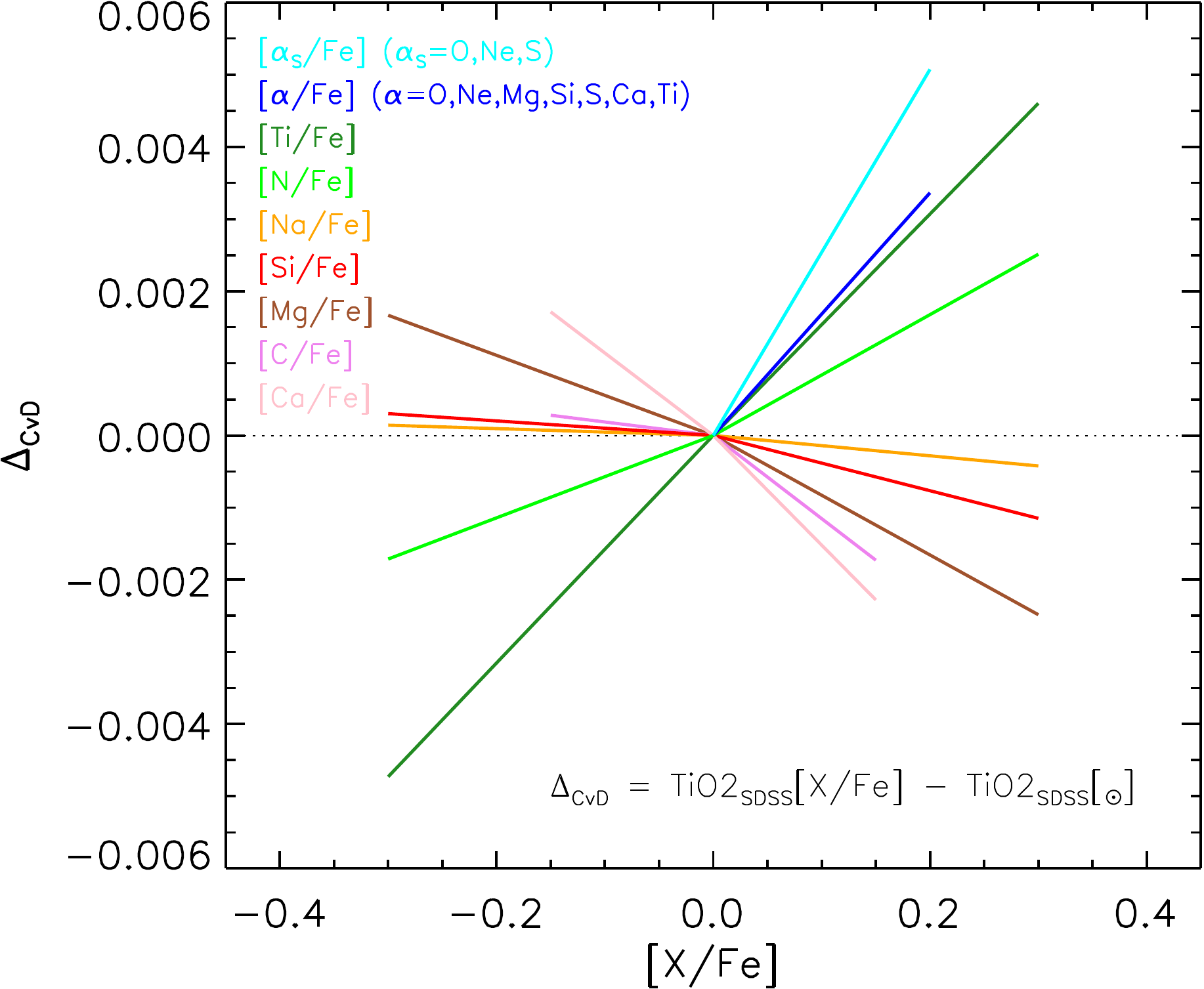}
  \caption{Difference in TiO2$_{\rm SDSS}$ (i.e., between [X/Fe] enhanced models and solar models), for various choices of X (as labelled).  All line strengths were computed using the Conroy \& van Dokkum (2012) model spectra (Chabrier IMF, solar metallicity, and an age of 13.5 Gyr) smoothed to our resolution of 300 km~s$^{-1}$. 
  }
  \label{DTiO2sdss}
\end{figure}

\begin{figure*}
  \centering
\includegraphics[width=0.95\linewidth]{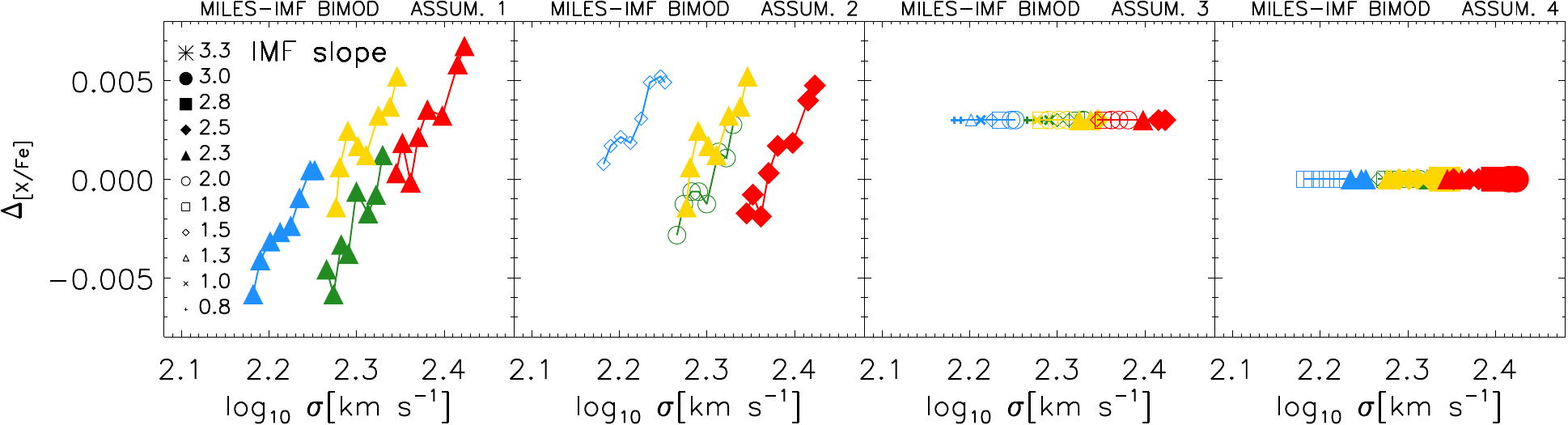}
\caption{Difference between the measured TiO2$_{\rm SDSS}$ and the value predicted by a model (with age, metallicity and [$\alpha$/Fe] given by Figure~\ref{MilesBI}) due to variation in [X/Fe] enhancements (i.e., not to the IMF) for the four models outlined in Table~\ref{tab:assumptions}.
  Left panel (ASSUMPTION~1):  All galaxies have the same IMF (we set the slope to $\Gamma_b$=2.3) whatever their $\sigma_0$ and $L_r$;
  Second from left (ASSUMPTION~2): The IMF may depend on $\sigma_0$ and $L_r$ (for bins B00, B10, B11 and B21 we use $\Gamma_b$=1.5, 2.0, 2.3 and 2.5) but there are no IMF gradients within galaxies;
  Second from right (ASSUMPTION~3): The IMF varies within a galaxy and $\Delta_{\rm [X/Fe]} = 0.003$;
  Right panel (ASSUMPTION~4): The IMF varies within a galaxy and $\Delta_{\rm [X/Fe]} = 0$.
  Hence, in the two left hand panels, the IMF is fixed and $\Delta_{\rm [X/Fe]}$ is allowed to vary, whereas in the two right hand panels $\Delta_{\rm [X/Fe]}$ is fixed and the IMF is allowed to vary.  
  Different symbols show the closest-fitting IMF slope (corresponding to the legend shown in left-hand panel), but the actual value of all the inferred properties is got by interpolating between the best-fitting models. Recall that the central regions correspond to the higher velocity dispersion values.}
  \label{TiOoptions}
\end{figure*}

\begin{figure}
  \centering
  \includegraphics[width=1\linewidth]{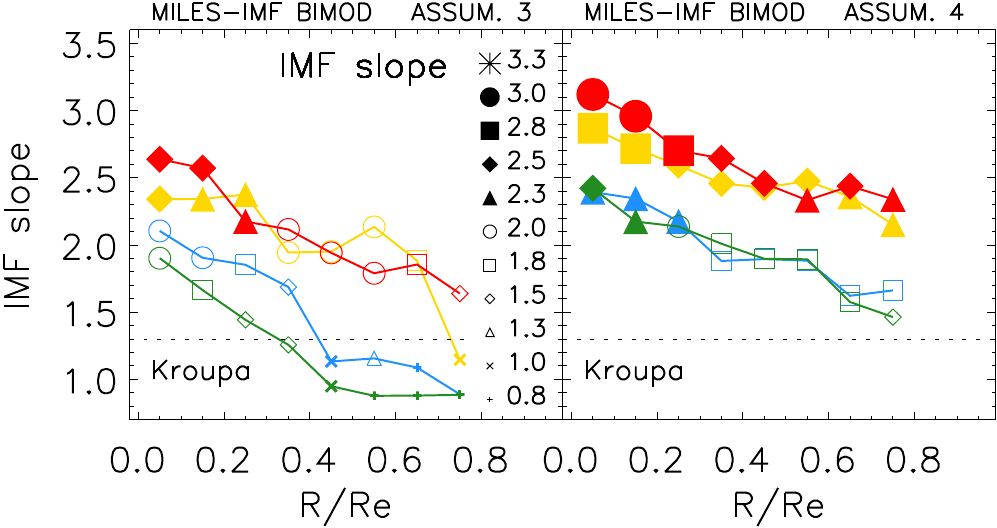}
  \vspace{-0.5cm}   
  \caption{Inferred IMF gradients in our four bins under ASSUMPTIONS 3 (left) and 4 (right). Different symbols represent different IMFs as in Figure \ref{TiOoptions}, but the actual value of IMF slope shown is got by interpolating between the two best-fitting IMF models. }
  \label{IMFgrad}
\end{figure}

To address the question of whether, in addition to age, metallicity or [$\alpha$/Fe] gradients within a galaxy, there are IMF gradients as well, we turn to the TiO2$_{\rm SDSS}$ index.  This is because Ti related indices are sensitive to the abundance of low-mass stars, so are indicators of the IMF slope.  Recent work \citep{LaBarbera2013, LaBarbera2016, MN2015, TW2017} has shown that, with some care, TiO2$_{\rm SDSS}$-[MgFe] can be used as an IMF diagnostic.  (We discuss other Ti-related indices in the Appendix.) However, note  these features are  effectively sensitive to stars between $\sim$0.2-0.3 M$_\odot$ to $\sim$0.9-1.0  M$_\odot$ \citep[see][]{Spiniello2014, MN2019}  but they are neither sensitive to stellar remnants, nor to very low-mass stars ($\sim$ 0.1  M$_\odot$). Therefore, the M/L ratio cannot be totally constraint with this kind of analysis. This is an intrinsic limitation of the stellar population analysis of old stellar populations which is also present in previous work in the literature using TiO as an IMF indicator.

Figure~\ref{TiO2MilesBI} illustrates the methodology.  As in the previous figures, symbols show our measurements and dotted lines show SSP age-metallicity grids for [$\alpha$/Fe]=0.25 and a wide range of IMFs.  The symbols show clear gradients in TiO2$_{\rm SDSS}$, and weaker, but significant differences between the four $\sigma_0$ and $L_r$ bins.  It is worth noting that the largest $\sigma_0$ bin (red symbols) lies far above the grids associated with Kroupa or Salpeter IMFs (recall that IMF-1.3 is similar to Kroupa; see Appendix about comparison with Salpeter IMF).  Indeed, except perhaps for the outermost regions of the lowest $\sigma_0$ bin (blue symbols) all our measurements lie far above these commonly used IMFs. Unfortunately, turning these comparisons into statements about the IMF is complicated because TiO2$_{\rm SDSS}$ line strengths depend on the underlying abundances as well as the IMF, as we now discuss.

Figure~\ref{DTiO2sdss} shows how TiO2$_{\rm SDSS}$ changes when [X/Fe] varies from its solar value, for a number of choices of X.  All line strengths were computed using the \cite{CvD2012} model spectra (Chabrier IMF, solar metallicity, and an age of 13.5 Gyr) smoothed to our resolution of 300 km~s$^{-1}$.  Clearly, TiO2$_{\rm SDSS}$ is not simply related to any one X (e.g. [TiO/Fe]).  E.g., recall that [$\alpha$/Fe] is a combination of a number of different elements: the blue line shows how TiO2$_{\rm SDSS}$ varies with [$\alpha$/Fe] (note that [$\alpha$/Fe]$\ne 0$ for all the objects in our sample).

Unfortunately, the \cite{CvD2012} and MILES spectra return different TiO2$_{\rm SDSS}$-[$\alpha$/Fe] scalings.  E.g., when [$\alpha$/Fe]$=0.2$ (as suggested by Figure~\ref{MilesBI}) then $\Delta_{\rm CvD}$TiO2$_{\rm SDSS} = 0.0035$ (blue line at [X/Fe] = 0.2 in Figure~\ref{DTiO2sdss}). If we use the MILES models (rather than \citealt{CvD2012}) with the same age and metallicity (but a Kroupa IMF, which is very similar to Chabrier) then $\Delta_{\rm MILES}$TiO2$_{\rm SDSS}$ changes by 0.0019 (instead of 0.0035) between [$\alpha$/Fe]=0.2 and [$\alpha$/Fe]=0.  Differences for other line indices can be even more dramatic.  For example, the dependence of the measured [MgFe] index on [$\alpha$/Fe] from the \cite{CvD2012} spectra is completely different compared to MILES.  In principle, these differences depend on age, metallicity and IMF as well.

For all these reasons, we do not actually use the scalings shown in Figure~\ref{DTiO2sdss} in our analysis.  Rather, they merely serve to illustrate how one might try to interpret the measured $\Delta$TiO2$_{\rm SDSS}$ (i.e. observed - model) trends in terms of [X/Fe] enhancements. In what follows, we use $\Delta_{\rm [X/Fe]}$ to express the difference between the measured TiO2$_{\rm SDSS}$ from the stacked spectra and TiO2$_{\rm SDSS}$ from the MILES model (with age, metallicity and [$\alpha$/Fe] given by Figure~\ref{MilesBI}) due to variation in [X/Fe] enhancements (i.e., not to the IMF). Figure~\ref{TiOoptions} shows what these $\Delta_{\rm [X/Fe]}$ trends might be, and how they might be used to constrain the IMF.
  
\begin{table}
 \centering
 ASSUMED IMF OR $\Delta_{\rm [X/Fe]}$\\
 \begin{tabular}{ll}
  \hline
  ASSUM. 1       & Same IMF for all galaxies; no IMF gradients \\ 
  ASSUM. 2       & IMF can vary with $\sigma_0$ and $L_r$; no IMF gradients \\
  ASSUM. 3       & Variable IMF + gradients; $\Delta_{\rm [X/Fe]}$ = 0.003\\
  ASSUM. 4       & Variable IMF + gradients; $\Delta_{\rm [X/Fe]}$ = 0 \\
  \hline
  \hline
 \end{tabular}
 \caption{List of assumptions about IMF variations and the allowed enhancement in TiO2$_{\rm SDSS}$ with respect to different elements.}
 \label{tab:assumptions}
\end{table}

\begin{figure*}
  \centering
   \includegraphics[width=0.9\linewidth]{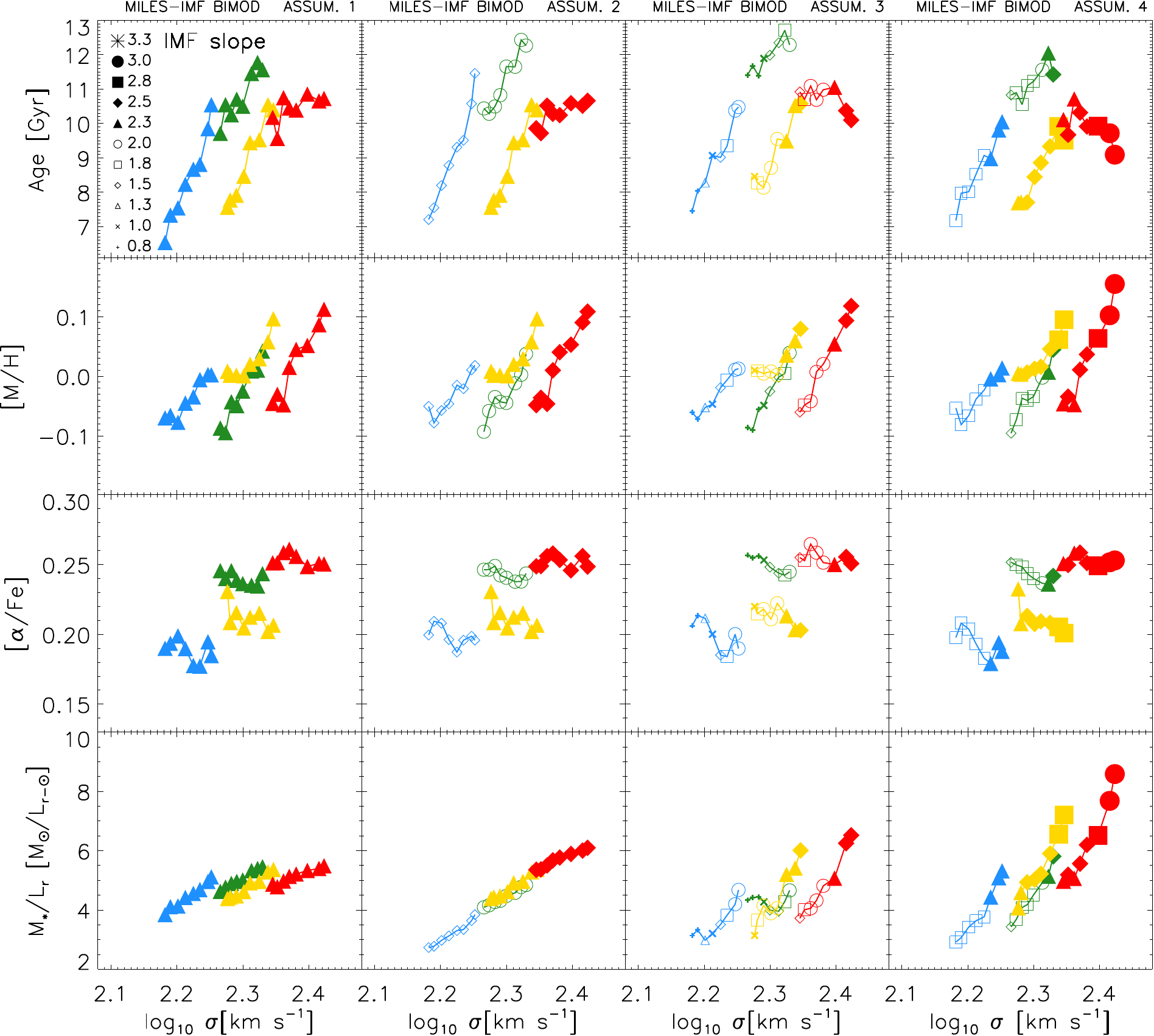}
   \caption{Inferred age, metallicity, [$\alpha$/Fe] and $M_*/L_r$ gradients in our $\sigma_0$ and $L_r$ bins associated with our four assumptions of Table~\ref{tab:assumptions}.
    Legend in the top left panel shows which symbol represents each IMF. Recall that the central regions correspond to the higher velocity dispersion values.}
  \label{ageZgrad}
\end{figure*}

First, suppose we assume that all galaxies have age, metallicity and [$\alpha$/Fe] as given by Figure~\ref{MilesBI}.  If we assume that all galaxies have the same IMF (hereafter ASSUMPTION~1), then the strong radial gradient in TiO2$_{\rm SDSS}$ shown in Figure~\ref{TiO2MilesBI} must be attributed to large gradients in $\Delta_{\rm [X/Fe]}$ since each IMF grid only spans a narrow range of TiO2$_{\rm SDSS}$ values.  The $\Delta_{\rm [X/Fe]}$ gradients associated with assuming the IMF has slope 2.3 for all galaxies are shown in the leftmost panel of Figure~\ref{TiOoptions}.  A milder version of this allows a different IMF for each $(L,\sigma_0)$ pair (hereafter ASSUMPTION 2) but does not allow gradients within a galaxy.  For the same reasons as before, the observed TiO2$_{\rm SDSS}$ gradients require gradients in $\Delta_{\rm [X/Fe]}$ which are shown in the panel that is second from left in Figure~\ref{TiOoptions}.  The symbols represent IMF slope:  for bins B00, B10, B11 and B21 we assumed this slope is 1.5, 2.0, 2.3 and 2.5.

ASSUMPTION~4 instead sets $\Delta_{\rm [X/Fe]}=0$ with respect to its [$\alpha$/Fe] value.  Then, Figure~\ref{TiO2MilesBI} implies large IMF gradients within a galaxy, as well as large IMF changes across the population.  The rightmost panel of Figure~\ref{TiOoptions} shows this case:  the fact that multiple symbols are needed for each $\sigma_0,L_r$ bin indicates that the IMF changes with radius.  Finally, ASSUMPTION~3 sets $\Delta_{\rm [X/Fe]}=0.003$; this implies slightly smaller IMF slopes but qualitatively similar gradients as for ASSUMPTION~4 (for each bin, there is a wide range of symbols in the second from right panel of Figure~\ref{TiOoptions}).

The IMF slopes and gradients associated with ASSUMPTIONS 3 and 4 are shown in Figure~\ref{IMFgrad}.  We use a different symbol for each IMF, as in the previous figure.  However, the IMF slope we show is got from interpolating between the two closest-fitting IMF models. While the IMFs are more bottom heavy for ASSUMPTION 4, in both panels, the IMF is more bottom heavy for large $\sigma_0$ or $L_r$, and it is more bottom heavy in the central regions.

\subsection{Age, metallicity and [$\alpha$/Fe] gradients}

We turn now to galaxy ages, metallicities, [$\alpha$/Fe] and $M_*/L_r$ values, paying particular attention to the robustness of our conclusions with respect to changes in the assumed $\Delta_{\rm [X/Fe]}$.  Figure~\ref{ageZgrad} shows the age, metallicity, [$\alpha$/Fe] and $M_*/L_r$ gradients associated with these scenarios. There are clear trends for age, metallicity and [$\alpha$/Fe] to increase with $\sigma_0$ across the population (bin B10 is peculiar, as we discuss shortly).  Whereas metallicity increases strongly towards the central regions in all four bins, this is less true for age:  moreover, ASSUMPTIONS~3 and~4, which allow for IMF gradients, show weaker age gradients (especially for bins B10 and B21).  Indeed, for Bin B21 (red symbols) there is almost a degeneracy between age and IMF, with the central regions (largest $\sigma$ values) prefering younger ages but more bottom-heavy IMFs.  In general, gradients in [$\alpha$/Fe] are weak, with a tendency to increase towards the outer regions. This is consistent with previous work \cite[e.g.][]{Vaughan2018a}. However, we find that this trend depends on the models used:  [$\alpha$/Fe] increases towards the center when using the TMJ models (see Appendix~A2).

Perhaps the most striking point is the anomalous behavior of the green symbols (B10). While age and metallicity tend to increase as $\sigma_0$ increases (from B00 to B11 to B21) the galaxies in bin B10 (green symbols) are the oldest, even though they are neither the most luminous, nor the ones with the largest $\sigma_0$.  They also have [$\alpha$/Fe] similar to B21 (red symbols) and enhanced relative to B11 (yellow), even though their $\sigma_0$ is the same as B11. These results are clearly visible in all panels.  Note that, because we were careful to restrict the sample to $z\le 0.08$, this age difference is not a consequence of lookback time differences between the B10 and the other samples.  These anomalies are discussed in more detail in our companion paper (Paper~II).

\subsection{$M_*/L_r$ gradients with varying IMF}

\begin{figure}
  \centering
   \includegraphics[width=0.9\linewidth]{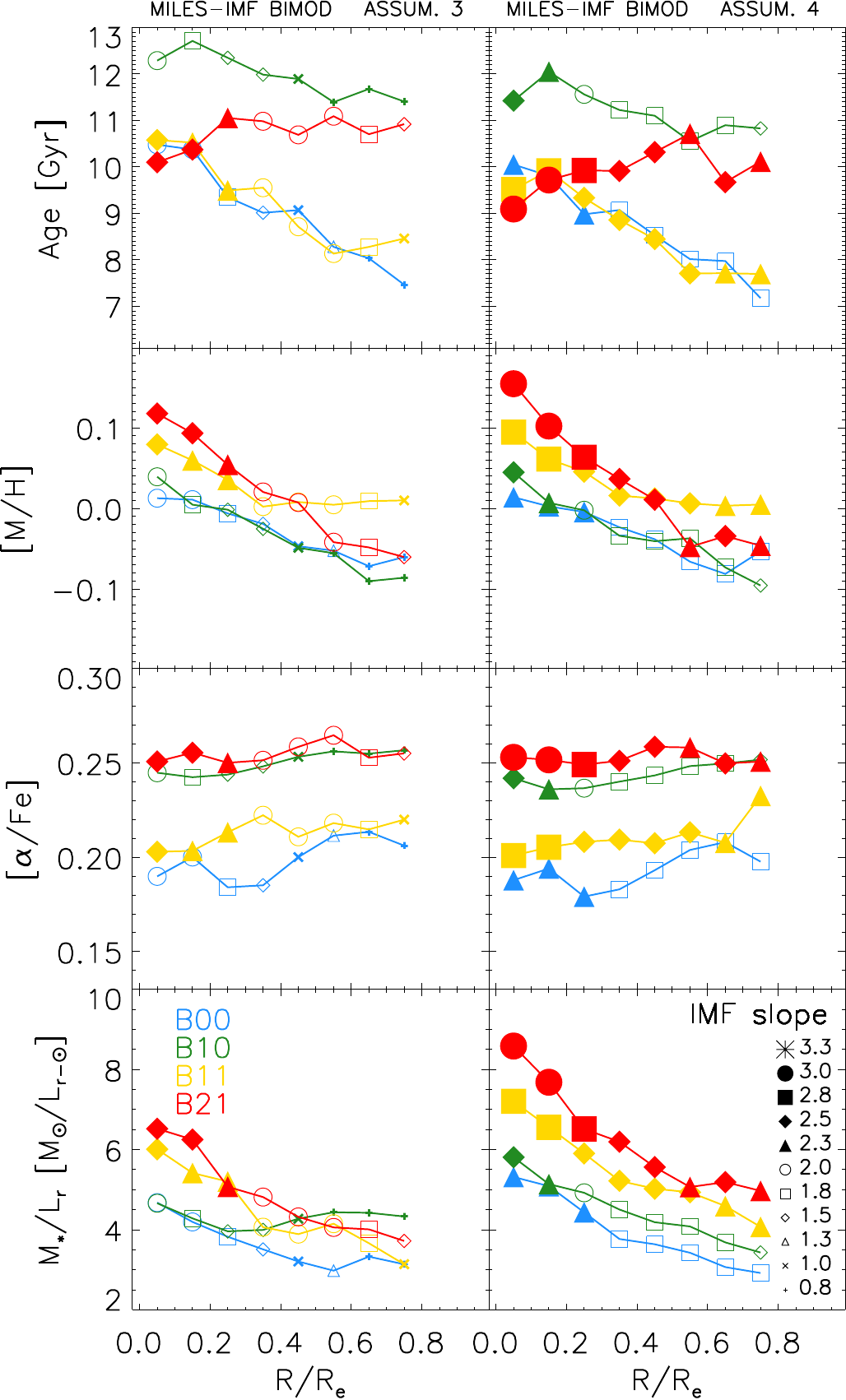}
  \caption{Inferred age, metallicity, [$\alpha$/Fe] and $M_*/L_r$ gradients  versus galactocentric distance in our $\sigma_0$ and $L_r$ bins associated with ASSUMPTION~3 (left) and ASSUMPTION~4 (right) (see text for details).}
  \label{allR}
\end{figure}

\begin{figure}
  \centering
  \includegraphics[width=0.49\linewidth]{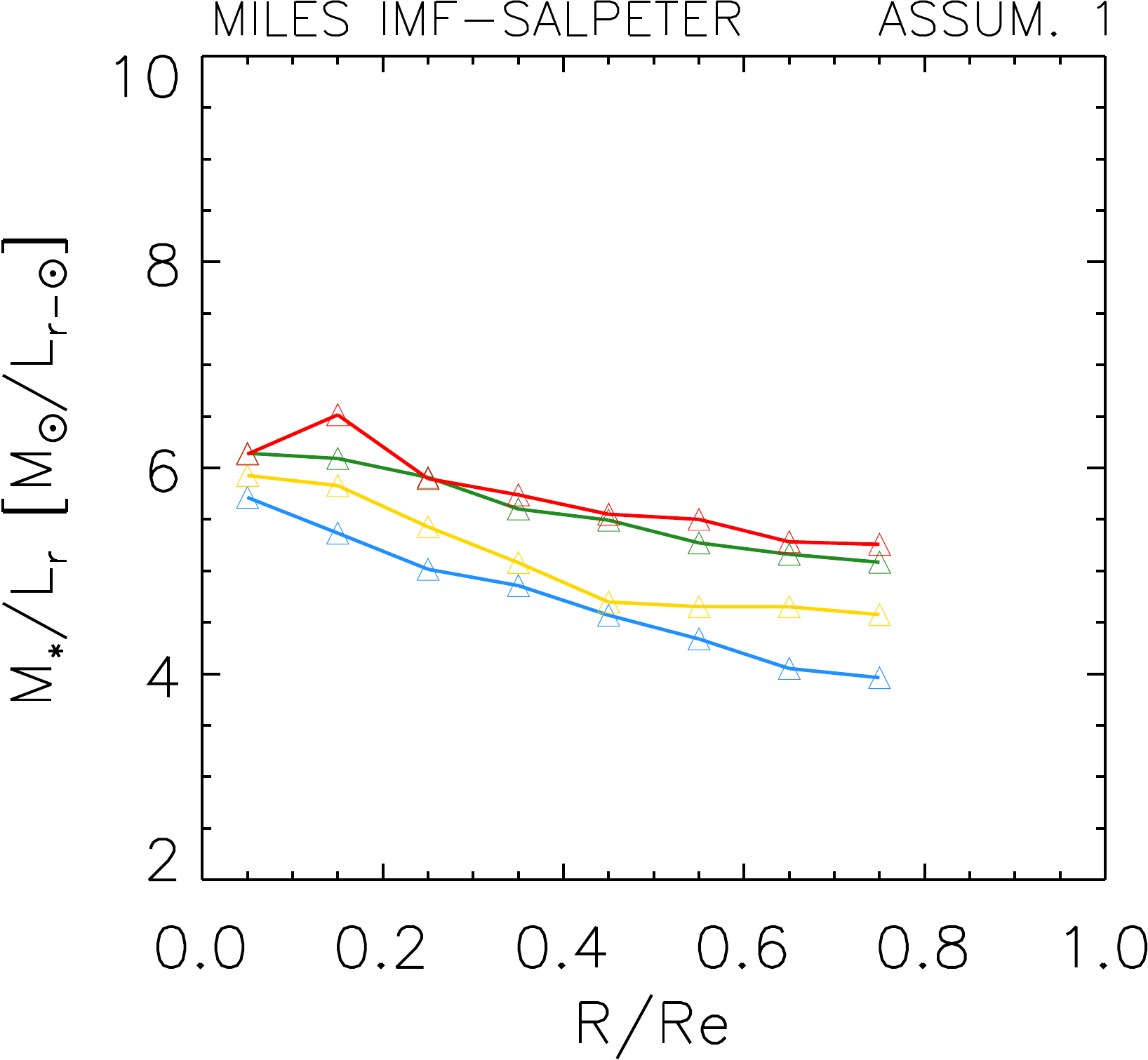}
  \includegraphics[width=0.49\linewidth]{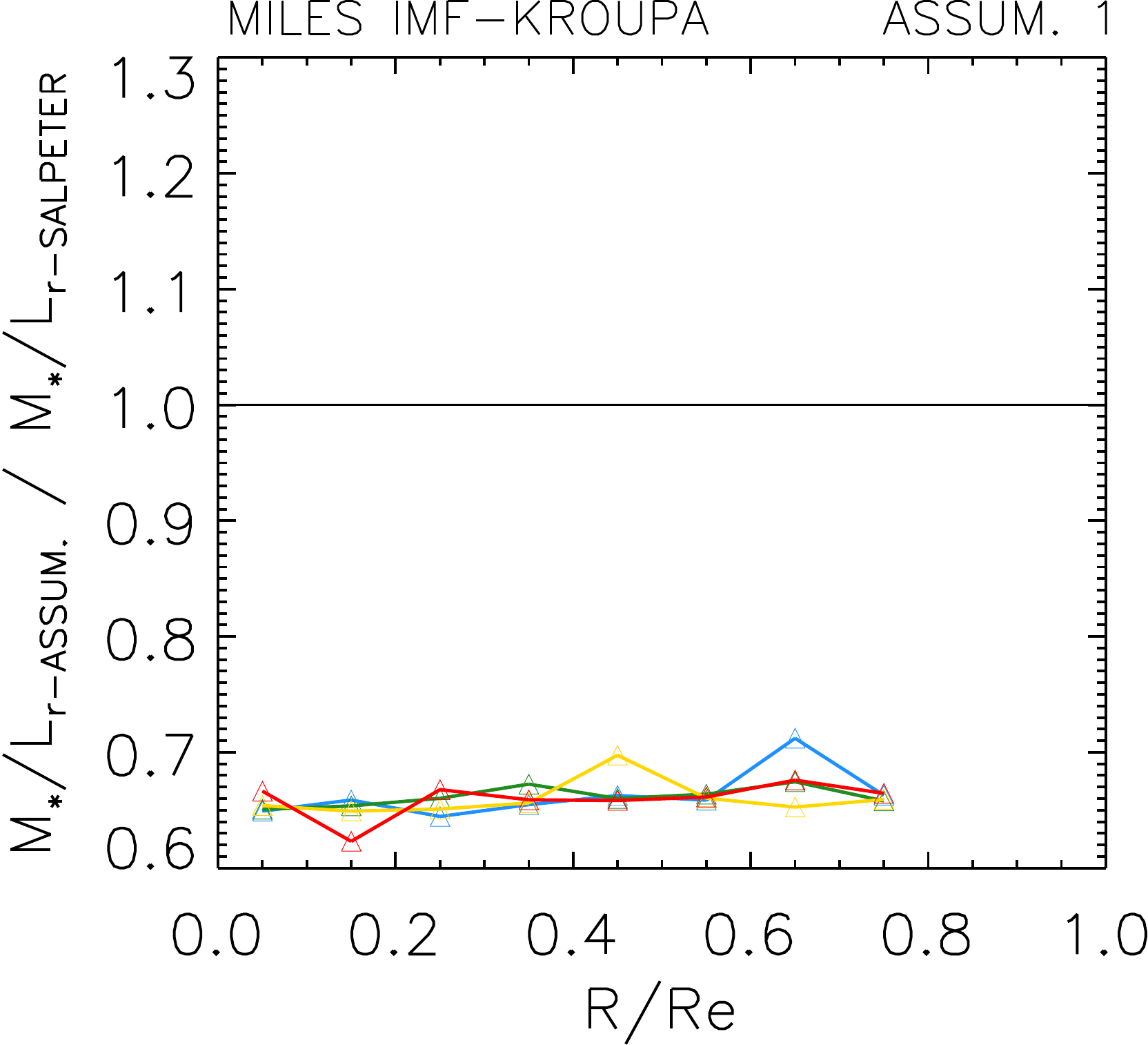}
  \vspace{-0.5cm}
  \caption{Left:  $M_*/L_r$ if the IMF is fixed to Salpeter on all scales for all galaxies. Right:  Ratio of $M_*/L_r$ to that for Salpeter, if the IMF is fixed to Kroupa for all scales for all galaxies.}
  \label{ratio2salp}
\end{figure}

\begin{figure}
  \centering
  \includegraphics[width=0.49\linewidth]{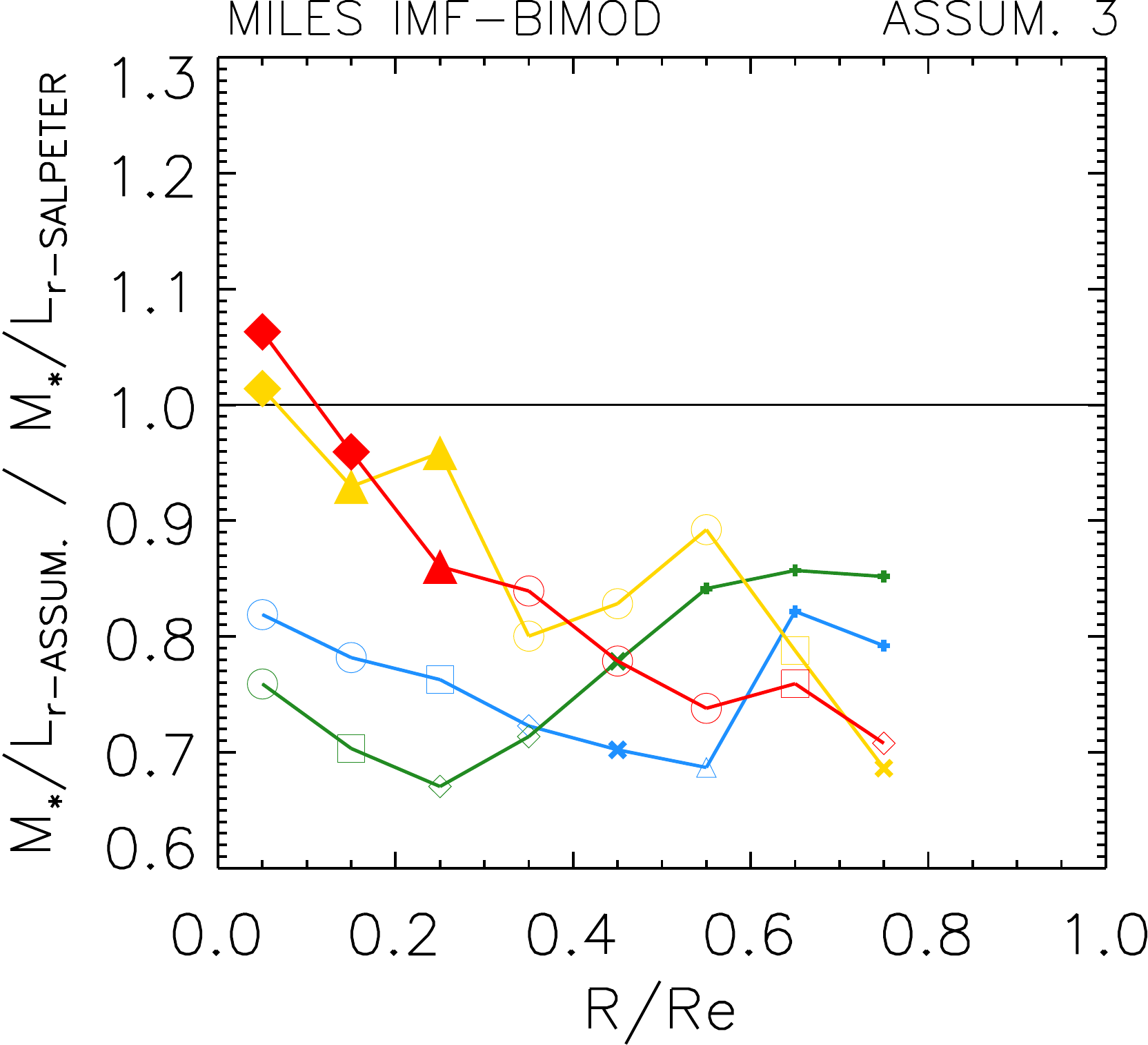}
  \includegraphics[width=0.49\linewidth]{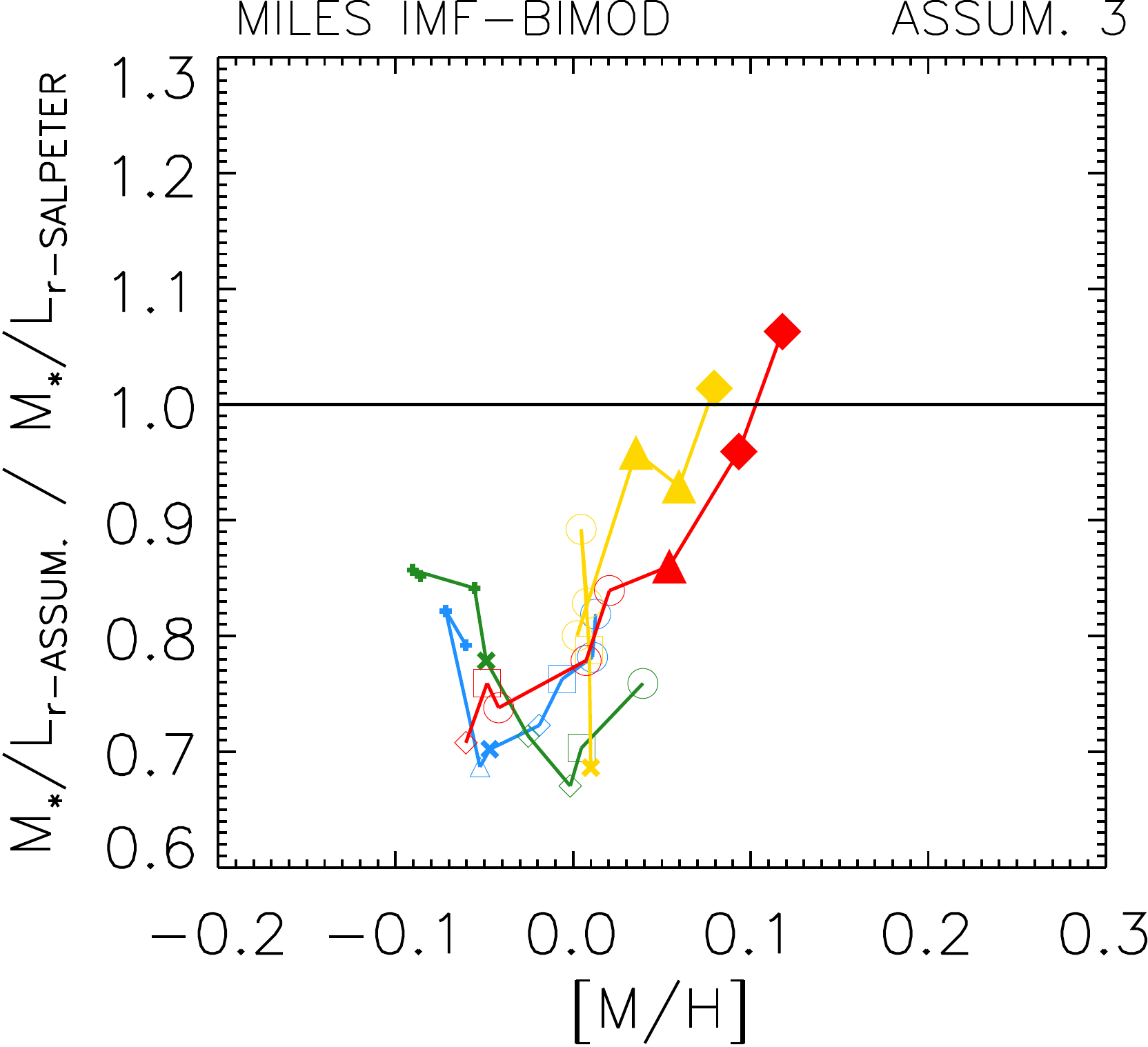}
  \vspace{-0.5cm}
  \caption{Ratio of $M_*/L_r$ shown in the bottom panels of Figure~\ref{allR} to that for Salpeter (left panel of Figure~\ref{ratio2salp}) as a function of radius (left) and metallicity (right).}
  \label{ratio2salp2}
\end{figure}

Finally, we turn to the question of $M_*/L_r$ gradients.  ASSUMPTIONS~1 and~2, in which there are no IMF gradients within galaxies, have the weakest $M_*/L_r$ gradients.  For these cases, $M_*/L_r$ decreases by less than 30\% from the inside out.  When the IMF is allowed to vary across the population but is fixed within a galaxy (ASSUMPTION~2), then $M_*/L_r$ is remarkably well-correlated with $\sigma$:  the same correlation with $\sigma$ applies within a galaxy and across the population (there is a single $M_*/L_r$-$\sigma$ correlation for all bins).  Allowing for IMF-gradients changes this:  each bin has its own $M_*/L_r$-$\sigma$ relation. In addition, allowing for IMF-gradients results in larger $M_*/L_r$ gradients (ASSUMPTIONS~3 and~4).  However, the increase is modest:  IMF-driven $M_*/L_r$ changes are never larger than a factor of $\sim 2$ out to 0.8 $R/R_e$.

Figure~\ref{allR} shows these gradients as a function of $R/R_e$ rather than $\sigma$ for ASSUMPTIONS~3 and~4 only. This format highlights the fact that although the galaxies in bin B10 have the oldest ages, largest [$\alpha$/Fe]-enhancements and lowest metallicities, their $M_*/L_r$ values are not extreme.  While their ages and [$\alpha$/Fe]-enhancements are like those in bin B21, their $M_*/L_r$ is lower, and much more like that for bin B00 (which has the same $L_r$).  This illustrates why it is important to fit for all stellar population properties, including the IMF, when determining $M_*/L_r$. The $M_*/L_r$ estimates for ASSUMPTION~3 are similar to but slightly smaller than for ASSUMPTION~4. (As discussed in Section~\ref{sec:MdMs}, ASSUMPTION~4 is problematic as it tends to produce $M_*$ estimates that exceed those based on the Jeans equation, whereas ASSUMPTION~3 is more reasonable.) 

To illustrate the difference between these $M_*/L_r$ values and those associated with the Kroupa or Salpeter IMFs which are commonly assumed, the left panel in Figure~\ref{ratio2salp} shows $M_*/L_r$ if we fix the IMF to be Salpeter for all galaxies (i.e. we estimated ages, metallicities, [$\alpha$/Fe] using ASSUMPTION~1 with IMF fixed to Salpeter). Note that in this case, the $M_*/L_r$ of bin B10 is more similar to that of bin B21 in contrast with the results of Figure~\ref{allR}. The right panel shows the result of repeating the analysis, but with the IMF fixed to Kroupa, and then dividing the $M_*/L_r$ which is returned by that for Salpeter (shown in the left hand panel).  This ratio is about 0.7 for all bins and scales, indicating that the $M_*/L_r$ gradient for Kroupa is just like that for Salpeter -- only the overall normalization is smaller.

Figure \ref{ratio2salp2} shows the ratio of the $M_*/L_r$ values in the bottom right panels of Figure~\ref{allR} (i.e. ASSUMPTION~3) to that shown in the  left panel of  Figure \ref{ratio2salp}  (i.e. fixed Salpeter IMF) versus radius (left panel) and  metallicity (right panel). This ratio is scale-dependent for the ellipticals with the largest $L_r$ and $\sigma_0$: in the central regions the $M_*/L_r$ estimate from our ASSUMPTION~3 is similar (or slightly higher) than that inferred by using a Salpeter IMF, decreasing to a Kroupa-like value by $\sim 0.8$ $R/R_e$. On the other hand, for lower $L_r$ and $\sigma_0$ ellipticals, $M_*/L_r$ of the central regions is similar to the outskirts and consistent with the value inferred if one uses a Kroupa IMF.  It is worth noting that the Kroupa value is a reasonable approximation to $M_*/L_r$ at $0.8\,R/R_e$ for {\em all} our bins (except B10), despite the fact that the IMF itself is {\em not} the same in all our bins (Figure~\ref{IMFgrad}).  

The right hand panel of Figure~\ref{ratio2salp2} shows that this ratio is tightly correlated with metallicity, and varies only weakly across the population. This agrees with previous work: E.g. \cite{MN2015b} argue that the IMF slope is approximately equal to $2.2 + 3\,$[M/H] for the BiModal models we are using here.

\subsection{Comparison with previous work}\label{sec:lit}
We now compare our results with previous work.  A direct comparison is not straight forward due to differences in methodology which include:
 sample selection;
 stellar population models used;
 IMF parametrization used;
 full spectral fitting versus Lick index measurements; 
etc. We will therefore only focus on works studying early-type galaxies (ETGs) where radial IMF variations have been allowed.

\cite{MN2015} present one of the first attempts to measure radial variations in the IMFs of ETGs.  They obtained spectra for 3 ETGs (two having $ \rm \sigma_0 = 300~km s^{-1}$, comparable to our B21-galaxies -- and one with $ \rm \sigma_0 = 100~km s^{-1}$, which is about $2\times$ smaller than the smallest $\sigma_0$ we probe). They used the MILES models with a bimodal parametrization of the IMF, which is one reason why we do as well.  For the two massive galaxies they found a significant IMF gradient (with IMF slope in the range $\Gamma_b=3.0-1.9$ out to 0.8 $ \rm R_e$).  This is consistent with our ASSUMPTION~4 estimates for B21 galaxies.  For their third (less massive) galaxy they found a rather flat IMF profile, whereas we see no significant difference in gradient strength at smaller $\sigma_0$.  (Whether this is because of the mismatch in the lowest $\sigma_0$ we probe is unclear.)  Using a similar approach, \cite{MN2015c} presented constraints on the IMF of a relic galaxy ($ \rm \sigma_0 = 430~km s^{-1}$), showing that it is bottom heavy at all radii ($\Gamma_b=3.0-2.5$). In both cases they obtain relatively flat age gradients, negative metallicity gradients and flat or negative $\alpha$-enhancement, in qualitative agreement with our ASSUMPTION~4 results, and slightly more bottom heavy IMFs than our ASSUMPTION~3 results.  In addition, we have already noted that we agree with \cite{MN2015b} that the IMF slope is tightly correlated with metallicity. However, as we discuss in Appendix~A, the precise scaling with [M/H] is model dependent.

Similar results were presented in \cite{LaBarbera2016} and \cite{LaBarbera2017}, where the MILES models were used to infer SSP parameters from long slit spectra of three massive ETGs ($ \rm \sigma_0 \sim 300~km s^{-1}$).  Flat, old age profiles (t$\sim$ 10 Gyr), strong, negative metallicity gradients and milder $\alpha$-enhanced gradients were found. They argue that the IMF slope seems to be better parametrized by a bimodal power law, is bottom heavy in the central regions, and decreases significantly with galactocentric distance ($\Gamma_b=3.0-1$), i.e., they find stronger IMF gradients than those from our ASSUMPTION~4. 

Strong IMF gradients with bottom-heavy IMFs in the central regions where also found by \cite{vD2017} after analysisng Keck spectra of 6 massive ETGs with updated \cite{CvD2012} models. Their results are broadly consistent with ours, except that they find a larger $M_*/L$ gradient than we do (they estimate $M_*/L$ changes by more than a factor of $3$ between the center and half the half light radius). We note that their IMF was parametrized by two logarithmic slopes set as free parameters, while we use a bimodal IMF with a fixed slope at masses larger than 0.6~$\rm M\sun$.  Finally, \cite{Sarzi2018} report a strong IMF gradient for M87 from interpreting MUSE integral field data with bimodal IMF MILES models. The implied $M_*/L$ change is milder (not more than a factor of $\sim 2$), in agreement with our ASSUMPTION~3. A significant IMF gradient in M87 was also found by \cite{Oldham&Auger2018} using multiple dynamical tracer populations to model the dark and luminous mass structure simultaneously.

All previous results are based on samples of galaxies at least ten times smaller than ours. \cite{Parikh2018}, on the other hand, used a sample with a similar size to ours ($\sim$300 galaxies). However, they binned their galaxies in mass bins and they included S0s in their sample, complicating the comparison with our results. (See Paper~II for why excluding S0s is important.) In addition, their IMF measurement is based on infrared indices (NaI and FeH) and a different set of SP models \citep{Thomas2011, Maraston2011, Villaume2017} with a unimodal IMF. Despite the differences in the methodology, their age, metallicity and $\alpha$-enhancement gradients in the most massive bin are consistent with our work.  The IMF gradients they derive when using NaI and the \cite{Maraston2011} models are in agreement with our ASSUMPTION 3, while they are slightly larger when obtained with NaI and the \cite{Villaume2017} models and inconsistent with the values based on FeH regardeless of the models used (i.e. they derive positive gradients). Using a Bayesian analysis, negative IMF slope gradients were also reported by \cite{Zhou2019} for their two most massive bins, while they found flat or even positive profiles for the less massive ETGs.

Finally, in contrast to previous work, we find an inverted age gradient for the most massive galaxies (bin B21):  their centers are slightly younger.  Inverted gradients, at the high mass end, were also reported by \cite{Zibetti2019}, which appeared while our work was being refereed.

To summarize, most previous work on IMF gradients has concluded that the central regions of massive ETGs are more bottom-heavy than Salpeter, but drop steeply at larger radii.  Our analysis based on ASSUMPTIONS 3 and 4 reproduces both trends.  However, ASSUMPTION~3 produces less bottom-heavy IMFs and slightly weaker gradients than ASSUMPTION~4.  As we show below, ASSUMPTION~4 IMFs are problematic as they tend to produce $M_*$ estimates that exceed those based on the Jeans equation, whereas ASSUMPTION~3 is more reasonable.

\section{Effect of gradients on stellar population and dynamical mass estimates}\label{sec:MdMs}

The $M_*/L_r$ gradients shown in the previous section are considerably smaller than those quoted by \cite{vD2017}.  Nevertheless, the analysis in \cite{Bernardi2018b} suggests that they are just large enough to matter for dynamical (Jeans-equation based) estimates of the stellar mass.  Therefore, we now check if gradients are necessary to reconcile stellar population and dynamical mass estimates.

\begin{figure}
 \centering
  \includegraphics[width=0.9\linewidth]{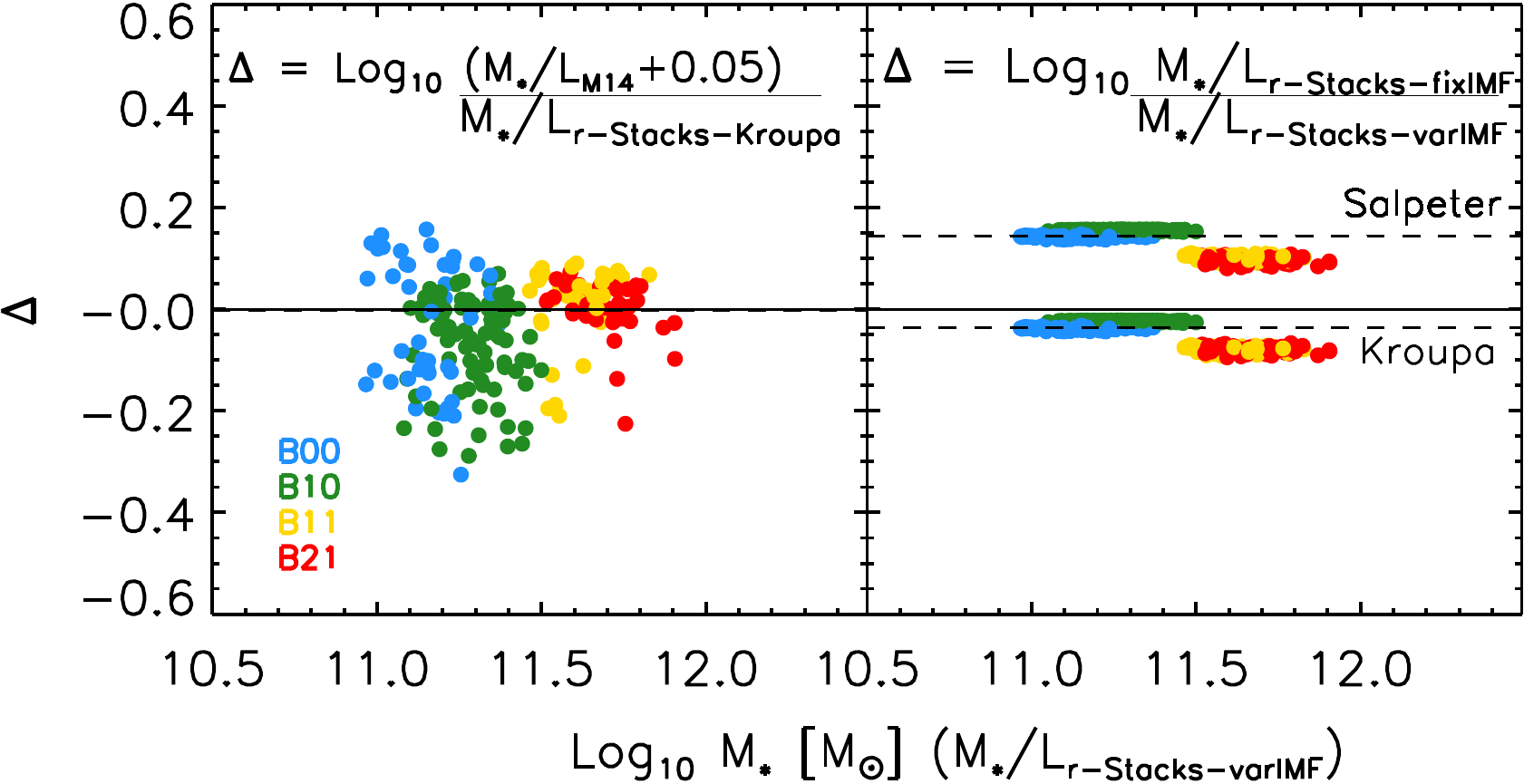}
  \caption{Comparison of integrated $M_*/L_r$ estimates. Left hand panel shows the values from Mendel et al. (2014) (shifted by 0.05 since Chabrier IMF based values differ from those from Kroupa IMF by 0.05~dex). Right hand panel shows the ratio of estimates from ASSUMPTION~1 (same IMF for all galaxies;  lower and upper set of symbols show results for fixed Kroupa and Salpeter IMFs) to ASSUMPTION~3 (IMF varies within a galaxy and across the population).}
  \label{fig:DMs}
\end{figure}

\begin{figure*}
 \centering
 \includegraphics[width=0.9\linewidth]{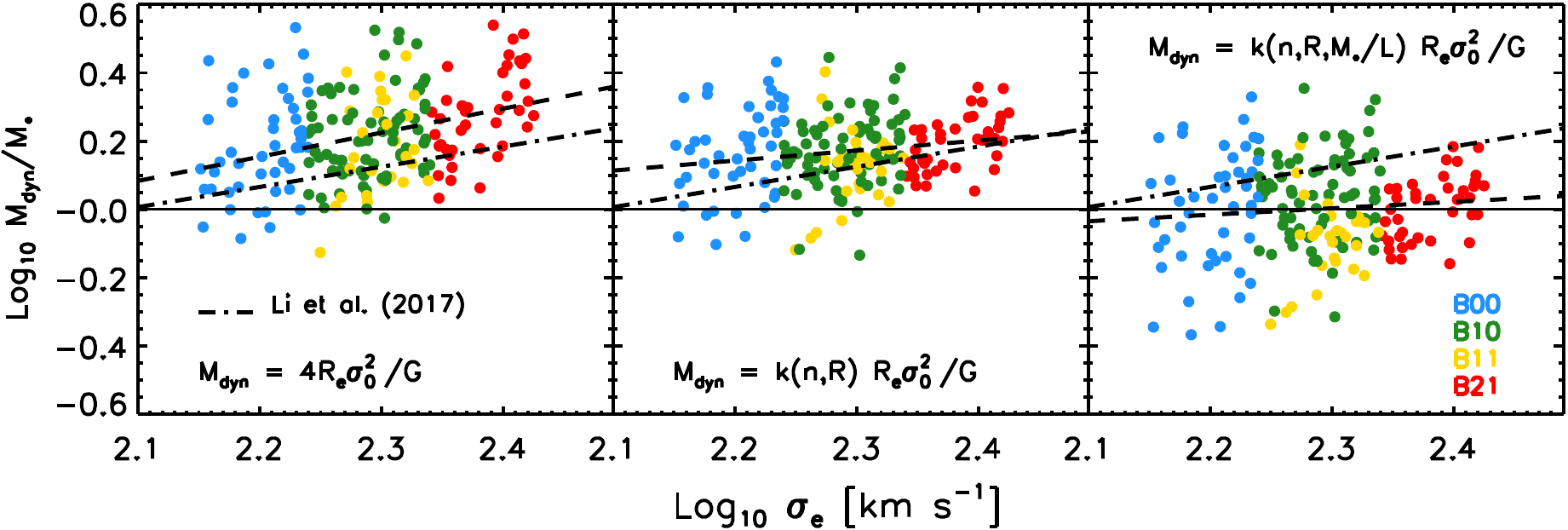}
  \caption{Comparison of three dynamical mass estimates with our stellar mass estimate from ASSUMPTION~3, in which IMF gradients contribute to $M_*/L_r$ gradients, versus $\sigma_e$.  All three panels have $M_{\rm dyn}\propto R_e\sigma_0^2/G$ with the proportionality constant the same for all galaxies (left), dependent on the light profile shape (middle) and on the product of the light and stellar mass-to-light ratio profiles which were used to estimate $M_*$ (right).  Dashed line shows how the ratio $M_{\rm dyn}/M_*$ scales with $\sigma_e$, and dot-dashed line shows the scaling reported by Li et al. (2017), offset slightly to account for the fact that the Salpeter IMF (their fiducial choice) has $\Delta=0.14$ in the right hand panel of our Figure~\ref{fig:DMs}. Dashed line is much shallower in the right hand panel:  self-consistently accounting for gradients with estimating $M_{\rm dyn}$ and $M_*$ removes the correlation between $M_{\rm dyn}/M_*$ and $\sigma_e$.}
  \label{fig:DMdMs}
\end{figure*}

\begin{figure}
 \centering
  \includegraphics[width=0.9\linewidth]{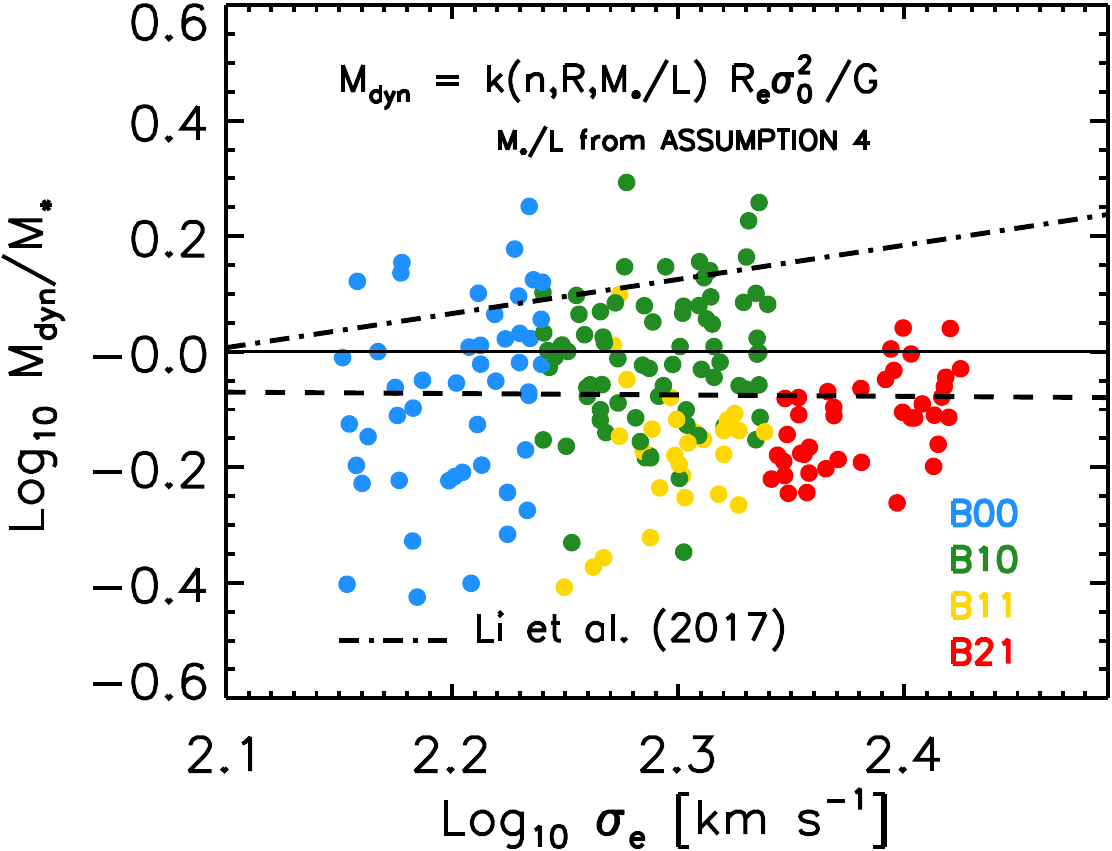}
  \caption{Same as right hand panel of Figure~\ref{fig:DMdMs} but using $M_*/L_r$ estimates from ASSUMPTION~4 instead of ASSUMPTION~3.  Here the SSP $M_*$ estimates tend to exceed the Jeans equation $M_{\rm dyn}$ estimates.}
  \label{fig:DMdMs4}
\end{figure}

We define the stellar mass of a galaxy as the sum over its circularized surface brightness profile (truncated at 7R$_e$) weighted by the $M_*/L_r$ profile corresponding to its bin in $L_r$ and $\sigma_0$ (see, e.g., bottom panels of Figure~\ref{allR}):  
\begin{equation}
  M_* \equiv 2\pi \sum_i \Delta R_i R_i\,I(R_i) \,(M_*/L_r)_i.
  \label{eq:Mstar}
\end{equation}
As a result, the $M_*$ estimate depends on what we assumed about the IMF.  Since our $M_*/L_r$ estimates only extend to about $R_e$ or so, and Figure~\ref{ratio2salp2} suggests that the Kroupa value is a reasonable approximation at $0.8\,R/R_e$ for all our bins except B10, we simply assume this remains true beyond $R_e$, and set $M_*/L_r$ to be that for Kroupa at large $R$.  We define an integrated or global mass-to-light ratio by dividing this $M_*$ estimate by the total luminosity $L_r$ (which is given by the same sum above but without the $M_*/L_r$ weight).

To begin, we first check that our integrated $M_*/L_r$ estimates are consistent with previous work (comparing integrated $M_*/L_r$ rather than $M_*$ estimates themselves removes systematics associated with the total luminosity $L_r$, see \citealt[][]{Bernardi2013,Bernardi2017a,Bernardi2017b,Fischer2017}). \cite{Mendel2014} provide $M_*$ and $L_r$, and hence $M_*/L_r$ estimates for these galaxies which are based on the assumption that the IMF is Chabrier across the population and that there are no gradients.  The closest IMF in our study is Kroupa, and it is well-known that $M_*/L_r$ from Chabrier differs from that for Kroupa by 0.05~dex \cite[e.g.][]{Bernardi10}.  Therefore, the left hand panel of Figure~\ref{fig:DMs} shows the ratio of the Mendel estimates, shifted by this amount, to our ASSUMPTION~1 Kroupa estimates.  For ease of comparison with the results which follow, we show (the log of) this ratio as a function of $M_*$ estimate returned by using the $M_*/L_r$ profiles of ASSUMPTION~3 (in which the IMF varies within a galaxy and across the population). The agreement with \cite{Mendel2014} is rather good considering that our estimate for each galaxy is obtained by using the average $M_*/L_r$ profile for its bin (this contributes to some of the scatter in Figure~\ref{fig:DMs}).

Having established consistency with the literature, the right hand panel compares our ASSUMPTION~1 estimates for two different IMF choices -- Kroupa and Salpeter -- with those from our ASSUMPTION~3.  If gradients did not matter, then the symbols would lie along the dashed lines shown.  The larger $L_r$ bins (orange and red) are clearly offset below:  for these more massive objects, gradients act to slightly increase the estimated stellar mass.  The effect is small because gradients are mainly present in the inner regions which contribute less than half the mass.  

Gradients are expected to affect dynamical mass estimates more dramatically \citep{Bernardi2018b}.  On dimensional grounds $M_{\rm dyn}\propto R_e\sigma^2/G$, so the estimate depends on the scale on which $\sigma$ is measured and the constant of proportionality.  In what follows, we use the central value $\sigma_0$ which is typically about 12\% larger than $\sigma_e$, the value averaged within the projected radius $R_e$ (e.g. Figure~\ref{fig:sigmaR}), and we explore three choices for the proportionality factor. Since this factor depends on the Sersic profile, here we show galaxies with {\tt FLAG$\_$FIT=1} ($\sim 64 \%$ of our E sample), i.e. whose photometry is better described by a single Sersic profile (see \citealt{Fischer2019} for details). Finally, it is conventional \citep{Cappellari2013b,Li17} to show the ratio $M_{\rm dyn}/M_*$ as a function of $\sigma_e$.

In Figure~\ref{fig:DMdMs}, we always use $M_*$ from ASSUMPTION~3; i.e., our $M_*$ values include the effects of gradients.  In the left hand panel we set $M_{\rm dyn} = 4R_e\sigma_0^2/G \approx 5R_e\sigma_e^2/G$ \cite[e.g.][]{McDermid2015}.  Notice that, on average $\log_{10}(M_{\rm dyn}/M_*)\approx 0.2$~dex.  Comparison with Figure~\ref{fig:DMs} suggests that, $M_{\rm dyn}$ would approximately equal $M_*$ if the IMF were Salpeter.  The dashed line shows that, in fact, $M_{\rm dyn}/M_*$ is not constant, but increases with $\sigma_e$.  For comparison, the dot-dashed line shows the $M_{\rm dyn}/M_*$-$\sigma_e$ scaling determined by \cite{Li17} (which ignored the effects of IMF gradients), shifted slightly to crudely account for the fact that, for their default IMF (Salpeter) $\Delta \sim 0.14$ in Figure~\ref{fig:DMs}.  The dashed lines is shifted upwards by $~\sim 0.1$~dex compared to the dot-dashed line.

The middle panel sets $M_{\rm dyn} = k(n, R)\, R_e\sigma_0^2/G$, with $k(n,R)$ given by Table~1 of \cite{Bernardi2018a}, where $n$ is the Sersic index of a single-component fit to the light profile and $R=0.1R_e$.  This $M_{\rm dyn}$ estimate accounts for the fact that galaxies have different light profiles, but assumes that $M_*/L$ is constant.  It further assumes that velocity dispersions are isotropic and the mass on sufficiently small scales is dominated by stellar rather than dark matter, and therefore normalizes the resulting Jeans-equation estimate to match the observed $\sigma_0$.  (Thus, $M_{\rm dyn}$ for each galaxy uses its $R_e$, $\sigma_0$, and light profile.)  These $M_{\rm dyn}$ values also are about 0.2~dex larger than our ASSUMPTION~3 based stellar mass estimates, and the scaling with $\sigma_e$ is still present. In this case the $M_{\rm dyn}/M_*$-$\sigma_e$ correlation is in good agreement with \cite{Li17}. This establishes consistency between ours and previous work, which has driven many to conclude that the IMF is Salpeter, or even super-Salpeter, at large $\sigma_e$ \cite[e.g.][]{Cappellari2013b, Li17}.  I.e., previous work argues that the discrepancy between the ATLAS$^{\rm 3D}$ $M_{\rm dyn}$ estimator and $M_*$ (based on a fixed IMF) is removed by increasing $M_*$  (i.e., by changing the default IMF choice in a $\sigma_e$-dependent way).  However, as we noted in the context of Figure~\ref{ratio2salp}, our analysis suggests that Salpeter-like bottom-heavy IMFs are only really seen in the central regions, so it is not obvious that this reconciliation of $M_{\rm dyn}$ and $M_*$ is self-consistent.

Of course, this comparison is unfair since ASSUMPTION~3 indicates that $M_*/L_r$ gradients are present, especially in the more massive galaxies.  For these galaxies, the IMF may be Salpeter-like in the central regions, but (except for bin B10) it is more Kroupa-like at $R_e$ (and, we assume, beyond).  The right hand panel shows the result of including the $M_*/L_r$ gradients shown in Figure~\ref{allR} -- and otherwise following the same methodology which was used for the middle panel -- when estimating $M_{\rm dyn}$ (see \citealt{Bernardi2018b}).  In this case, $M_{\rm dyn}$ for each galaxy uses its own $R_e$, $\sigma_0$, and light profile shape, but all galaxies in a bin have the same $M_*/L_r$ profile (c.f. Figure~\ref{allR}).  Comparison with the other two panels shows that accounting for $M_*/L_r$ gradients reduces the $M_{\rm dyn}$ estimate by $\sim 0.2$ dex and brings it into good agreement with the stellar population based $M_*$.  Moreover, the correlation with $\sigma_e$ is removed (dashed line is much flatter).

Figure~\ref{fig:DMdMs4} (similar to the right hand panel of Figure~\ref{fig:DMdMs}) compares $M_*$ and $M_{\rm dyn}$ using $M_*/L_r$ estimates from ASSUMPTION~4. In this case the $M_*$ estimates tend to exceed the Jeans-equation based estimates of $M_{\rm dyn}$.

Thus, our analysis shows that accounting self-consistently for the same gradients when estimating both $M_*$ and $M_{\rm dyn}$ and assuming a limited range of IMF and $\Delta_{\rm [X/Fe]}$ variations, brings the two into agreement by reducing $M_{\rm dyn}$ significantly and increasing $M_*$ slightly, rather than from increasing $M_*$ and leaving $M_{\rm dyn}$ unchanged.  This is a different resolution of the $M_*$-$M_{\rm dyn}$ discrepancy than has been followed in the recent literature. We can now specify the stellar mass scale identified by \cite{Bernardi2011}, $2\times 10^{11}M_\odot$ if the IMF were Chabrier, without also specifying an IMF.  The offset of $\sim 0.05$~dex from Chabrier to Kroupa combined with the $\sim 0.05$~dex offset between Kroupa and our variable IMF estimate (right hand panel of Figure~\ref{fig:DMs}), suggest that this scale is more like $3\times 10^{11}M_\odot$.

We end this section with two words of caution.  First, the IMF and $M_*/L_r$ values we derive are light- rather than mass-weighted estimates.  Second, they are strongly dependent on the SSP models used.  While our general age and  metallicity trends are quite robust regardless of the model (even for the peculiar bin B10), the IMF and $M_*/L_r$ values obtained using (our extension of) the TMJ models are significantly larger than those shown here (see Appendix~A for a more detailed comparison, and Figures~\ref{IMFgradTMJ} and~\ref{TMJgradients} in particular).  This raises the question of whether it is possible to constrain the absolute value of $M_*/L_r$.  This may be possible because the larger (IMF and) $M_*/L_r$ values associated with the TMJ models imply larger $M_*$ values.  This would not be problematic were it not for the fact that these models also produce strong gradients (Figure~\ref{TMJgradients}), the effect of which is to decrease the associated $M_{\rm dyn}$.  As a result, they have $M_* > M_{\rm dyn}$, which is unreasonable.  Thus, it may be that requiring $M_*\approx M_{\rm dyn}$ provides a useful constraint on single stellar population models.

\section{Conclusions}\label{sec:conclusions}
We measured a number of Lick indices (Table~\ref{tab:Lick}) using stacked spectra of $\sim 300$ MaNGA elliptical galaxies at $z \le 0.08$ (Figures~\ref{fig:spec}).  Each stack covers a narrow range in $\sigma_0$ and $L_r$ (Figure~\ref{fig:sample} and Table~\ref{tab:bin}).  In each bin, we found significant radial gradients in the line strengths (Figure~\ref{fig:lickR}).  We used SSP models to interpret the differences between bins, and the radial gradients within each bin (Figures~\ref{MilesBI} and~\ref{TiO2MilesBI}). 

\begin{itemize}
  
\item Age, metallicity and [$\alpha$/Fe] generally increase with $\sigma_0$ (and $L_r$).  However, galaxies with $\sigma_0$ between about $200-250$~km~s$^{-1}$ and $M_r$ between $-21.5$ and $-22.5$ (approximately $5-20\times 10^{10}M_\odot$) tend to be anomalously old, metal poor and [$\alpha$/Fe]-enhanced (Figure~\ref{ageZgrad}). We discuss a plausible explanation for this in our companion paper (Paper~II).

\item Except for the most massive bin (B21), galaxies are older towards the center. In all bins, metallicity increases towards the center. Whether or not they are more [$\alpha$/Fe]-enhanced towards the center is model-dependent:  MILES-based models have much weaker gradients than TMJ (compare Figures~\ref{allR} and~\ref{TMJgradients}).  These conclusions are qualitatively unchanged if we assume all galaxies have the same IMF, or if the IMF can vary across the population but is fixed within a galaxy (i.e. there are no IMF gradients), or if we allow IMF gradients (Figure~\ref{ageZgrad}).  However, age gradients are weaker if we allow IMF gradients.

\item When IMF variations are required (because $\Delta_{\rm [X/Fe]}$ variations are limited), the data indicate that the IMF is increasingly bottom-heavy (has a steeper slope) than the often-used Kroupa IMF towards the central regions (Figure~\ref{IMFgrad}) and tends to be more bottom-heavy for the largest $L_r$ and $\sigma_0$ galaxies.

\item It is important to fit for all stellar population properties, including the IMF, when determining $M_*/L$.  The $M_*/L$ ratio, and its dependence on distance from the center, is sensitive to IMF variations and other population gradients. Analyses which do not allow IMF-gradients imply a $\sim 30\%$ increase in $M_*/L_r$ from $R_e$ to the central regions (Figures~\ref{allR} and~\ref{ratio2salp}), and suggest that $M_*/L_r$ is the same function of $\sigma$ within a galaxy as it is of $\sigma_0$ across the population (Figures~\ref{ageZgrad}). If IMF-gradients are allowed, then this difference can be as large as a factor of 2, and the scaling with $\sigma$ within a galaxy differs from that across the population (Figures~\ref{ageZgrad} and~\ref{allR}).  We find a factor of $\sim 2$ decrease from the central regions to $R_e$ for the largest $L_r$ and $\sigma_0$ galaxies. However, at lower $L_r$ and $\sigma_0$, the IMF is shallower and the $M_*/L_r$ of central regions is similar to the outskirts (Figures~\ref{allR} and~\ref{ratio2salp2}).

\item Although the $M_*/L_r$ gradients we find are weaker than some recent estimates \citep{vD2017}, they are strong enough to impact Jeans-equation analyses of the dynamical mass of elliptical galaxies.  Ignoring gradients makes $M_{\rm dyn}$ about 0.2~dex larger than the stellar population estimate of the stellar mass (Figure~\ref{fig:DMs}). Accounting self-consistently for these gradients when estimating both $M_*$ and $M_{\rm dyn}$ brings the two into good agreement (Figure~\ref{fig:DMdMs}): gradients reduce $M_{\rm dyn}$ by $\sim 0.2$ dex while only slightly increasing the $M_*$ inferred using a Kroupa IMF.  This is a different resolution of the $M_*$-$M_{\rm dyn}$ discrepancy that has been followed in the recent literature where $M_*$ is increased while leaving $M_{\rm dyn}$ unchanged. In addition, requiring $M_*\le M_{\rm dyn}$ provides a useful constraint on single stellar population models.
    
\end{itemize}

As \cite{Bernardi2018b} note, our resolution of the $M_*$-$M_{\rm dyn}$ discrepancy affects estimates of the stellar mass density: the larger values implied by previous $M_{\rm dyn}$ analyses will result in overestimates.  Now that we have better determined how $M_*/L$ gradients vary over the population, it would be interesting to revisit Jeans equation analyses with these more realistic gradients to constrain the dark matter content and anisotropic velocity dispersions in elliptical galaxies (e.g. \citealt{Chae2018, Chae2019}).

\section*{Acknowledgements}

We are grateful to K. Westfall and T. Parikh for many helpful discussions about the MaNGA data, to C. Maraston, D. Thomas, A. Vazdekis and G. Worthey for correspondence about their SSP models, to I. Ferreras for his helpful comments about relevant prior work, and to the referee for a helpful report. This work was supported in part by NSF grant AST-1816330.

Funding for the Sloan Digital Sky Survey IV has been provided by the Alfred P. Sloan Foundation, the U.S. Department of Energy Office of Science, and the Participating Institutions. SDSS acknowledges support and resources from the Center for High-Performance Computing at the University of Utah. The SDSS web site is www.sdss.org.

SDSS is managed by the Astrophysical Research Consortium for the Participating Institutions of the SDSS Collaboration including the Brazilian Participation Group, the Carnegie Institution for Science, Carnegie Mellon University, the Chilean Participation Group, the French Participation Group, Harvard-Smithsonian Center for Astrophysics, Instituto de Astrof{\'i}sica de Canarias, The Johns Hopkins University, Kavli Institute for the Physics and Mathematics of the Universe (IPMU) / University of Tokyo, Lawrence Berkeley National Laboratory, Leibniz Institut f{\"u}r Astrophysik Potsdam (AIP), Max-Planck-Institut f{\"u}r Astronomie (MPIA Heidelberg), Max-Planck-Institut f{\"u}r Astrophysik (MPA Garching), Max-Planck-Institut f{\"u}r Extraterrestrische Physik (MPE), National Astronomical Observatories of China, New Mexico State University, New York University, University of Notre Dame, Observat{\'o}rio Nacional / MCTI, The Ohio State University, Pennsylvania State University, Shanghai Astronomical Observatory, United Kingdom Participation Group, Universidad Nacional Aut{\'o}noma de M{\'e}xico, University of Arizona, University of Colorado Boulder, University of Oxford, University of Portsmouth, University of Utah, University of Virginia, University of Washington, University of Wisconsin, Vanderbilt University, and Yale University.





\bibliographystyle{mnras}
\bibliography{biblio} 




\appendix

\section{Other SSPs and IMF-indicators}
The main text presented results based on the MILES-Padova SSPs with a wide range of IMFs (what they call BiModal), and used TiO2$_{\rm SDSS}$ as the primary IMF-diagnostic.  Here we show what is possible with other SSPs, and other TiO indicators.

\begin{figure}
  \centering
  \includegraphics[width=0.9\linewidth]{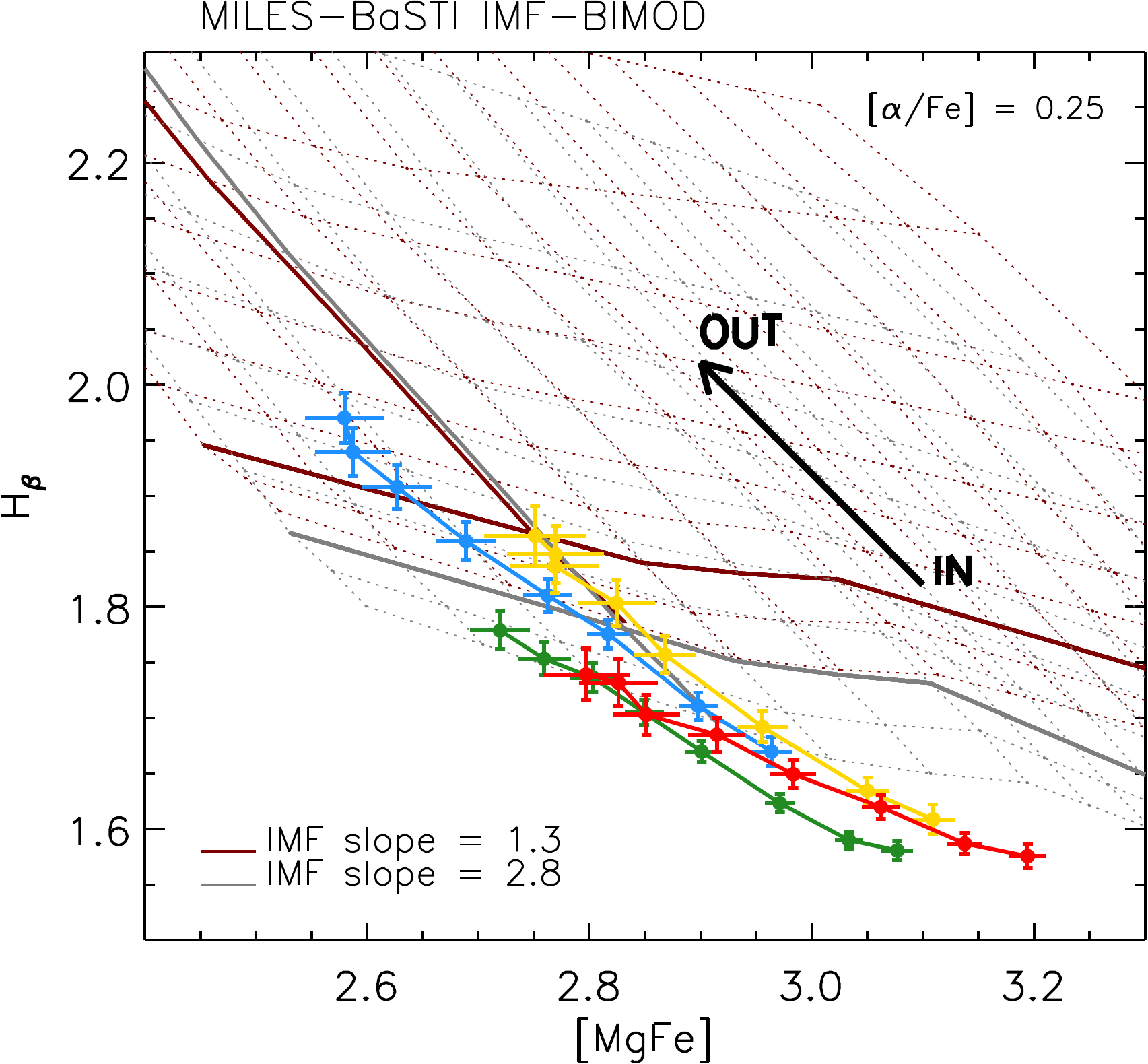}
  \caption{Same as top panel of Figure~\ref{MilesBI} but for the MILES-BaSTI models. Ages returned by these models are unphysically old.}
  \label{BaSTI}
\end{figure}

\begin{figure*}
  \centering
  \includegraphics[width=0.32\linewidth]{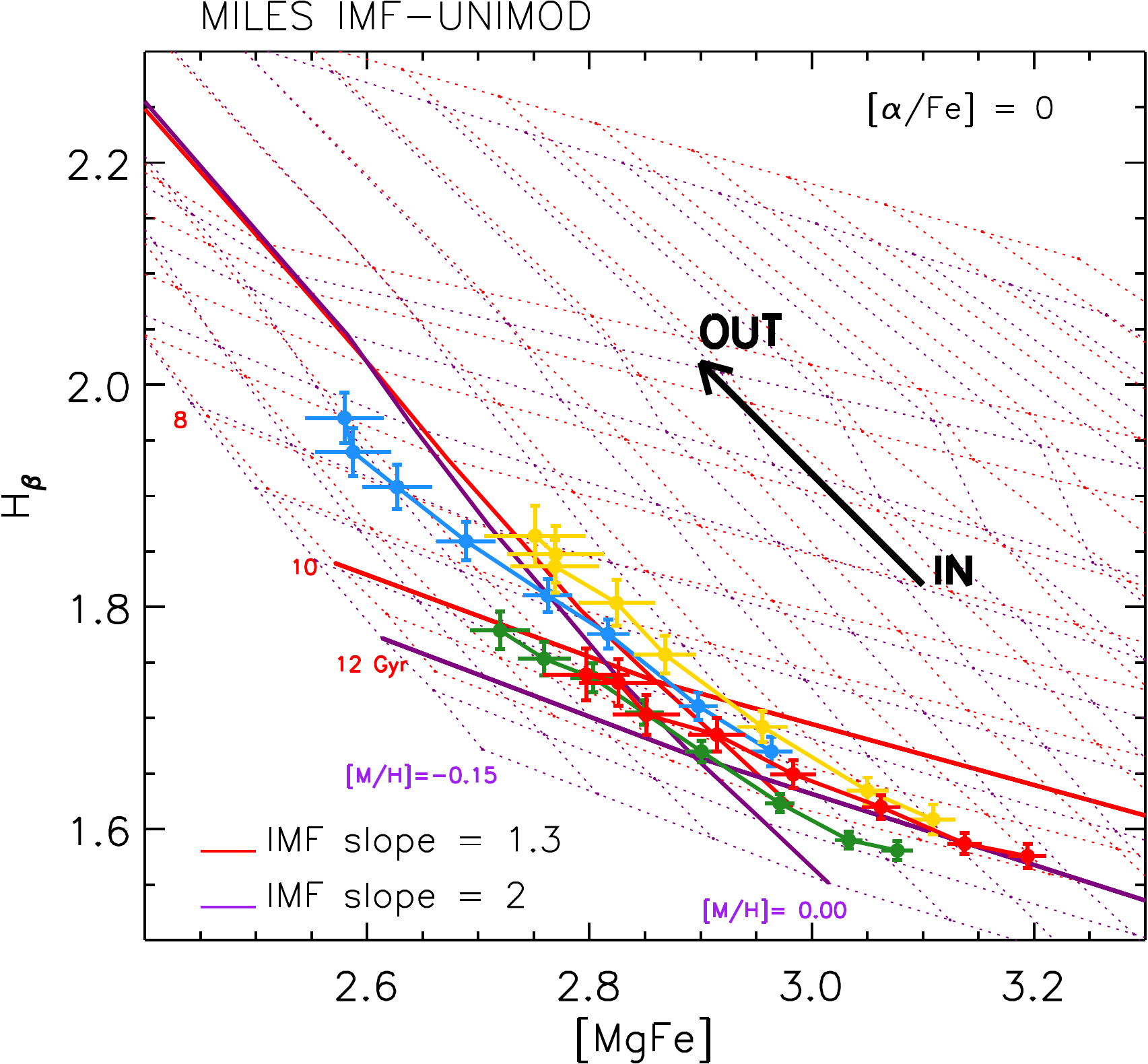}
  \includegraphics[width=0.32\linewidth]{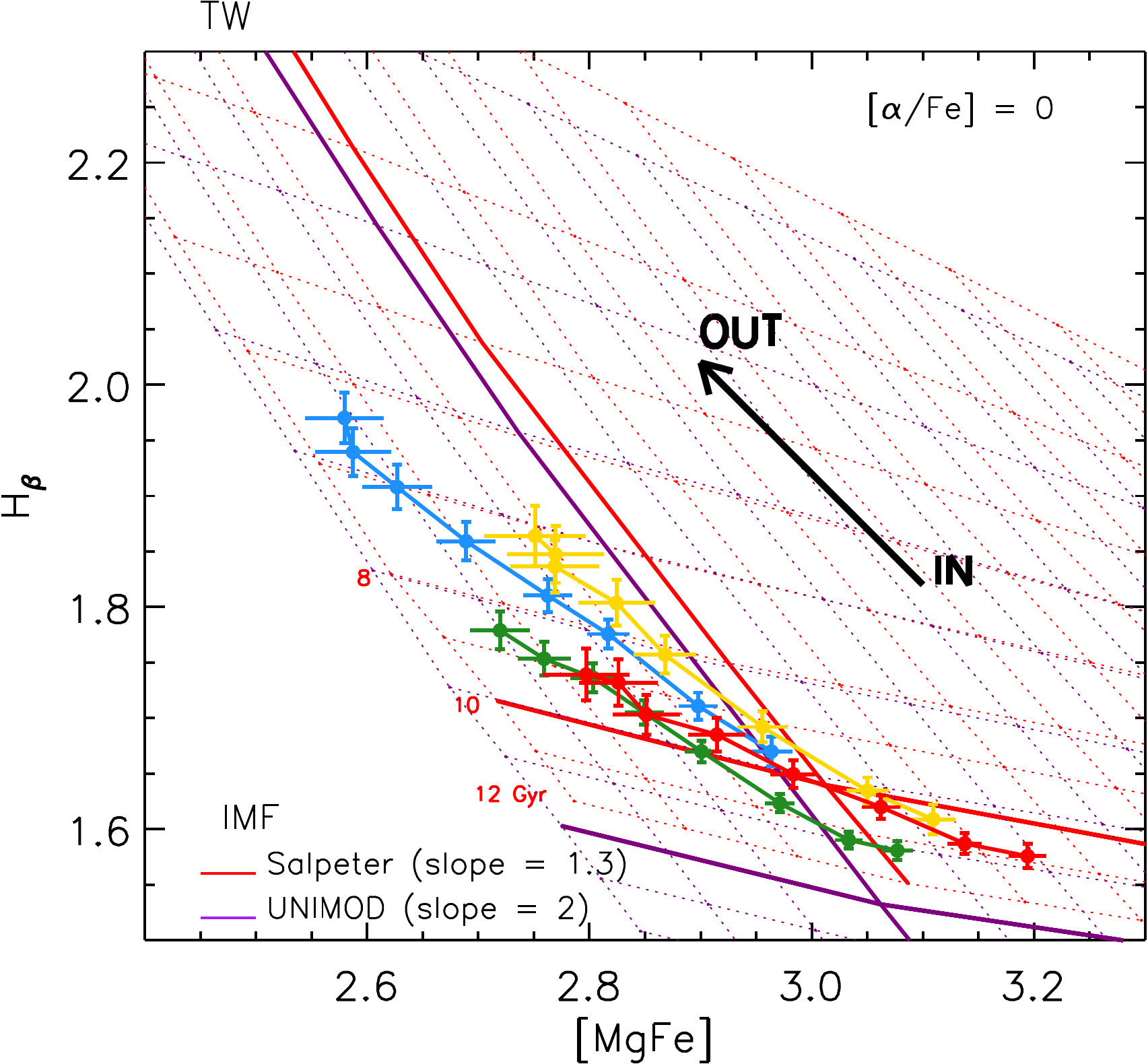}
  \includegraphics[width=0.32\linewidth]{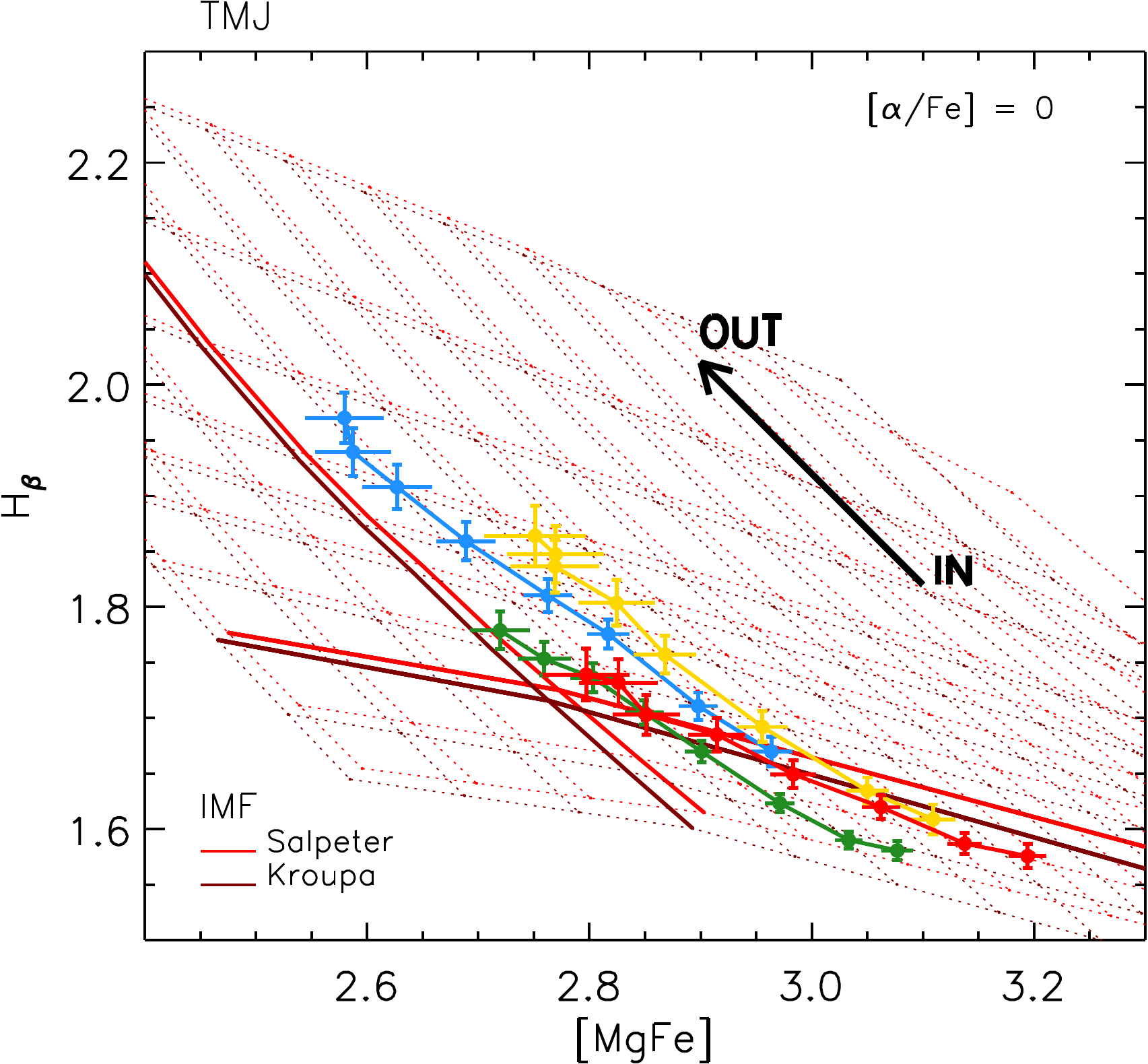}
  \includegraphics[width=0.32\linewidth]{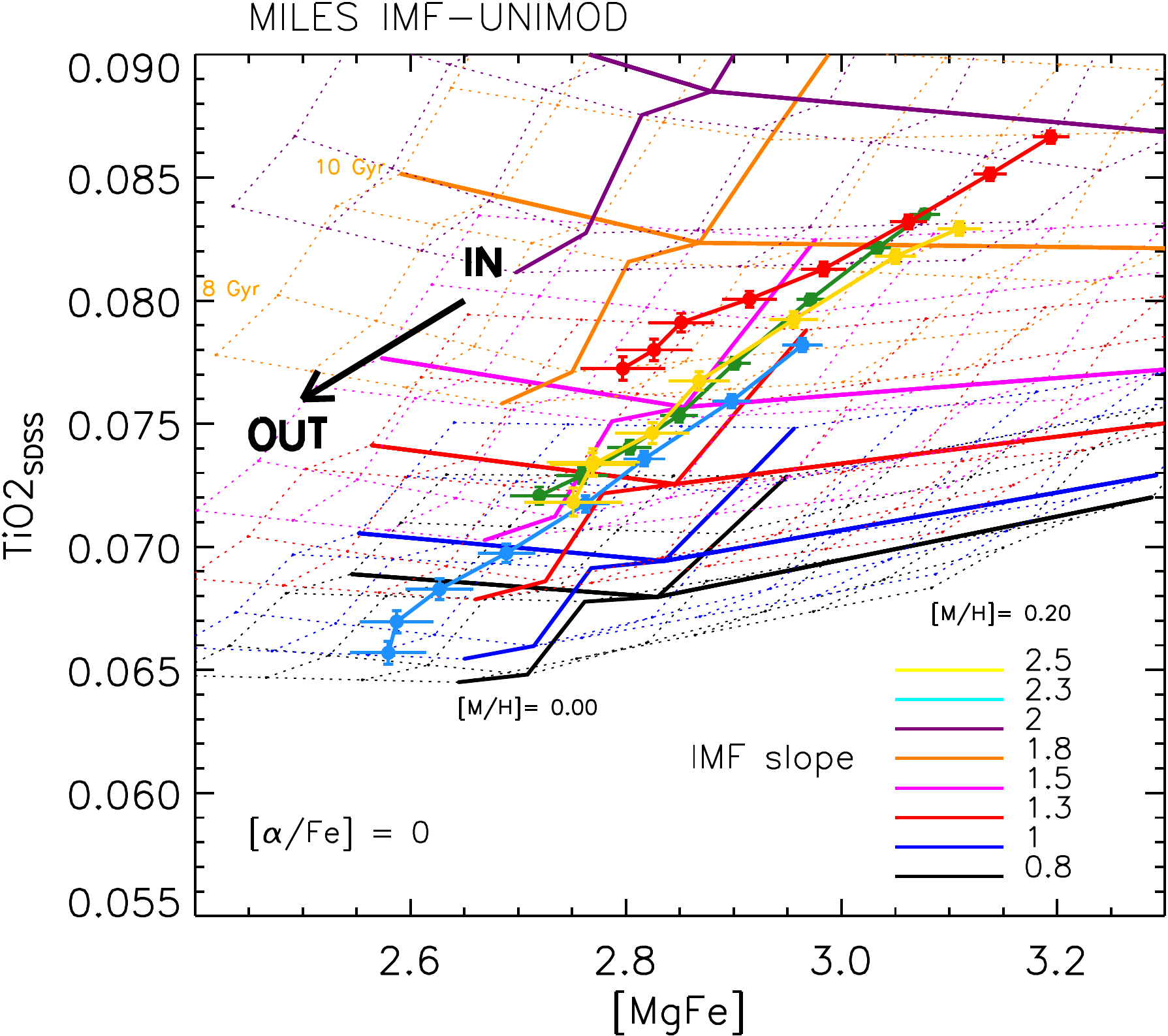}
  \includegraphics[width=0.32\linewidth]{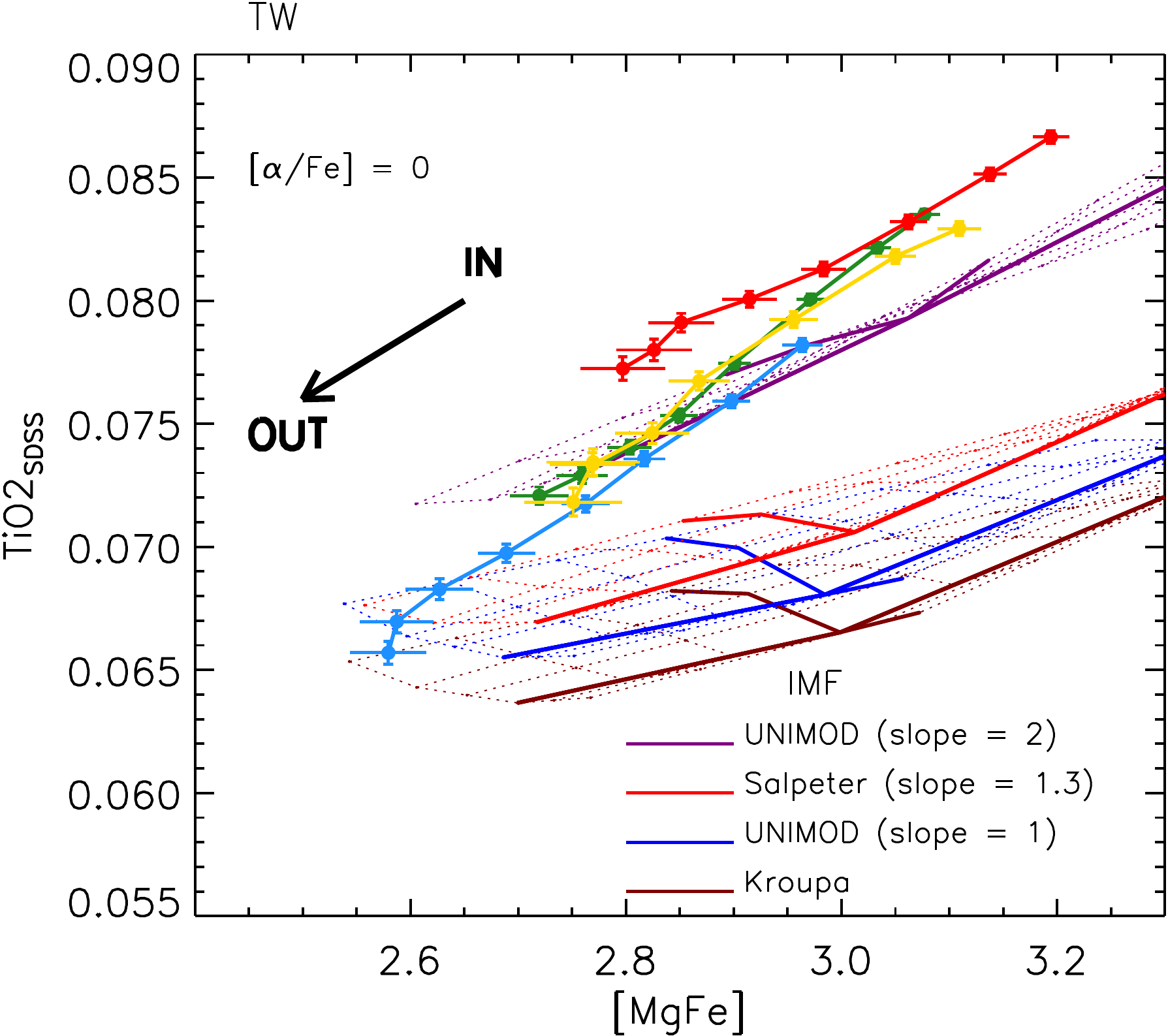}
  \includegraphics[width=0.32\linewidth]{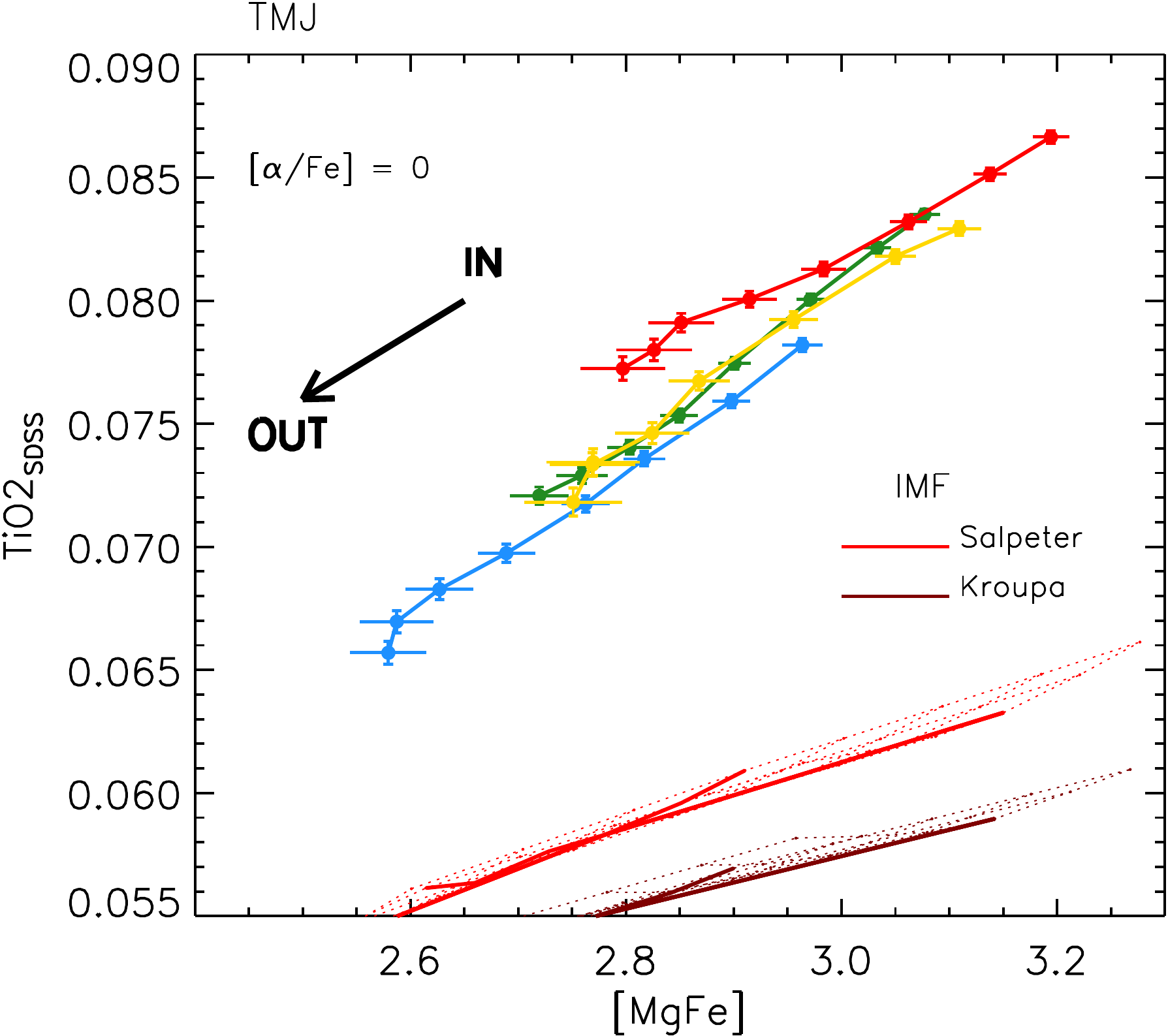}
  \caption{Same as Figures~\ref{MilesBI} and~\ref{TiO2MilesBI} but for the MILES-Padova UniModal models (left), the TW models (middle) with the same IMFs, and the TMJ models (right).}
  \label{compare}
\end{figure*}

\begin{figure*}
  \centering
  \includegraphics[width=0.32\linewidth]{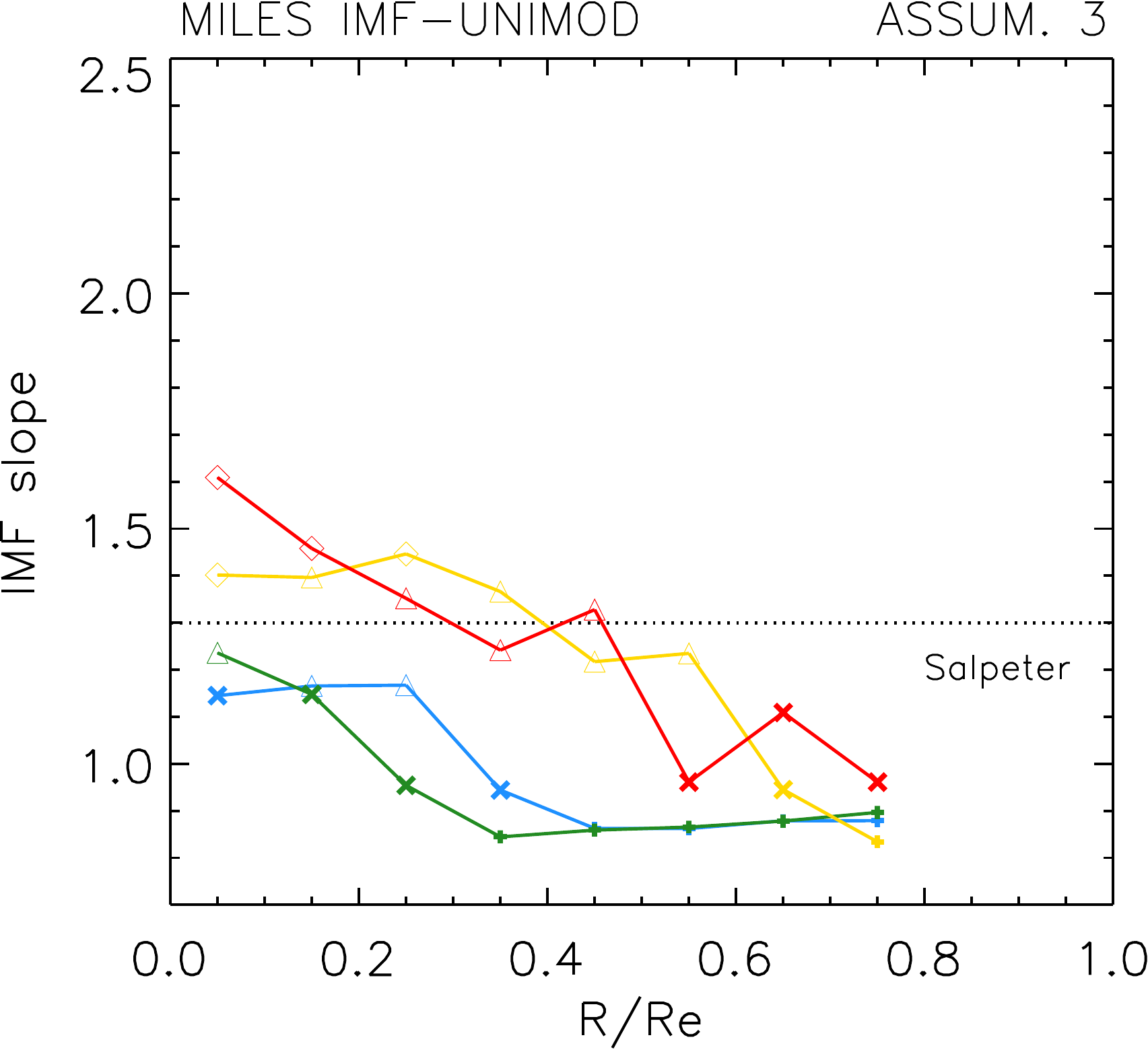}
  \includegraphics[width=0.32\linewidth]{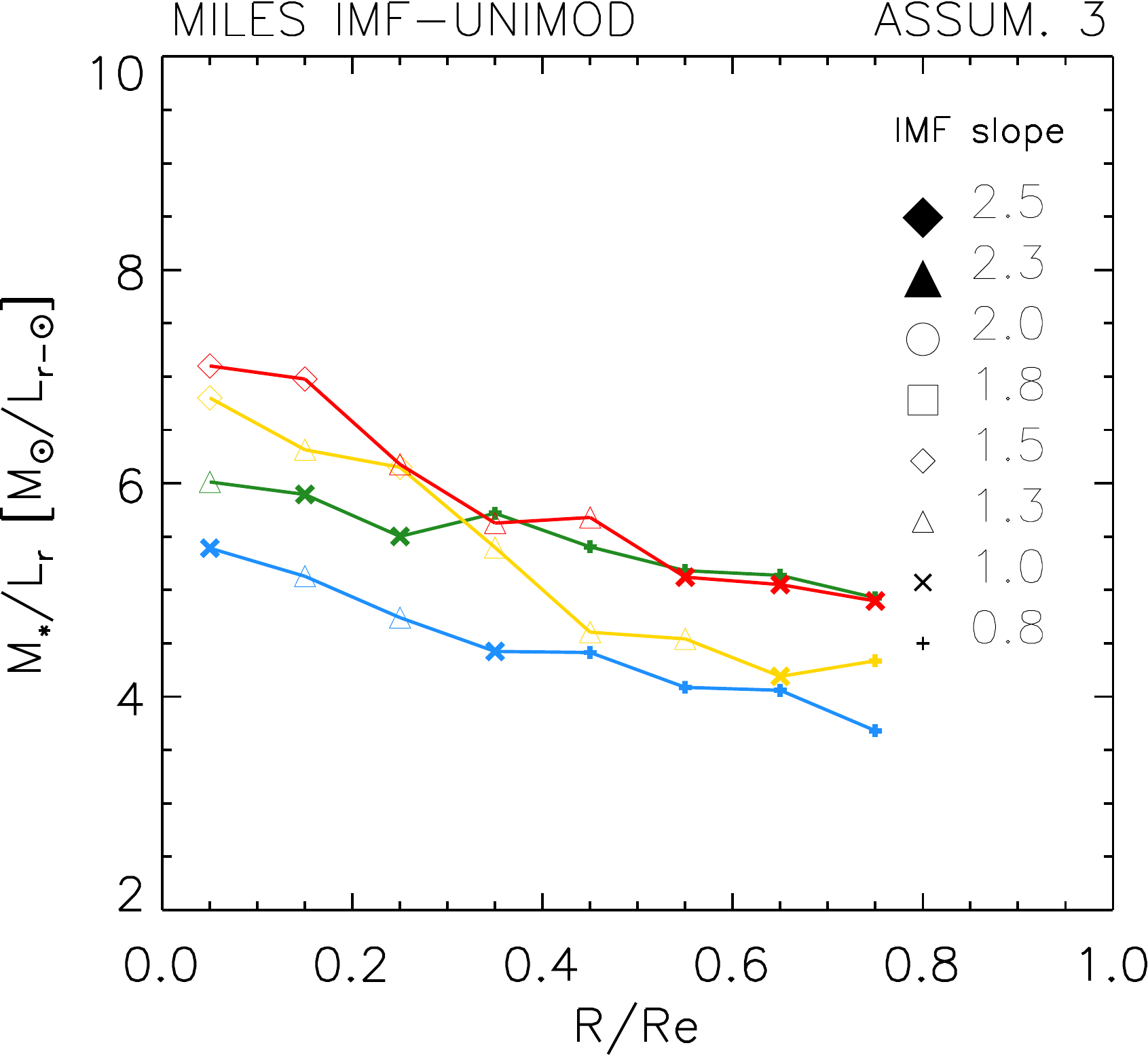}
  \includegraphics[width=0.32\linewidth]{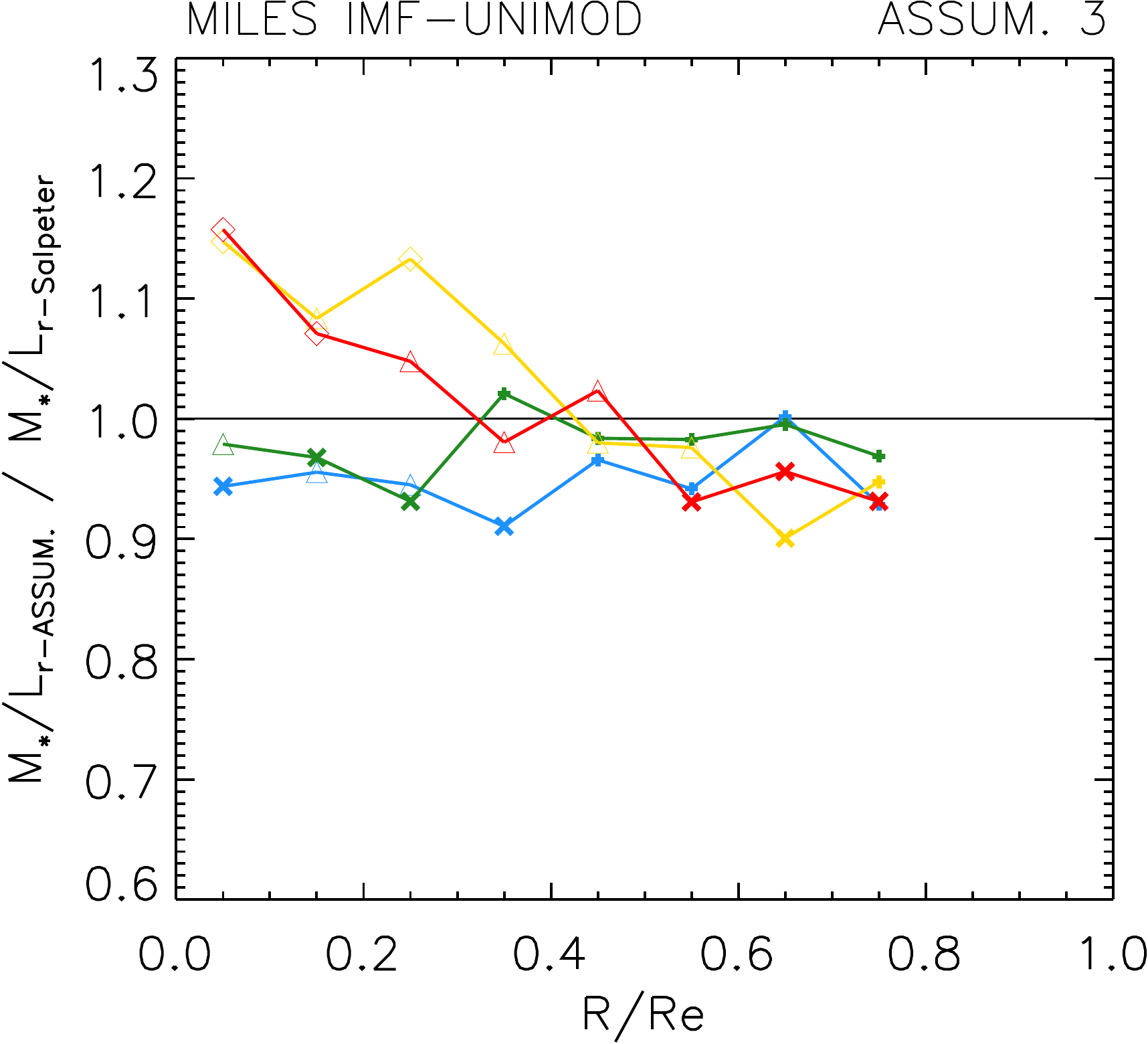}
  \caption{Left: IMF slope gradients (similar to left panel of Figure \ref{IMFgrad}) inferred by using a varible IMF (ASSUMPTION~3) but using the MILES-Padova UniModal (instead of BiModal) models. Middle: Corresponding $M_*/L_r$ gradients (similar to bottom right panel of Figure~\ref{allR}). Right: Ratio of $M_*/L_r$ inferred by using a variable IMF to that for Salpeter (similar to left panel of Figure~\ref{ratio2salp2}).}
  \label{IMFgradUNI}
\end{figure*}

\subsection{Other SSPs}
Figure~\ref{BaSTI} shows the H$_\beta$-[MgFe] grids associated with the MILES-BaSTI as opposed to the MILES-Padova models.  These grids assign ages in excess of 15~Gyrs to the oldest galaxies in our sample.  This is unreasonable, and is why we work with the MILES-Padova models in the main text.

Figure~\ref{compare} compares the MILES-Padova UniModal, \cite{TW2017} and \cite{TMJ2011} models (TW and TMJ hereafter), all for [$\alpha$/Fe]=0.  Although it is known that [$\alpha$/Fe]$\ne$0, for TW only [$\alpha$/Fe]=0 grids are currently available.  Fortunately, changing [$\alpha$/Fe] does not make a dramatic difference for the ages and metallicities.

The MILES and TW grids are for the same two IMFs:  the top-center panel shows that the TW grids imply sub-solar metallicities for all but the center-most parts of all four bins.  The bottom left panel shows that the MILES Uni-Modal grids cover our measurements; one of the reasons why we used the BiModal grids in the main text is that the BiModal grids provide more closely spaced IMFs. The bottom-center panel shows that the TW grids just barely cover our measurements; unfortunately, only [$\alpha$/Fe]=0 is available for these models, but we know that [$\alpha$/Fe]=0 is not realistic.  This is why we have not used these models further.  Note, in addition, that these TiO2$_{\rm SDSS}$-[MgFe] grids slope up and to the right; this is qualitatively different from the MILES models.

The TMJ models are available only for two IMFs:  Kroupa and Salpeter.  Comparison of the top right panel of Figure~\ref{compare} with Figure~\ref{MilesBI} shows that the TMJ models return older ages and higher metallicities (we show shortly that they also return higher [$\alpha$/Fe]).  Nevertheless, the relative differences between the $(\sigma_0,L)$ bins is similar.  In particular, galaxies in the intermediate $\sigma_0$ but small $L_r$ bin (green symbols) are found to be older than those in the two bins having higher-$L_r$ (yellow and red symbols).

\begin{figure*}
  \centering
  \includegraphics[width=0.32\linewidth]{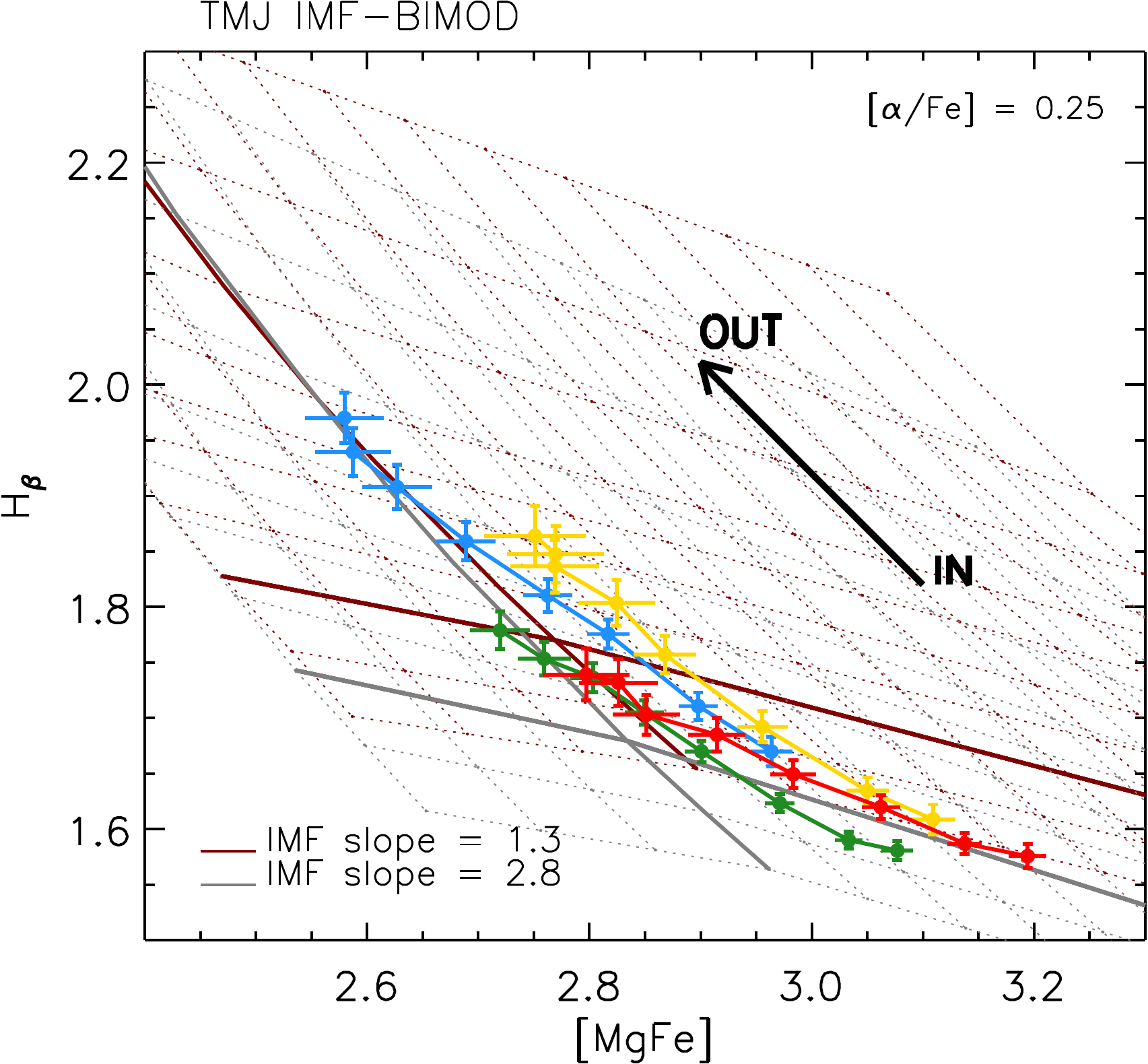}
  \includegraphics[width=0.32\linewidth]{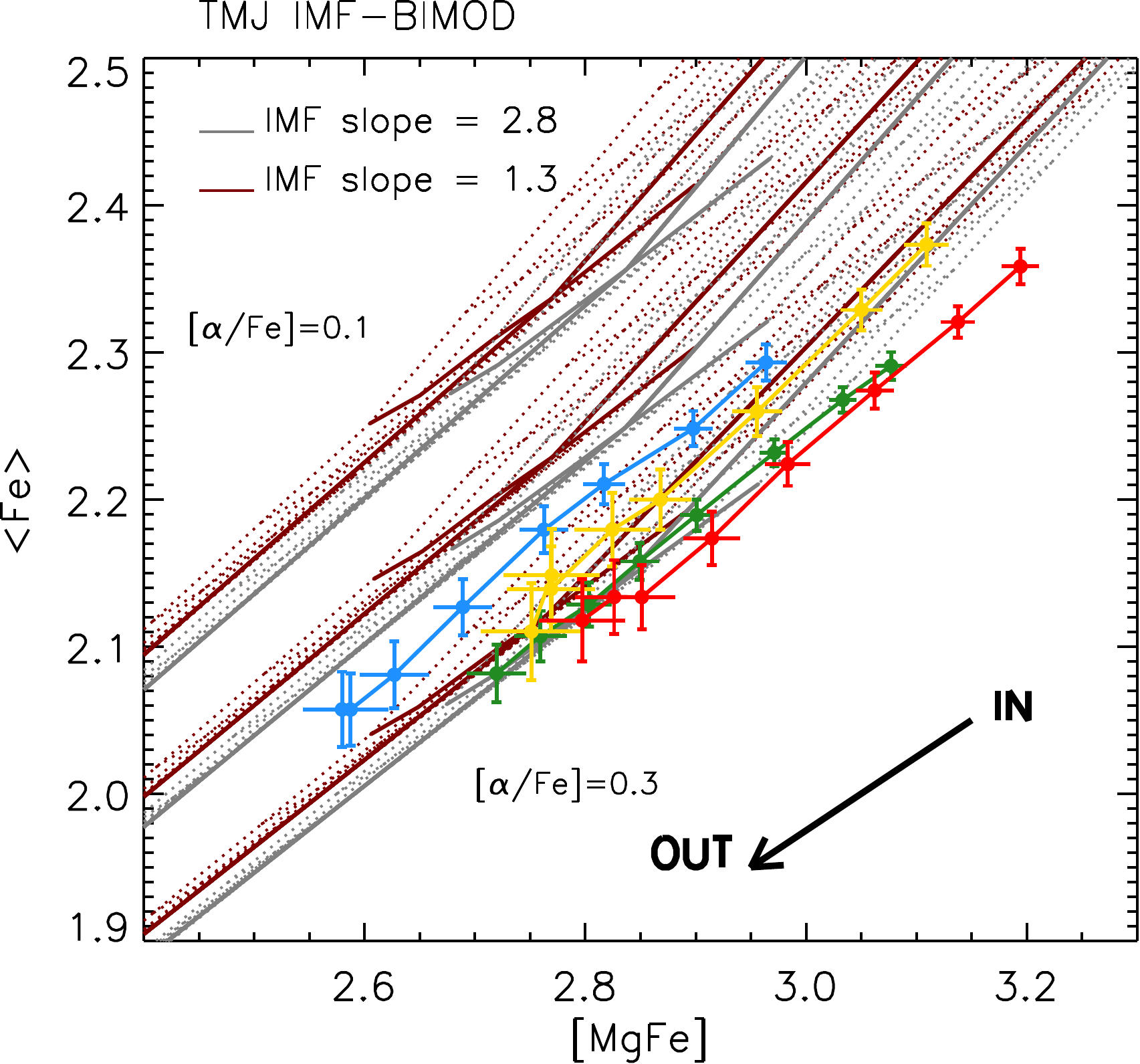}
  \includegraphics[width=0.32\linewidth]{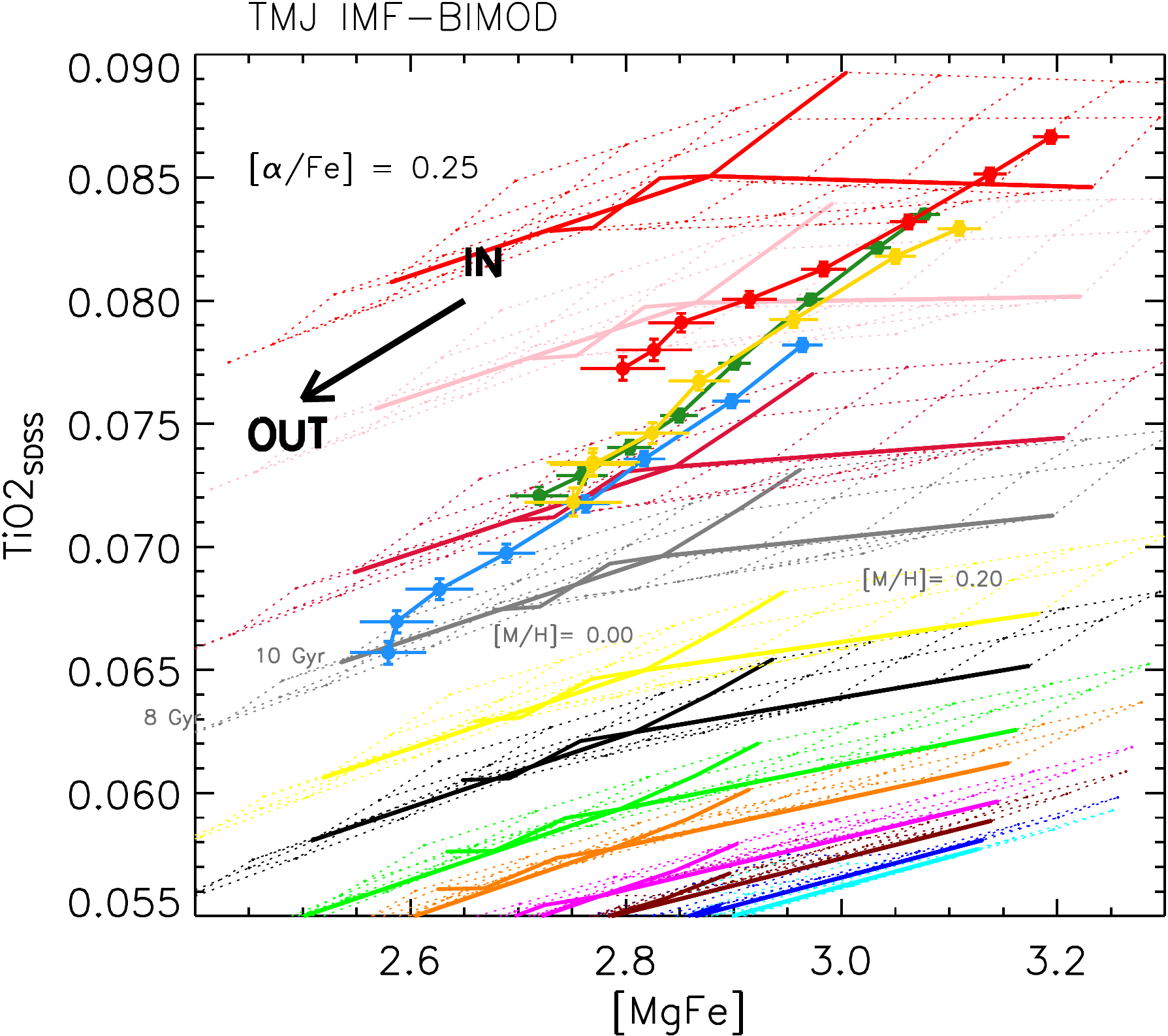}
  \caption{Lick-index diagnostic diagrams with TMJ-model grids obtained by extending TMJ-Kroupa to other BiModal IMFs using the MILES-Padova BiModal IMF models presented in the main text.}
  \label{TMJcheat}
\end{figure*}

\begin{figure}
  \centering
  \includegraphics[width=0.49\linewidth]{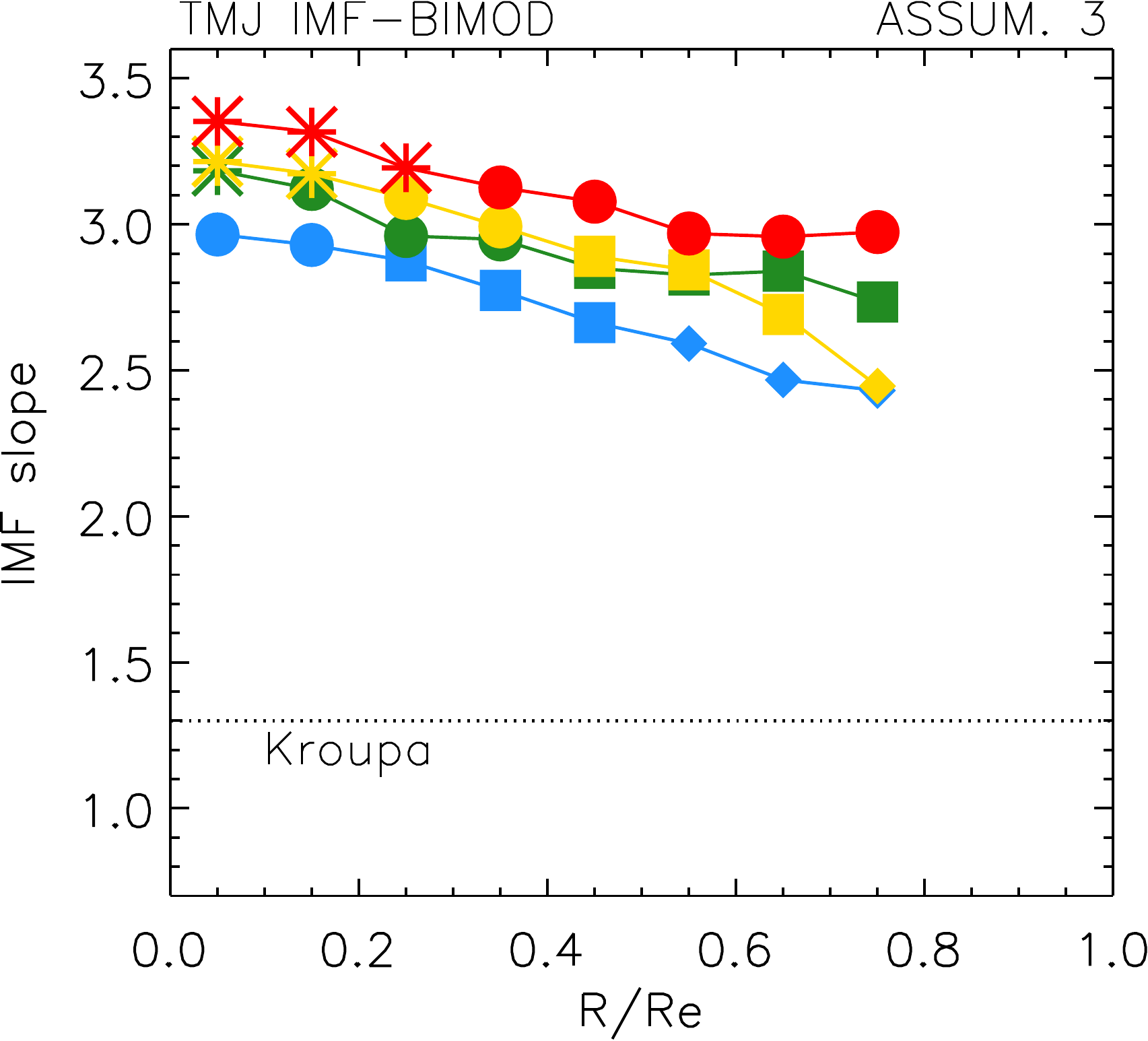}
  \includegraphics[width=0.49\linewidth]{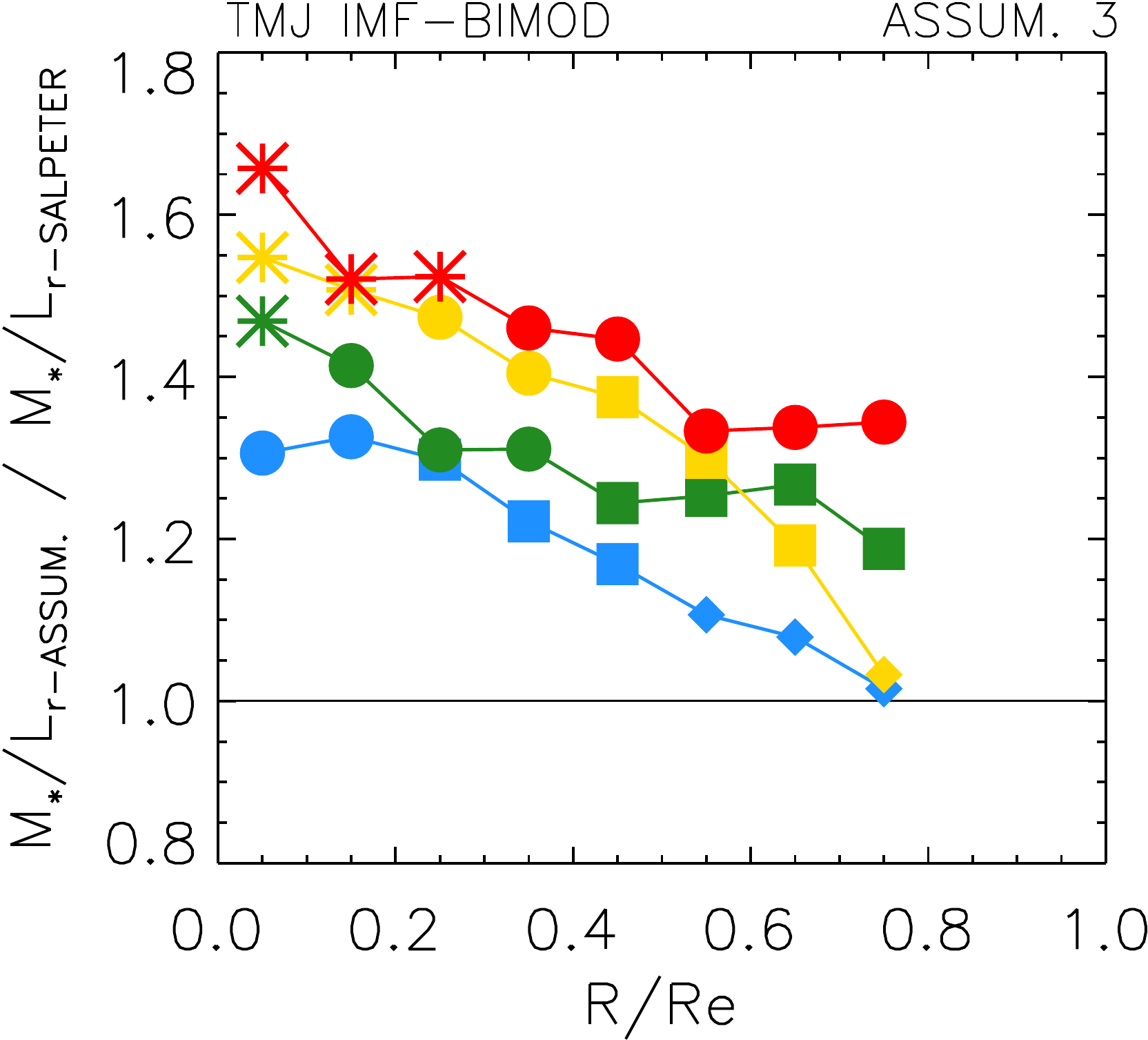}
  \caption{Inferred gradients of the IMF slope (left panel) and the ratio of $M_*/L_r$ to that for Salpeter (right panel) for our four bins assuming $\Delta_{\rm [X/Fe]}= 0.003$ but using the MILES-extended TMJ models (same as left panel of Figure \ref{IMFgrad}).}
  \label{IMFgradTMJ}
\end{figure}

\begin{figure}
  \centering
 \includegraphics[width=0.9\linewidth]{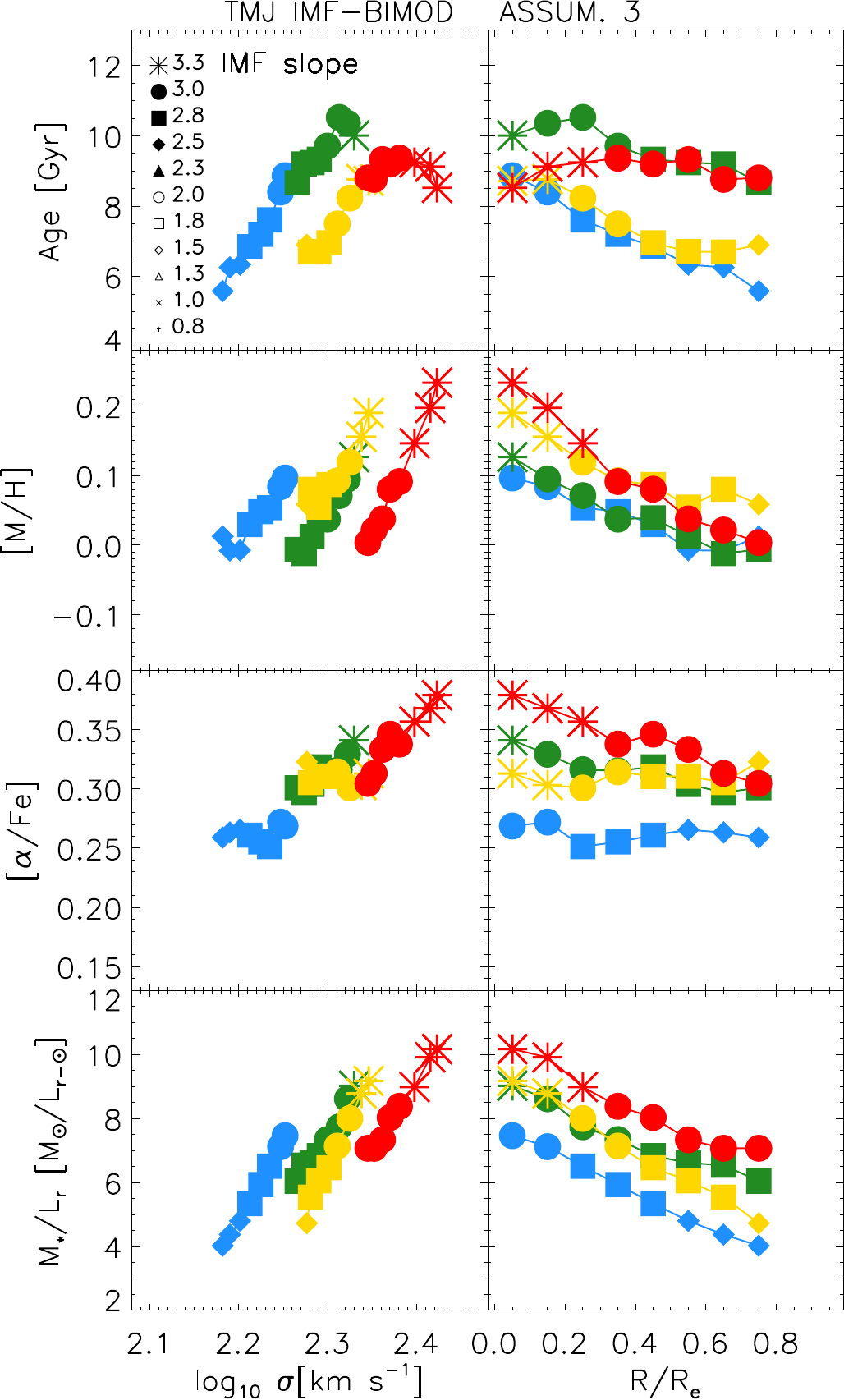}
  \caption{Inferred age, metallicity, [$\alpha$/Fe] and $M_*/L_r$ versus $\sigma$ (left panels) and radius (right panels) for our $\sigma_0$ and $L_r$ bins associated with ASSUMPTIONS~3, i.e. assuming that $\Delta_{\rm [X/Fe]}= 0.003$, for the MILES-extended-TMJ models.}
 \label{TMJgradients}
\end{figure}

The bottom panel shows TiO2$_{\rm SDSS}$-[MgFe].  Here, both TMJ model grids lie so far from the measurements that it is difficult to justify using the TMJ models without somehow incorporating additional IMFs.  If these TMJ-Kroupa grids agreed with the MILES-BiModal-1.3 grids shown in the main text, then there would be some justification for assuming that the MILES-based results are likely to be unbiased.  Unfortunately, this is not the case:  even though the Kroupa and MILES-BiModal-1.3 IMFs are the same, the TMJ-Kroupa and MILES-BiModal-1.3 model grids differ.  Not only are they offset in TiO2$_{\rm SDSS}$ strength, but the response of TiO2$_{\rm SDSS}$ strength to changes in age and metallicity is weaker for TMJ:  as a result, each TMJ grid spans a narrower range of TiO2$_{\rm SDSS}$ values.

For completeness, Figure~\ref{IMFgradUNI} shows some of the results obtained using the MILES-Padova UniModal (instead of BiModal) models: (left) the IMF slope gradients inferred by using a varible IMF (i.e. ASSUMPTION~3); (middle) the corresponding $M_*/L_r$ gradients; (right) the ratio of $M_*/L_r$ to that for Salpeter. The inferred properties are similar to those obtained from the BiModal models except that the $M_*/L_r$ tend to be lower, and the IMF slopes shallower (compare with Figures~\ref{allR} and \ref{ratio2salp2}).  At $R/R_e=0.8$, the IMFs are all shallower than Salpeter (left) but the $M_*/L_r$ values are not very different from Salpeter (right).  In addition, at $R/R_e\sim 0.4$ the IMFs in the two low-$L$ bins (blue and green) are different from the two at higher-$L$ (yellow and red), but the $M_*/L_r$ values are similar.  Clearly, as \cite{LaBarbera2013,Ferreras2013} have emphasized, gradients in IMF slope alone are not good indicators of $M_*/L_r$ gradients.

\subsection{MILES-extended TMJ models}\label{sec:MilesTMJ}
In an attempt to extend the reach of the TMJ models, we have transferred the MILES IMF grids to TMJ as follows.  We find the offsets of each MILES grid point from its respective MILES-BiModal-1.3 point and apply these to the TMJ-Kroupa point.  We then scale the range covered by each grid by the same factor by which MILES and TMJ differ for Kroupa.  The two panels in Figure~\ref{TMJcheat} show the result, but we caution that because this procedure is not fully self consistent, the results which follow are suggestive only.  This is why they are not in the main text.  Comparison with right-hand panels of Figure~\ref{compare} shows that now the extended-TMJ models completely cover the H$_\beta$-[MgFe] and TiO2$_{\rm SDSS}$-[MgFe] measurements.  From these we can infer age, metallicity, [$\alpha$/Fe] and IMF-gradients (although we have used the various assumptions listed in Table~\ref{tab:assumptions} about how $\Delta_{\rm [X/Fe]}$ varies across the population, here we show the results from ASSUMPTION~3), and finally, $M_*/L_r$ gradients, shown in Figures \ref{IMFgradTMJ}--\ref{TMJgradients}. (Of course, the $M_*/L_r$ values actually come from MILES, but, at least for the Kroupa and Salpeter IMFs, they are similar to TMJ.)

\begin{figure*}
  \centering
  \includegraphics[width=0.3\linewidth]{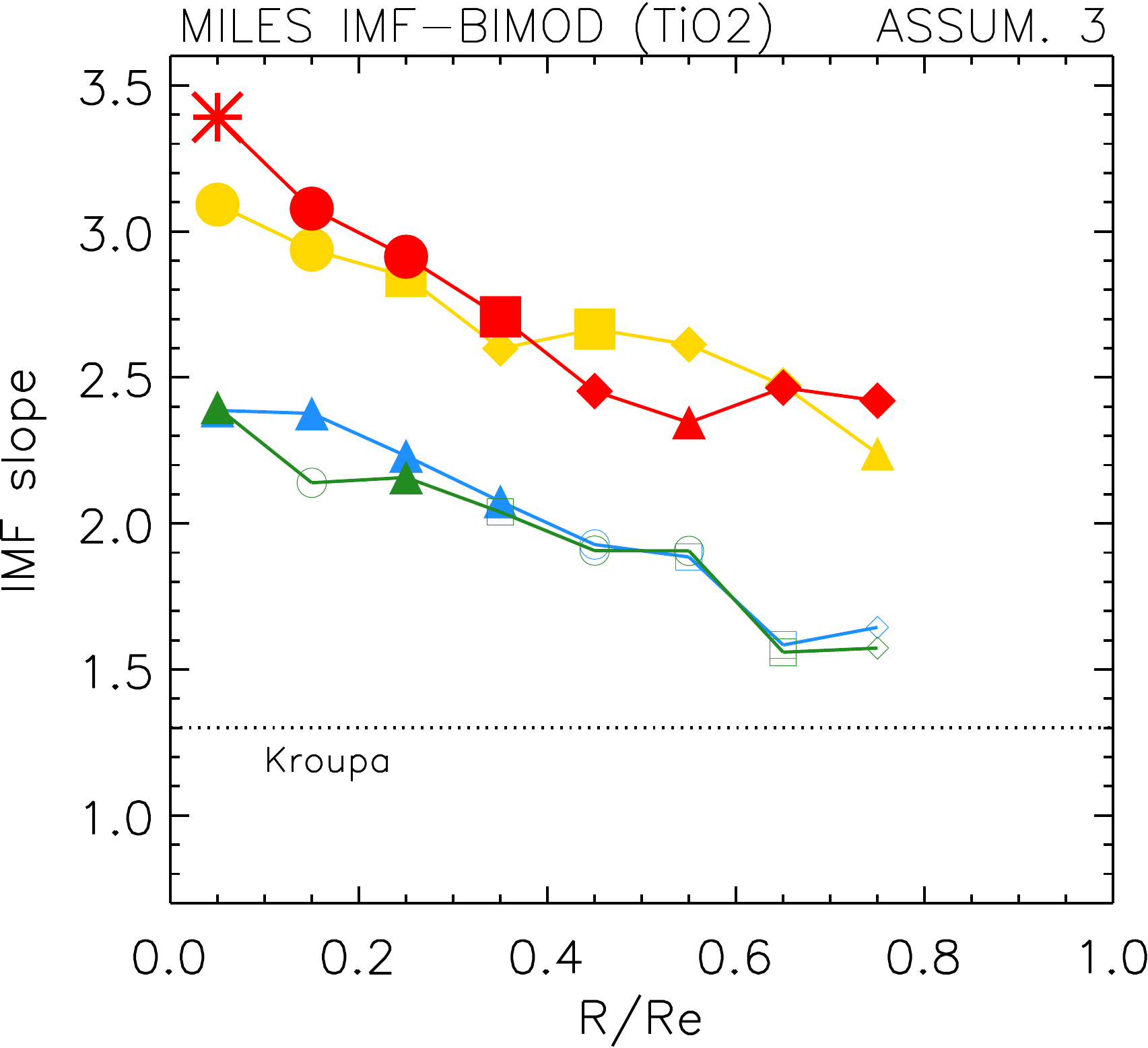}
  \includegraphics[width=0.3\linewidth]{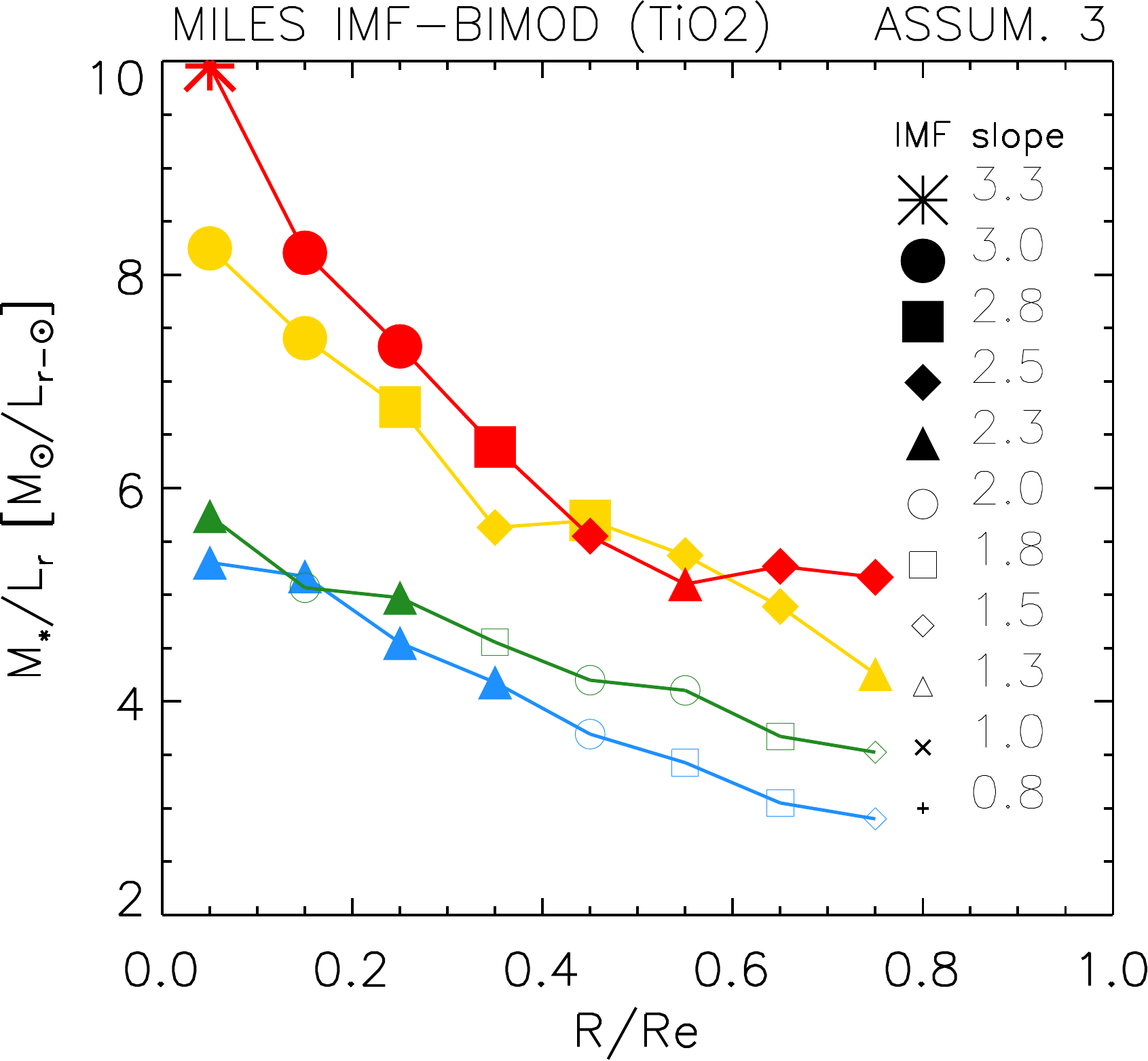}
  \includegraphics[width=0.3\linewidth]{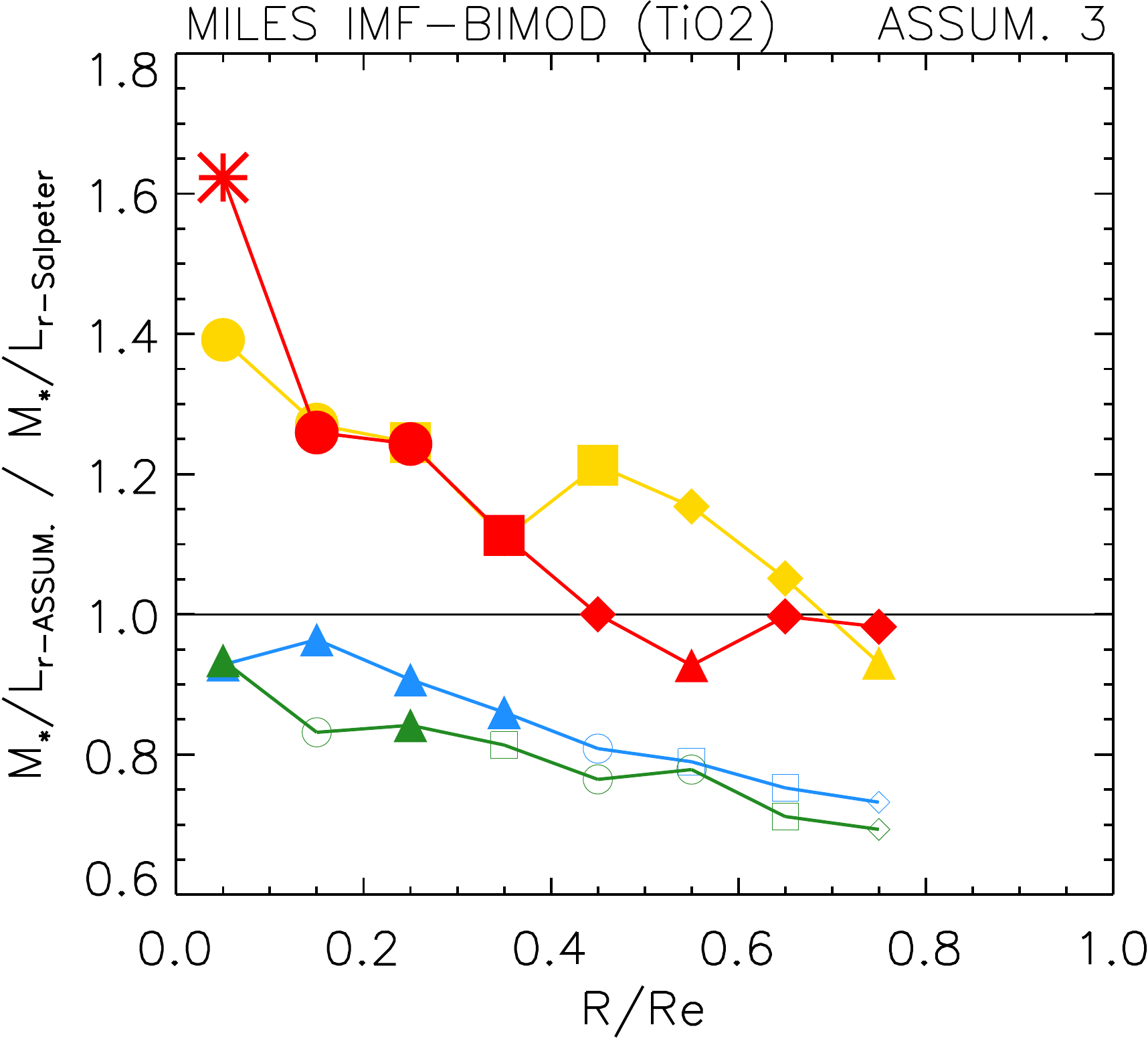}
  \caption{Left: the IMF slope gradients (same as in Figure \ref{IMFgrad}) but using TiO2 (instead of TiO2$_{\rm SDSS}$) as the IMF sensitive index. Middle: $M_*/L_r$ gradients from ASSUMPTION 3 using TiO2 (instead of TiO2$_{\rm SDSS}$). Right: the ratio of $M_*/L_r$ inferred by using TiO2 to that for Salpeter.}
  \label{IMFgradTiO2}
\end{figure*}

\begin{figure*}
  \centering
  \includegraphics[width=0.3\linewidth]{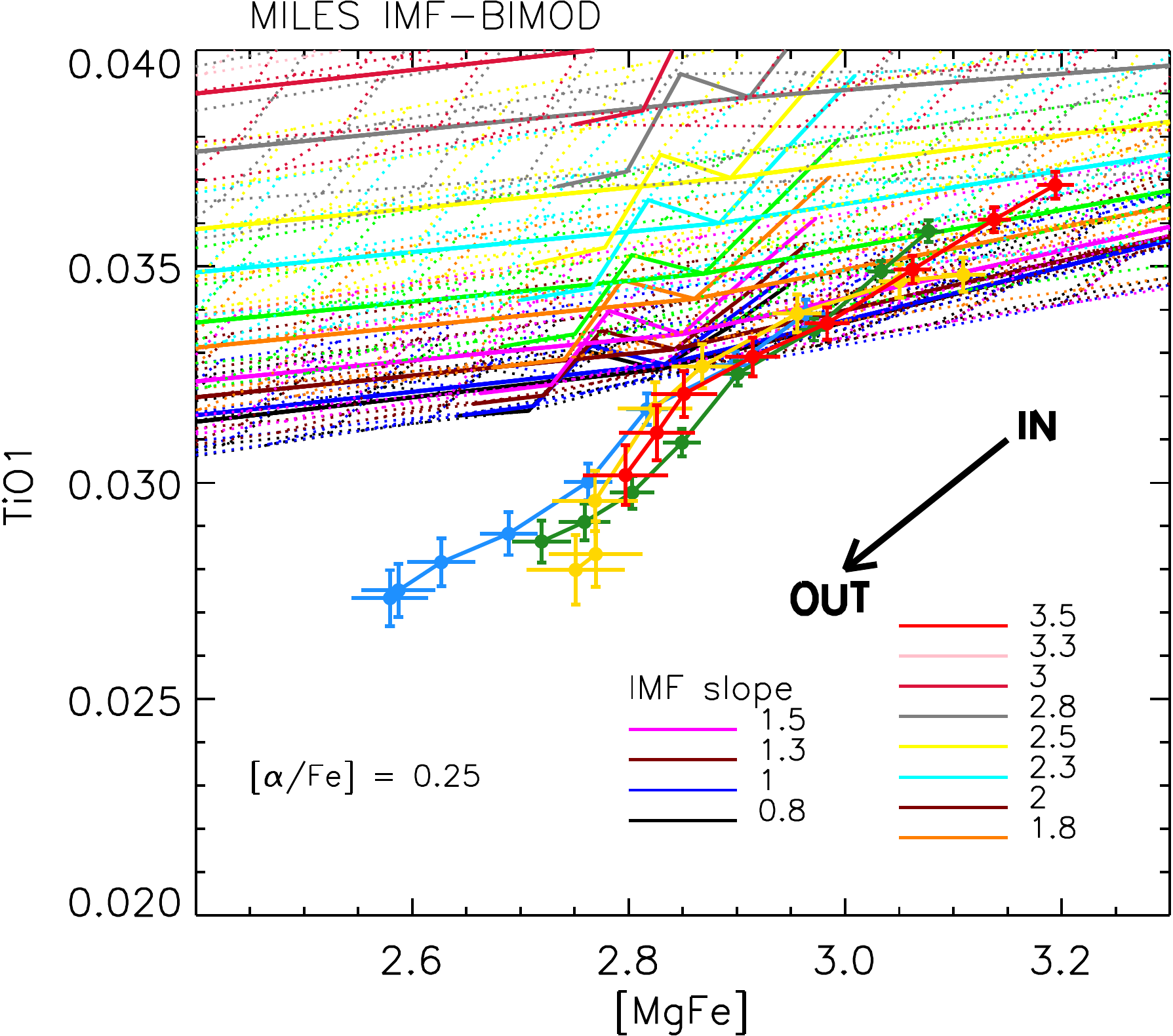}
  \includegraphics[width=0.3\linewidth]{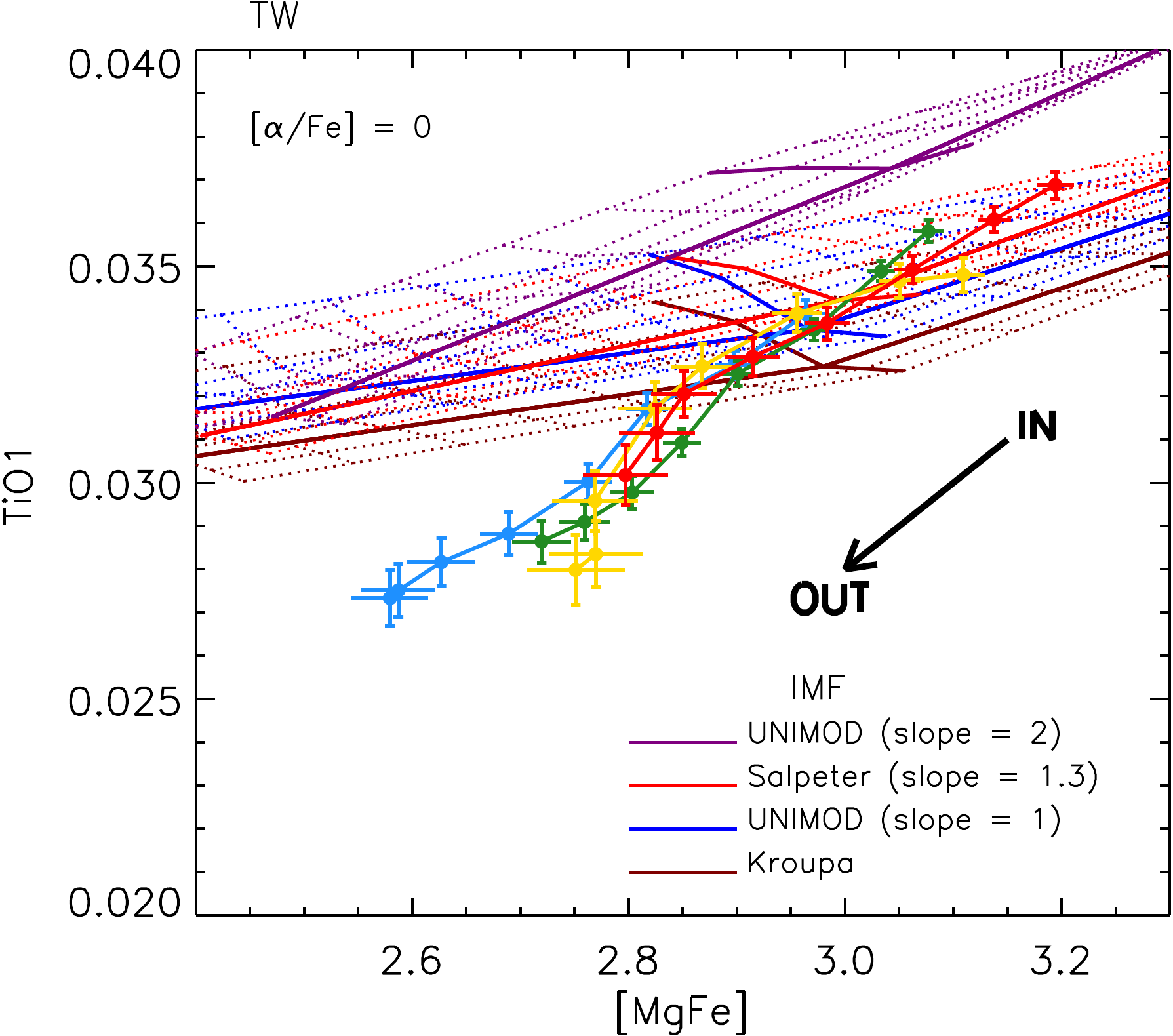}
  \includegraphics[width=0.3\linewidth]{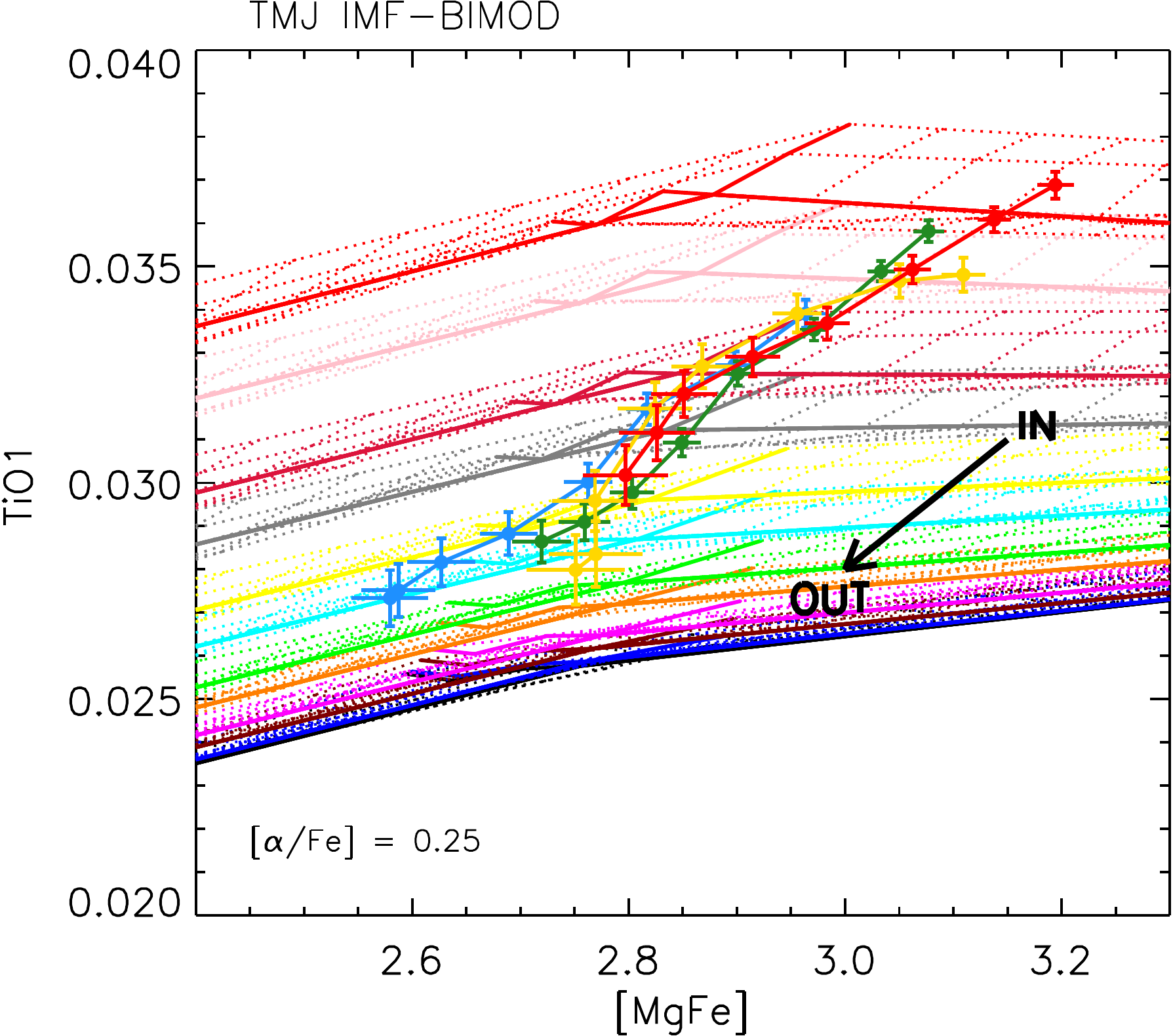}
  \caption{TiO1-[MgFe] index diagram, showing gradients in line strength in each bin (thick bold arrow indicates the direction of increasing galactocentric distance), and differences between bins.  Dotted lines show age-metallicity grids for [$\alpha$/Fe]=0.25 and a range of IMF slopes, to illustrate how this diagram can be used to discriminate between IMFs.  Left panel uses the same MILES-Padova models shown in the main text; middle panel shows the TW models; right panel shows the TMJ models, extended to other IMFs using the method described in Appendix~\ref{sec:MilesTMJ}.}
  \label{compareTiO1}
\end{figure*}

The biggest qualitative difference with respect to the MILES models is [$\alpha$/Fe]:  Here the gradients are much stronger;  [$\alpha$/Fe] correlates tightly with $\sigma$, both within a galaxy and across the population. In addition, the inferred ages are younger, and the $M_*/L_r$ values and IMF slopes are significantly larger than the MILES-based ones presented in the main text.  As we note in the main text, the larger $M_*/L_r$ values would not be problematic if these models did not produce strong gradients (Figure~\ref{TMJgradients}), the effect of which is to decrease the associated $M_{\rm dyn}$.  As a result, they have $M_* > M_{\rm dyn}$, which is unreasonable.  This is another reason why we did not present these extended-TMJ models in the main text.

\subsection{Other TiO indicators}\label{sec:TiO1}
The main text used TiO2$_{\rm SDSS}$ \cite[defined in][]{LaBarbera2013} as an IMF indicator.  This is a modified version of the TiO2 index of \cite{Trager1998}, in which the red sideband is re-defined to minimize deviation between the models and data (see Appendix A of \citealt{LaBarbera2013} for a more detailed discussion).  Although recent IMF studies focus on TiO2$_{\rm SDSS}$ instead of TiO2 (e.g., \citealt{MN2015,TW2017}), we have repeated all the analyses in the main text using TiO2 and TiO1 as well.

\medskip
\noindent{\bf TiO2:}  The TiO2 analysis returns higher $\Delta_{\rm [X/Fe]}$ and higher IMF slopes (left panel in Figure \ref{IMFgradTiO2}), which translate into larger $M_*/L_r$ values (middle and left panels in Figure~\ref{IMFgradTiO2}), with overall slightly larger $M_*/L_r$ gradients.

\medskip
\noindent{\bf TiO1:}  The TiO1-based analysis is more complicated.
Figure~\ref{compareTiO1} shows the TiO1-[MgFe] grids for the same MILES-Padova models shown in the main text, for the TW models, and for the TMJ models discussed in the previous section (recall that, for these models, only the Kroupa and Salpeter IMFs are really from TMJ; all the other IMFs were obtained by shifting and scaling the MILES-models).
The most striking point is that the MILES-Padova models, which cover the observed range of TiO2$_{\rm SDSS}$, all lie well-above most of the measurements.  The discrepancy cannot be removed by appealing to [X/Fe] enhancements: for the same IMF to match TiO1 and TiO2$_{\rm SDSS}$ requires different [X/Fe] enhancements, which is unphysical.  This is also true of the TW models.  In contrast, the TMJ models fare much better.  This is because the starting point for these models -- the Kroupa and Salpeter models -- lie below the TiO1 measurements, just as they do for TiO2$_{\rm SDSS}$ (compare Figure~\ref{compare}).  The net result is that TiO1 implies slightly lower IMF slopes but otherwise the same trends as the TiO2$_{\rm SDSS}$ trends presented earlier.



\bsp	
\label{lastpage}
\end{document}